\documentclass[nonblindrev, copyedit]{informs3}       % for review, not blinded
\makeatletter
\def\footnoterule{\kern-3\p@
  \hrule \@width 2in \kern 2.6\p@} % the \hrule is .4pt high
\makeatother
\usepackage{natbib,graphicx}

\usepackage[colorlinks,citecolor=blue,urlcolor=blue,bookmarks=false,hypertexnames=true]{hyperref} 
\usepackage{bbm}
\usepackage{url}
\usepackage{xpatch}
\usepackage[table, svgnames, dvipsnames]{xcolor}
\usepackage{color, colortbl}
\usepackage{stackengine}
\definecolor{Gray}{gray}{0.8}
\usepackage{lipsum}
\usepackage{array}
\newcolumntype{L}[1]{>{\raggedright\let\newline\\\arraybackslash\hspace{0pt}}m{#1}}
\newcolumntype{C}[1]{>{\centering\let\newline\\\arraybackslash\hspace{0pt}}m{#1}}
\newcolumntype{R}[1]{>{\raggedleft\let\newline\\\arraybackslash\hspace{0pt}}m{#1}}

\usepackage{caption}
\usepackage{pdfpages}
\usepackage{algorithm}
\usepackage{csquotes}
\usepackage{dsfont}
\usepackage{subfig}
\usepackage{amssymb}
\usepackage{epstopdf}
\usepackage{multirow}
\usepackage{lscape}
\usepackage{makecell}
\usepackage{tabularx}
\usepackage{multicol}
\usepackage{fancyhdr}
\usepackage[flushleft]{threeparttable}
\usepackage{setspace}
\usepackage{array, booktabs}
\usepackage{color}
\usepackage{tikz}

\usetikzlibrary{arrows,automata}

\usepackage[margin=1in]{geometry}

\usepackage{setspace}

\usepackage[normalem]{ulem}

\usepackage{float}

\usetikzlibrary{calc,arrows,matrix,positioning}

\usepackage{multirow}

\usepackage{algpseudocode}
\usepackage{adjustbox}

\algnewcommand{\IIf}[1]{\State\algorithmicif\ #1\ \algorithmicthen}
\algnewcommand{\EndIIf}{\unskip\ \algorithmicend\ \algorithmicif}

\bibpunct[, ]{(}{)}{,}{a}{}{,}

\def\bibfont{\small}

\def\bibhang{24pt}

\TheoremsNumberedThrough

\EquationsNumberedThrough

\begin{document}
\RUNAUTHOR{D. Pirayesh Neghab }
\RUNTITLE{Title}

\TITLE{Explaining Exchange Rate Forecasts with Macroeconomic Fundamentals Using Interpretive Machine Learning}
%\date{}





\ARTICLEAUTHORS{

\AUTHOR  {Davood Pirayesh Neghab$^{a}$, Mucahit Cevik$^{a,}$\footnote{Corresponding author can be reached at:
{mcevik@torontomu.ca}}, M. I. M. Wahab$^{a}$} \vspace{2mm} \AFF{$^{a}$Department of Mechanical and Industrial Engineering,
Toronto Metropolitan University, Toronto, Canada 
}

% \AUTHOR  {} \AFF{ 
% }
% \AUTHOR  {} \AFF{
% }
}
	
%Aim, Background, Methodology, Results, Conclusion.

\ABSTRACT{
%Background
The complexity and ambiguity of financial and economic systems, along with frequent changes in the economic environment, have made it difficult to make precise predictions that are supported by theory-consistent explanations.
Interpreting the prediction models used for forecasting important macroeconomic indicators is highly valuable for understanding relations among different factors, increasing trust towards the prediction models, and making predictions more actionable.
%Aim 
In this study, we develop a fundamental-based model for the Canadian–U.S. dollar exchange rate within an interpretative framework. 
%Methodology
We propose a comprehensive approach using machine learning to predict the exchange rate and employ interpretability methods to accurately analyze the relationships among macroeconomic variables. 
Moreover, we implement an ablation study based on the output of the interpretations to improve the predictive accuracy of the models. 
%Results
Our empirical results show that crude oil, as Canada's main commodity export, is the leading factor that determines the exchange rate dynamics with time-varying effects. 
The changes in the sign and magnitude of the contributions of crude oil to the exchange rate are consistent with significant events in the commodity and energy markets and the evolution of the crude oil trend in Canada. 
Gold and the TSX stock index are found to be the second and third most important variables that influence the exchange rate.
%Conclusion
Accordingly, this analysis provides trustworthy and practical insights for policymakers and economists and accurate knowledge about the predictive model's decisions, which are supported by theoretical considerations. 
}

	
\maketitle

\section{Introduction}
%%%%%%%%%%%%%%%%%%%%%%%%%%%%%%%%%%%%%%%%%%%%	
\makeatletter
\let\normalsize\relax
\let\@currsize\normalsize
\makeatother

% \begin{figure}[t]
%     \centering 
%     \subfloat{\includegraphics[width=.85\textwidth]{Figure/Oil_Events.png}}\\
%     \caption{Political, environmental, financial, and technological developments that have a significant impact on oil prices between 1990 and 2020 (sourced from \url{www.cer-rec.gc.ca}).}
% \end{figure}

\begin{figure}[!ht]
    \centering \subfloat{\includegraphics[width=.9\textwidth]{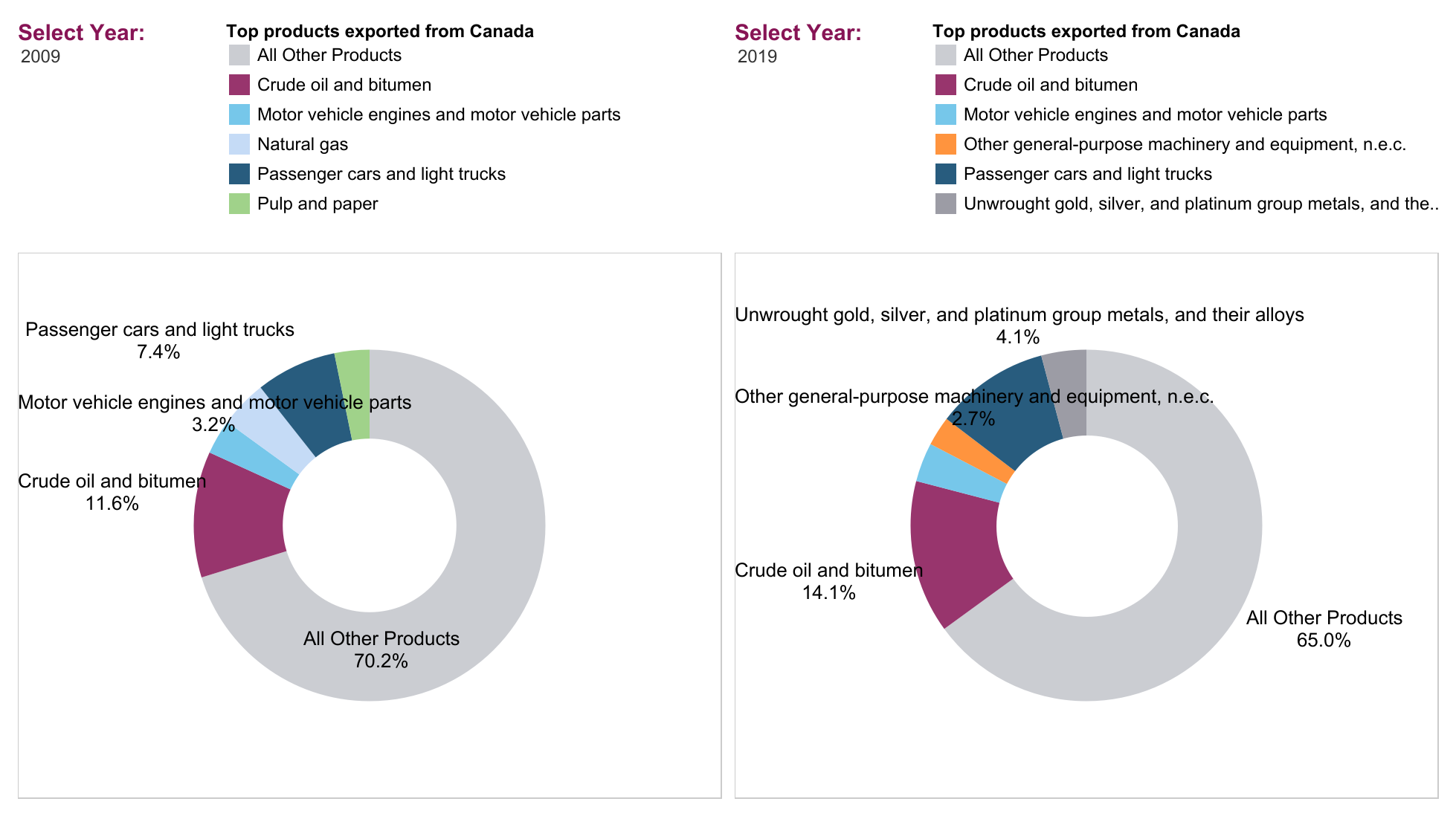}}\\
    \caption{Top exported products from Canada in 2009 vs. 2019 (Source: \url{www.cer-rec.gc.ca}).}\label{fig:oil2009-2019}
\end{figure}

The exchange rate is a key factor in economic growth and a useful measure of uncertainty about the economic environment~\citep{gala2007real}, and it quickly adjusts to macroeconomic trends and monetary policies~\citep{rosa2011high}. 
In response to the need for high-quality forecasts for exchange rates, over the last two decades, a significant amount of work has established fundamental-based approaches to competently modelling exchange rate dynamics~\citep{kurita2022canadian}. 
In this regard, based on the importance of a wide range of variables that are well studied and documented, commodity prices play a major role in determining the exchange rates for economies where primary commodities constitute a considerable share of exports. 
Among the commodity price/exchange rate pairings, the crude oil price and the Canadian-U.S. dollar exchange rate have gained more traction due to the evolution of Canada's oil export and its anticipated consequential effect on economic development. 
Crude oil exports from Canada have increased in value significantly in recent years, accounting for 14.1\% of total exports in 2019 (see Figure~\ref{fig:oil2009-2019}). 
Following this evolution, as of 2021, Canada has become the largest supplier of crude oil to the U.S., constituting 61\% of the imported gross crude oil to the U.S.~\citep{eia}).  

Empirical evidence shows that there exist dynamic relationships among commodity prices and exchange rates~\citep{buetzer2012global}.
% According to  empirical evidence, commodity prices impact the exchange rate with dynamic relationships~\citep{buetzer2012global}. 
Consequently, several studies have investigated the nexus between crude oil prices and exchange rates and established time-varying relationships~\citep{chen2007oil, ferraro2015can, han2018oil}. 
However, the dynamics identified by traditional methods in these investigations are not directly explained by standard theoretical considerations as a result of being potentially driven by other macroeconomic
variables~\citep{beckmann2020relationship}. 

To develop econometrically well-specified and interpretable exchange rate models, several major considerations need to be addressed concurrently. (1)~\textit{Nonlinearity}: while the majority of the existing literature employs formal tests of linearity between crude oil prices and exchange rates~\citep{chen2007oil, reboredo2012modelling, atems2015exchange}, %and some limited classical nonlinear modelling~\citep{akram2004oil, de2016nonlinear, xu2019dynamic}.
there is strong evidence that the nature of the relationship among macroeconomic variables is nonlinear due to the irregularities of the movements of economic indicators~\citep{park2022interpretable}. (2)~\textit{Multicollinearity}: most of the existing models that relate exchange rates to a fundamental variable are univariate and exclude other variables from the analysis, though such variables directly or indirectly influence the relationship~\citep{beckmann2020relationship}. 
Therefore, multicollinearity exists as a consequence of including multiple explanatory variables in multivariate models. 
(3)~\textit{Noise}:
in rapidly growing commodity markets, inevitable financialization adds additional noise to the time series as a result of boosting speculative behaviour in these markets, such as crude oil, leading to a higher noise-to-pattern ratio~\citep{han2018oil}. (4)~\textit{Time-variation}: empirical evidence suggests that the effects of fundamentals on exchange rates are time-varying, implying that the exchange rates respond asymmetrically to different time-specific changes and shocks in fundamentals~\citep{kumar2019asymmetric}. (5)~\textit{Backward-forward looking analysis}: disentangling in-sample and out-of-sample empirical analyses can be an issue; that is, past relationships that are successfully explained by a model do not always represent the most accurate theoretical underpinnings for future predictions~\citep{zhang2013links}. Consequently, it is critical to contemporaneously address these issues to reach a reliable and proper modelling of exchange rates. %which are also outlined by~\citet{beckmann2020relationship}.

\noindent \paragraph{\textbf{Motivation}}
In the exchange rate modelling, two prominent research areas exist that partially address the aforementioned issues: the first research stream mainly investigates the impact of commodity prices on exchange rates using classic linear models, such as ordinary least squares (OLS), and some limited nonlinear models that are unable to suggest a systematic prediction ability in the presence of multiple predictors~\citep{ferraro2015can, kohlscheen2017walk, beckmann2022we}. 
Some of these studies also focused on dependency and time-varying relationships using traditional techniques, such as correlation and copulas with restrictive assumptions, allowing for easier, yet insufficient interpretations~\citep{reboredo2012modelling}. 
Another stream of recent research utilizes machine learning to address the challenges in forecasting, but these studies are primarily focused on prediction and they typically do not offer an analysis of how fundamentals impact exchange rates~\citep{islam2020foreign, abedin2021deep}. 
In this regard, we aim to address the issues mentioned above by exploiting the predictive power of machine learning along with recent advances in interpretability methods. 
New interpretability methods have been found to be highly effective and are well-suited to trace the relationships between external covariates and target variables. 
As such, in a retrospective-prospective empirical study on historical observations and future predictions, we shed light on the contribution of crude oil prices, coupled with other macroeconomic fundamentals, to exchange rate dynamics.

\begin{figure}[!ht]
    \centering
    \includegraphics[width=.65\textwidth]{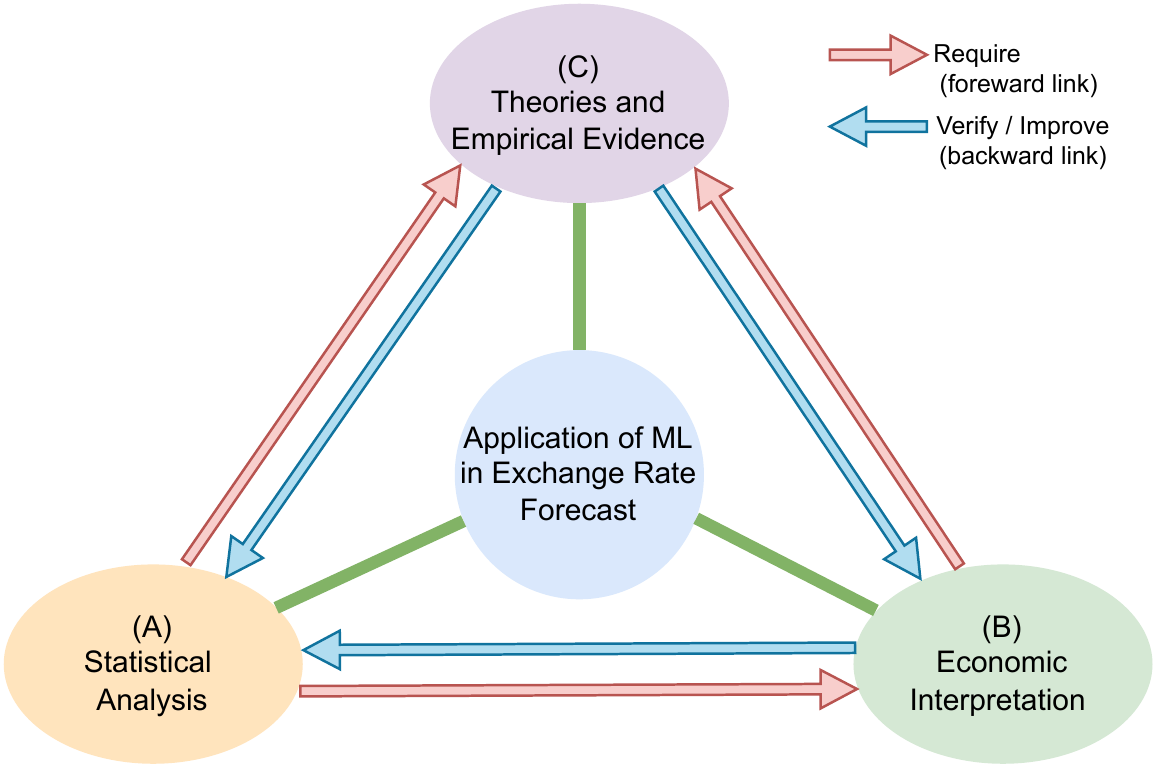}\\
    \caption{Machine learning application framework for exchange rate prediction.} \label{fig:framework}
\end{figure}

\paragraph{\textbf{Research objectives}}
In this study, our main focus is on the Canadian-U.S. dollar exchange rate and the factors influencing it. 
Because the performance of models and the insights gained from interpretations are highly dependent on the exchange rates, training/testing period, and forecast horizon~\citep{rossi2013exchange}, we aim to provide a comprehensive and general approach for applying machine learning in exchange rate modelling with three major aspects: (A) Statistical Analysis, (B) Economic Interpretation, and (C) Theories and Empirical Evidence. 
Figure~\ref{fig:framework} shows the suggested framework that is employed in this study for theory-consistent and fundamental-based exchange rate modelling with an interpretative approach. 

We consider an extensive set of machine learning methods to model exchange rate dynamics using the macroeconomic variables for statistical analysis in step~A. 
Unlike previous studies that have mainly focused on developing deep learning-based forecasting models and comparing them with similar methods, we analyze the performance of a well-diversified set of methods that includes new linear, tree-based, and deep learning models. Following step~A, we interpret the outcome of the prediction models in step~B.
% and extend the existing studies. 
We use the recent advanced interpretability methods to gain knowledge about the model's decision, which in turn increases the transparency of machine learning, minimizes inaccurate decisions, and promotes practitioner trust to the proper level. 
Lastly, in step~C, we revisit the recent evidence and theories on the relationships among exchange rates and other variables with an emphasis on crude oil prices in Canada as a major crude oil exporter.

As indicated by backward blue links in Figure~\ref{fig:framework}, interpretations are corroborated by known theories and existing causal inferences (e.g., from C to B). 
This procedure confirms the accuracy of the economic interpretations through a retrospective-prospective practice with regard to country-, time-, and sample-specific forecasting. 
Accordingly, to support the effectiveness of the interpretations, we focus on relationships that are explained by the in-sample best-fitted model and on some instances of predictions that tie in with important events in the oil market and the evolution of trends in Canadian crude oil exports. 
Regarding the link from B to A, the outcomes of the interpretations in B help us detect the most influential input variables and achieve competitive performance in less complex models by considering only the important contributors to the exchange rate. 
To this end, we conduct an ablation study to examine the performance of the models by eliminating the less important variables and to understand the statistical contribution of the important variables to the models. 
The other model improvement is conducted through the link from C to A, in which we take into account the theoretical and empirical evidence from the literature on autoregressive models. 
Hence, we incorporate the lags of the exchange rate into the set of fundamental input variables and leverage their predictive power to increase the statistical performance of models in practice.
% \footnote{An example of A to C is related to directional predictions using new methods and the comparison of the results with early theories such as the random walk theory. In this case, the models attempt to achieve an accuracy of more than 50\% in the exchange rate trend classification problem.}.

\paragraph{\textbf{Contribution}} We summarize the contributions of our study as follows:
\begin{itemize}
    \item To the best of our knowledge, this is the first study that applies machine learning to exchange rate forecasts using macroeconomic variables with a special focus on the impact of crude oil prices on Canadian exchange rates. Furthermore, it uses interpretability methods, including feature importance and SHapley Additive exPlanations (SHAP), to gain a cohesive understanding of models' decisions. The suggested theory-consistent and interpretative machine learning approach provides a comprehensive framework for further applications of machine learning in finance and economics.  
    
    \item In financial and economic fundamental-based modelling, we outline several issues that are inherently contemporaneous when analyzing the relationships, namely, \textit{nonlinearity}, \textit{multicollinearity},  \textit{noise}, \textit{time-variation}, and \textit{backward-forward looking analysis}. %We address the first three  issues by choosing a set of competitive models to achieve an adequate model with minimum forecasting errors, and use  interpretability methods such as feature importance and SHAP for other issues.
    By effectively addressing these issues, we enhance the applicability of machine learning in this domain, which is mainly used for prediction in previous studies.
    
    \item We conduct a detailed numerical study that provides important insights into the foreign exchange rate prediction.
    % From our empirical results, we obtain the following implications: 
    For instance, we find that tree-based models outperform other methods in terms of prediction performance. However, these models are largely ignored in recent similar studies that focus on deep learning-based methods. 
    During Covid-19, relationships are successfully explained by new linear models. 
    In regards to the interpretative viewpoint, we extend the analysis from traditional univariate models to multivariate models that include the effect of other macroeconomic fundamentals. 
    Nonetheless, the crude oil price is a consistent, albeit time-varying, important variable. 
    We also find that gold is another commodity that impacts the Canadian exchange rate as a result of recently increased exports from Canada.

\end{itemize}

\paragraph{\textbf{Organization of the paper}} The remainder of the paper is organized as follows. Section~\ref{sec:litreview} reviews the related literature on exchange rate prediction models. 
Section~\ref{sec:models} presents the details of the selected machine learning models followed by interpretability methods. % in Section~\ref{sec:interpret}. 
Section~\ref{sec:empirical} explains the empirical framework and outlines the data, prediction process, performance metrics, and results. Finally, Section~\ref{sec:conclusion} concludes the paper.

%%%%%%%%%%%%%%%%%%%%%%%%%%%%%%%%%%%%%%%%%%%%	
%%%%%%%%%%%%%%%%%%%%%%%%%%%%%%%%%%%%%%%%%%%%	
\section{Literature Review}\label{sec:litreview}
%%%%%%%%%%%%%%%%%%%%%%%%%%%%%%%%%%%%%%%%%%%%	
%%%%%%%%%%%%%%%%%%%%%%%%%%%%%%%%%%%%%%%%%%%%
Forecasting exchange rates has been challenging task mainly due to irregularities in the movement patterns of economic indicators~\citep{park2022interpretable}. 
In this regard, a long line of research has detailed the significance of the historical behaviour of exchange rates in economic performance and econometric analysis. 
These studies mostly employ statistical models that predict future values based on past  observations~\citep{epaphra2016modeling}. 
In particular, autoregressive models and their extensions have been frequently used to model exchange rate prediction. 
These techniques typically require stationary time series, and hence they mostly focus on exchange rate returns and volatility~\citep{pong2004forecasting, christian2013forecasting, barunik2016modeling, bartsch2019economic}. 

% typically focus on modelling exchange rate  returns and volatility when using these techniques since they need stationary time series~\citep{pong2004forecasting, christian2013forecasting, barunik2016modeling, bartsch2019economic}. 

% Some researchers demonstrate that traditional econometric and time series techniques are not reliable in practice due to the unrealistic assumptions that apply to these methods~\citep{kilian2003so}. 
% According to this viewpoint, economic models attempt to link a particular exchange rate to other variables to capture  dynamics that complicate the forecast~\citep{engel2015factor}.
% Consequently, in several studies, researchers have considered methods that employ information about underlying fundamentals and macroeconomic variables~\citep{cheung2005empirical, stock2006forecasting}. 

% \subsection{Macroeconomic Fundamentals}
% \subsection{Traditional methods}
Several studies have proposed methods to relate exchange rates to macroeconomic fundamentals since 1970~\citep{cheung2005empirical}. 
They posit that the expectations of future exchange rates depend on variables, e.g., money supplies, real incomes, trade balances, inflation rates, and interest rates~\citep{taylor1993discretion, molodtsova2009out, lansing2017explaining, you2020forecasting}. 
A rich body of literature expands on these theoretical models by incorporating further macroeconomic variables that have some predictive power for exchange rates, such as GDP~\citep{antwi2020effect}, consumer price~\citep{evans2012exchange}, productivity rate~\citep{wright2008bayesian}, unemployment rate~\citep{bacchetta2010can},  asset rate~\citep{cheung2005exchange}, net foreign assets~\citep{della2012predictive}, and the stock market return~\citep{mun2012joint}. 
We refer the reader to \citep{rossi2013exchange} for a more detailed review of economic predictors of exchange rates that have been used in the literature.

A stream of literature examines exchange rate behaviour in economies, where primary commodities account for a significant portion of exports~\citep{chen2003commodity, issa2008turning, cayen2010drives, djemo2021predicting}. 
In early studies,~\citet{amano1998exchange} and \citet{chen2007oil} report a robust relationship between the price of crude oil and the exchange rate in different countries and suggest that crude oil prices have significant forecasting power. 
\citet{beckmann2020relationship} review the theoretical and empirical research on this relationship. 
In this regard, a few studies focus on the Canadian-U.S. dollar exchange rate forecast using crude oil as Canada's major commodity export~\citep{chen2002exchange}. 
\citet{ferraro2015can} empirically show that crude oil can be used as a strong predictor for the Canadian exchange rate at a daily frequency. 
They demonstrate similar results for Norway and Australia and other commodity prices/exchange rate pairs, such as gold prices/the South African rand-U.S. dollar exchange rate and copper prices/the Chilean peso-U.S. dollar exchange rate. 

Several studies either explain or forecast the exchange rate using linear models and techniques~\citep{zhang2016exchange, beckmann2016forecasting}. However, the nature of the relationship is typically nonlinear and time-varying~\citep{akram2004oil, ferraro2015can}. 
Furthermore, financial time series transmit information with a high noise-to-pattern ratio which makes forecasting a complex task. 
Additionally, using several explanatory variables in forecasting lowers the predictive ability of the model, and that is caused by the multicollinearity problem~\citep{park2022interpretable}. 
In this respect, with the recent advances in computational power, machine learning methods have gained traction and have been widely employed to effectively address the common concerns with conventional approaches in economic and financial forecasting modelling.

% \subsection{Machine Learning}
% Recent advancements in hardware and data storage have widened the applicability of new approaches such as machine learning and deep learning to the fields of economics and finance~\citep{henrique2019literature}.
Some hybrid approaches to forecasting exchange rates have been proposed, including support vector regression-based hybrid~\citep{ince2006hybrid}, genetic algorithm-based hybrid~\citep{sarangi2022forex}, neural networks-based hybrid~\citep{nag2002forecasting}, and fuzzy logic-based hybrid~\citep{hernandez2021using}. 
\citet{yilmaz2021should} propose a hybrid model based on autoregressive integrated moving average (ARIMA) and long short-term memory (LSTM) to predict the Canadian dollar, Australian dollar, and British pound against the U.S. dollar in monthly terms. 
Their results show that the suggested model outperforms the alternatives, including random walk, support vector machine, and recurrent neural networks (RNNs). 
\citet{pradeepkumar2018soft} provide a comprehensive review of exchange rate prediction methods that used soft computing hybrid techniques between 1998 and 2017. 

% Some recent studies entirely rely on new methods because they are able to efficiently detect trends, seasonality, redundant variables, and high uncertainty. 
Machine learning has been extensively used in recent studies for various exchange rate prediction tasks.
Some researchers develop and apply machine learning to classify the direction of exchange rate changes~\citep{giles2001noisy, qian2010foreign, galeshchuk2017deep}. 
Considering the profitability of trading strategies, some studies evaluate the directional forecasting accuracy from an investor perspective based on tree-based models~\citep{adegboye2021machine} as well as RNNs~\citep{dautel2020forex, das2022deep}.  
% Several studies have directly predicted exchange rates using machine learning and deep learning. 
\citet{liu2017foreign} propose a machine learning model based on convolutional neural networks (CNNs) for predicting three exchange rates: EUR/USD, USD/JPY, and GBP/USD. 
They show the superiority of the suggested model compared to other deep learning models in long-term forecasting. 
\citet{islam2020foreign} present a model that combines the gated recurrent unit (GRU) and LSTM to forecast the future closing prices of exchange rates. 
In short-term prediction tasks, their proposed model outperforms all other models for high-frequency data. 
\citet{abedin2021deep} propose an ensemble deep learning model and predict several exchange rates during the pre-Covid-19 and Covid-19 periods. 
They show that the performance of the suggested model strongly depends on the volatility of the foreign currency markets and varies during these periods across currencies. 
\citet{sezer2020financial} review the application of deep learning to financial time series forecasting, while~\citet{islam2020review} provide a review of recent advancements specifically in exchange rate prediction.
However, these deep learning models are typical of black-box nature due to their complex structure, which limits the interpretability of their predictions~\citep{park2022interpretable}. 
%\citet{giles2001noisy} suggest neural network-based method to predict five different daily exchange rates. Their results show that the proposed method can address difficulties with non-stationarity, overfitting, and nose in data. 

% \subsection{Interpretative Frameworks}
A clear understanding of a model's predictions enhances the adoption of machine learning and reduces inaccurate and potentially dangerous decisions~\citep{ozyegen2022evaluation}. 
In this regard, depending on the nature of the prediction task, the interpretability methods are classified as (1)~intrinsic and post-hoc, (2)~global and local, and (3)~model-specific and model-agnostic~\citep{adadi2018peeking}.
Explainable methods have gained popularity in financial and economic modelling due to their significant benefits in increasing trust in the prediction models and making predictions more actionable. 
% Several recent studies have used machine learning and interpreted the results to provide new insight into prediction outcomes. 
\citet{kellner2022opening} propose a neural network-based quantile regression model for bank loan loss prediction problem, where they decompose the model's prediction into their relative feature importance using a measure based on gradient information. 
\citet{jabeur2021forecasting} and \citet{gao2022explainable} employ tree-based methods for predicting the gold price and crude oil price, respectively, and examine the impact of variables using SHAP analysis. 
\citet{park2022interpretable} present a deep learning model based on the
LSTM networks to forecast economic growth rates and crises. 
They derive economic patterns of growth and crisis using an explainable framework and the SHAP method.
%\citet{jabeur2021forecasting} apply six methods with the superiority of extreme gradient boosting over other advanced machine learning models to predict the gold price. They use SHAP values to investigate the importance of different features that affect gold prices. 

\citet{djemo2021predicting} build an ensemble method by combining the outcomes of five machine learning models to classify directional movements in exchange rates. 
Their feature importance results show that the inflation differential, the industrial production index, and the term of trade are the leading macroeconomic variables in the appreciation of the Japanese Yen against the South African Rand rate. 
\citet{liu2021investigation} propose a method that combines LSTM with random forest (RF) to predict exchange rates for G10 currencies during Covid-19. 
They use SHAP to explain the impact of non-pharmaceutical interventions and economic support policies on foreign exchange markets. 
Similar to \citet{djemo2021predicting}, \citet{su2022exchange} forecast exchange rate movements while using the lags of the exchange rate, rather than macroeconomics variables, as inputs in univariate models. 
They employ the accumulated local effects algorithm to investigate the importance of lagged data in exchange rate modelling.

%The former study is related to forecasting directional movements in exchange rates with limited feature importance discussion and no empirically evident interpretations, and the latter study does not use macroeconomic variables and is confined to the short period of the pandemic.   

While our paper has connections to studies such as \citep{ferraro2015can} and \citep{kurita2022canadian}, which use crude oil prices to forecast exchange rates using traditional models; differently from these works, we investigate this relationship in the presence of further macroeconomic variables using machine learning. 
In addition, our paper differs from \citep{djemo2021predicting}, \citep{liu2021investigation}, and \citep{su2022exchange} in several ways. 
\citet{djemo2021predicting} forecast directional movements in exchange rates with a limited feature set, which hinders an extensive analysis with model interpretations.
% importance discussion and no empirically evident interpretations, 
We investigate exchange rate forecasting with a comprehensive use of machine learning and analyze the subsequent interpretations by using empirical evidence to gain a clear understanding of the model predictions. 
\citet{liu2021investigation} do not use macroeconomic variables in modelling and their analysis is confined to the short period of the recent pandemic. 
Similarly, \citet{su2022exchange} do not consider macroeconomic variables and rely only on the lags of exchange rates in its predictions during the post-Covid-19 outbreak. 
In contrast to these studies, we propose fundamental-based models using macroeconomic variables and analyze their performance in different expansionary and stagnant economic cycles in addition to the Covid-19 period. 
Furthermore, we use interpretability methods to disentangle the in-sample and out-of-sample empirical analyses under time-varying relationships with attention to theoretical consistency. 
Finally, our paper extends these studies by utilizing the interpretation results to improve the predictions and prune the models through an ablation study. 
The relative positioning of our paper in relation to other relevant studies is shown in Table~\ref{tab:littable}. 
% along with their summary.
\renewcommand{\arraystretch}{1.15}
\begin{table}[!ht]
\centering
\caption{Summary of the most relevant studies in the literature.}
\label{tab:littable}
\resizebox{1\textwidth}{!}{
\begin{tabular}{L{3.3cm}L{2.6cm}L{3.5cm}L{2.5cm}L{2.4cm}L{2.7cm}}
\toprule
\textbf{Study} &  \textbf{Target variable} &  \textbf{Input type} & \textbf{ML model} & \textbf{Interpretation method} &  \textbf{Period/ frequency}\\
\midrule
%\citet{kellner2022opening}& Loss given default& Macroeconomic &QRNN, GMM, RT, BR, FLR &Gradient information& 2000–2016/-\\
%\midrule
\citet{jabeur2021forecasting}&Gold price& Macroeconomic, commodity, exchange rate& LR, NN, RF, LGBM, XGB, CatB&SHAP& Jan 1986-Dec 2019/monthly\\
\midrule
\citet{gao2022explainable}&Crude oil price&Commodity, exchange rate, digital currency, Web, Covid-related&MLR,  KNN, RF, XGB, LGBM, CatB, DNN& SHAP&Feb 2020-Jun
2022/daily\\
\midrule
\citet{park2022interpretable}&Economic growth/crisis&Economic, financial&LSTM, RF, ANN, LASSO &SHAP&1990-2019/quarterly\\
\midrule
\citet{djemo2021predicting}& Exchange rate movements&Macroeconomic&Ensemble&Feature importance&Jan
1991–May 2020/daily\\
\midrule
\citet{liu2021investigation}&Exchange rate&Nonpharmaceutical, economic support policy, Covid-related&LSTM-RF&SHAP&Jan 2019-Jan 2021/daily\\
\midrule
\citet{su2022exchange}&Exchange rate movements&lags of exchange rate &LGBM, LSTM, ANN&ALE&Dec 2019-Aug 2021/daily\\
\midrule
\midrule
\textit{Our study}&Exchange rate&Macroeconomic, Commodity& LGBM, ETR, XGB, RIDGE, LASSO, GRU&Feature importance, SHAP& Jan 2009-Dec 2021/daily, weekly\\
\bottomrule
\end{tabular}

}
\begin{tablenotes}
      \scriptsize
      \item %QRNN: Quantile Regression Neural Network, GMM: Gaussian Mixture Model, RT: Regression Tree, BT: Beta Regression, FLR: Fractional Logit Regression,
      LR: Linear Regression, NN: Neural Network, RF: Random Forest, LGBM: Light Gradient Boosting Machine, XGB: Extreme Gradient Boosting, CatB: Category Boosting, MLR: Multiple Linear Regression, KNN: K-Nearest Neighbor, DNN: Deep Neural Network, LSTM: Long Short Term Memory, ANN: Artificial Neural Network, ETR: Extra Tree Regression, LASSO: Least Absolute Shrinkage and Selection Operator, GRU: Gated Recurrent Unit, SHAP: SHapley Additive exPlanations, ALE: Accumulated Local Effect 
    \end{tablenotes}
\end{table}
%related to the papers by \citet{djemo2021predicting} and  \citet{liu2021investigation} with an extension to different expansion and Stagnation business cycles and more comprehensive use of machine learning and  subsequent interpretations. Using several machine learning models, we address the forecasting task with the nonlinearity issue and disentangle the in-sample and out-of-sample empirical analysis with time-varying relationships. 

%We also employ interpretability methods to gain a clear understanding of the models and use the results to improve the predictions and prune the models using ablation study.

%%%%%%%%%%%%%%%%%%%%%%%%%%%%%%%%%%%%%%%%%%%%	
%%%%%%%%%%%%%%%%%%%%%%%%%%%%%%%%%%%%%%%%%%%%
% \section{Forecasting Models}\label{sec:models}
\section{Machine Learning Methods}\label{sec:models}
%%%%%%%%%%%%%%%%%%%%%%%%%%%%%%%%%%%%%%%%%%%%
%%%%%%%%%%%%%%%%%%%%%%%%%%%%%%%%%%%%%%%%%%%%
% Machine learning models have become strong competitors to traditional statistical models for the forecasting tasks. 
% There have been significant developments in this field in recent years, both in terms of the quantity and variety of generated models as well as their theoretical understanding~\citep{ahmed2010empirical}. 
% Recent trends show that the economic/financial forecasting models are required to be validated statistically in parallel with economic interpretations and model development. 
% This greatly benefits practitioners by limiting their options and providing them with information on the advantages and disadvantages of the models that are currently available.

In this study, we consider several forecasting models in three main categories: (1)~Linear (penalized) regression, (2)~Tree-based models, and (3)~Deep learning. 
We choose representative models from each category for the exchange rate forecasting task.

Penalized regression methods are shown to produce better forecasts for non-stationary data containing cointegrated variables than conventional statistical forecasting techniques, such as OLS, AR(1), and ARIMA, which suffer from overfitting on high dimensional data with multicollinearity~\citep{smeekes2018macroeconomic}. 
We select the two most widely used models from this category: Least Absolute Shrinkage and Selection Operator (LASSO), and RIDGE.
LASSO adds a penalty term to the loss function equal to the absolute value of the magnitude of the coefficients that brings some of the coefficients to zero, effectively performing feature selection and enhancing the interpretability of the model.
On the other hand, RIDGE includes the square of the magnitude of coefficients in the loss function as a penalty, which helps prevent overfitting.

% performs feature selection by adding a regularization term to the loss function that shrinks the coefficients of less important features to zero.
% As for regularization, $\ell1$- and $\ell2$-regularization are used for LASSO and RIDGE, respectively.

% It successfully excels over competing solutions in a number of applications from the field of forecasting~\citep{januschowski2022forecasting}. 
Several recent studies show the relatively strong performance of tree-based methods over traditional forecasting methods and deep learning-based approaches~\citep{januschowski2022forecasting, ilic2021explainable}. 
Consequently, we use three tree-based models: Extra Trees Regressor (ETR), Extreme Gradient Boosting (XGB), and Light Gradient-Boosting Machine (LGBM).
ETR~\citep{geurts2006extremely} is a bagging ensemble that combines the predictions of several trees to help reduce the variance in predictions and enhance the model reliability.
XGB~\citep{chen2016xgboost} and LGBM~\citep{ke2017lightgbm} are boosting ensembles that sequentially train multiple decision trees, and each tree aims to correct the mistakes made by the previous tree.
Both methods have built-in feature engineering capabilities that handle important time series data features, such as trends, seasonality, and lags. LGBM is typically considered to be more suitable for larger datasets or higher dimensional features due to its efficient tree-building algorithm and simpler hyperparameter tuning process.
These models can be used for time series forecasting by training them on historical time series data that can be preprocessed to include the lagged target values as features along with external covariates, such as commodity prices and other economic indicators.

Numerous deep learning models have also been commonly employed for forecasting problems as a result of the growing accessibility of massive volumes of data and computing environments.
We consider a GRU model, a special type of RNN, as a representative deep learning model in our analysis.
GRU models use an internal memory cell to store past information and update those at each time step to allow capturing the temporal dependencies in the data~\citep{cho2014properties}.
GRUs are well-suited for time series forecasting as they can handle complex patterns and dependencies in the data.
Similar to other deep learning models, they typically require a large amount of training data and can be computationally expensive.
Additionally, the performance of a GRU model is greatly impacted by hyperparameters such as the number of layers, and number of units in the GRU layer.
Interpretable machine learning models play an important role in ensuring that economic/financial time series predictions are reliable, trustworthy, and can be used to make informed decisions~\citep{lundberg2017unified}.
The ability to interpret the predictions of these models can help experts to identify potential biases and errors in the data, and identify the most important features of the financial time series. 
This information can then be used to improve the quality of the data and to make more informed investment decisions.
We employ two different approaches to interpret the forecasting models: feature importance and SHAP. 
We briefly describe these two interpretability methods below.
% The output of a prediction model must be properly interpreted, which is of utmost importance. It enhances the level of practitioner confidence, offers suggestions for  model improvement, and aids in comprehending  the represented process~\citep{lundberg2017unified}. 
% The feature importance describes a method that orders the features according to their effectiveness at forecasting a given variable. 
% While SHAP is a post-hoc interpretability method that explains the outputs of a model by breaking down a prediction to show the contribution of each feature. 
% Below, we briefly describe the two methods used for interpretation.

\begin{itemize}
    \item \textit{Feature importance}: 
    This approach can be considered a global interpretability method, which focuses on understanding the overall behavior of a model, including how it makes predictions for the entire dataset.
    Feature importance quantifies the contribution of each feature to the overall prediction performance of a model.
    In decision trees, this can be determined in various ways, such as counting how often a feature is used for tree splitting, calculating the average reduction in impurity achieved by the feature, or measuring the average decrease in accuracy when the feature is permuted.
    Feature importance can be used to interpret time series forecasting models that include external covariates (e.g., crude oil price, inflation) and lag features~\citep{ozyegen2022evaluation}. 
    Specifically, we can obtain information on which variables, whether lagged values of the time series itself or external covariates, have the largest impact on the target outcome.
    The variables with the highest feature importance can be considered the most important predictors for the target.
    
    % there are several methods for determining the importance of various variables in a model. In light of these, for less complex models, intrinsic global methods with the goal of explaining the entire logic and reasoning of a model are introduced in the literature.
    % They discover the best features during training, making them model-specific~\citep{ozyegen2022evaluation}. For example, in linear models, coefficients are used to determine the importance of the associated variable, whereas mutual information as decision tree criteria is used to perform feature selection in decision tree models. For more complex models such as deep neural networks, the influence methods, which are model-agnostic are used to  change the input or internal components and record the changes in model performance~\citep{han2015deep}. In this study, we use feature importance to gain knowledge about past relationships between macroeconomic variables while training the models.

    \item \textit{SHAP}: 
    SHAP (SHapley Additive exPlanations) values are based on the concept of Shapley values from the game theory, where the idea is to calculate the contribution of each feature to the prediction as if it were negotiated among all the features~\citep{lundberg2017unified}.
    SHAP is a model-agnostic interpretability method. Hence it can be used to obtain a comparable measure of feature importance across different models.
    SHAP values provide an explanation for each prediction made by a model. Thus, it can be considered a local interpretability method.
    Accordingly, SHAP values can be generated for a large number of test instances, and resulting contribution values can be averaged to obtain global interpretations~\citep{ozyegen2022evaluation}. 
    For time series forecasting, SHAP values can be calculated to measure the contribution of each feature, including lagged values of the time series and external covariates, to a specific prediction, taking into account the interactions among features.
    SHAP values can be obtained for each time point in the forecast, and it can help understand how the model makes predictions and how the contributions of the different features change over time.
    Previous studies show that SHAP outperforms alternative perturbation-based interpretability methods on time series datasets by satisfying three desirable properties: local accuracy, missingness, and consistency~\citep{schlegel2019towards, ozyegen2022evaluation}.

    % The SHAP method, proposed by~\citet{lundberg2017unified}, falls into the post-hoc interpretability and model-agnostic category, which generates the explanation after model training. Furthermore, it provides local explanations with a more detailed picture of the model's decisions by calculating each input feature's contribution to the prediction. To explain the original complex model, this method learns a local linear model. Some studies have revealed that SHAP outperforms alternative perturbation-based model interpretability approaches on time series datasets by satisfying three desirable properties: local accuracy, missingness, and consistency~\citep{schlegel2019towards, ozyegen2022evaluation}. In this work, we use SHAP values to investigate the level of contribution of each variable to the forecasts, allowing us to detect how they change in future.
\end{itemize}

%%%%%%%%%%%%%%%%%%%%%%%%%%%%%%%%%%%%%%%%%%%%%%%%%%%%%%%%%%%%%%%%%%%%%%%%%%%%%%%%%%
%%%%%%%%%%%%%%%%%%%%%%%%%%%%%%%%%%%%%%%%%%%%%%%%%%%%%%%%%%%%%%%%%%%%%%%%%%%%%%%%%%
\section{Empirical Analysis}\label{sec:empirical}
%%%%%%%%%%%%%%%%%%%%%%%%%%%%%%%%%%%%%%%%%%%%%%%%%%%%%%%%%%%%%%%%%%%%%%%%%%%%%%%%%%
%%%%%%%%%%%%%%%%%%%%%%%%%%%%%%%%%%%%%%%%%%%%%%%%%%%%%%%%%%%%%%%%%%%%%%%%%%%%%%%%%%
In this section, we provide the details of the experiments to evaluate the performance of different methods in exchange rate prediction. 
First, we provide information on the data collected, the process of model training, and testing. 
% We then explain the metrics that are used to evaluate different methods and compare their forecasting performance. 
We then report the empirical results from our detailed numerical study.
Second, we first compare the performance of different forecasting methods.
Then, we provide a feature importance analysis and investigate the contribution of different variables to the forecasts using SHAP values.
% The numerical results start with predictions followed by the feature importance analysis and investigation of the contribution of different variables to the forecasts using SHAP values. 
Finally, we present additional results with ablation studies to better understand the impact of the most important variables on the exchange rate. 
We implement the forecasting methods in Python on a computer with an 8-core M1 CPU and 8 GB of RAM.

\subsection{Dataset}
Our dataset consists of 10 macroeconomic variables and the CAD/USD rate as the target variable and covers the period from the beginning of January 2009 to the end of December 2021. 
We obtain daily CAD/USD exchange rate values from the Bank of Canada. 
We use three databases, including the Organization for Economic Co-operation and Development (OECD), Yahoo Finance, and the Federal Reserve Bank of St. Louis (FRBSL), to collect 10 macroeconomic variables. 
Some of these variables are collected daily, and the rest are monthly, with 3,252 and 156 observations, respectively. 
Table~\ref{tab:data} summarizes the information of the collected data.

\renewcommand{\arraystretch}{1.25}
\begin{table}[!ht]
\centering
\caption{Summary information on CAD/USD exchange rate and the considered macroeconomic variables.}
\label{tab:data}
\resizebox{.915\textwidth}{!}{
\begin{threeparttable}
\begin{tabular}{L{3.0cm}L{8.5cm}L{2.5cm}L{3cm}}
\toprule
\textbf{Data} & \textbf{Description} & \textbf{Frequency} & \textbf{Source}\\
\midrule
CAD/USD & Canadian exchange rate  & Daily   & Bank of Canada\\ 
Oil& WTI spot price FOB (U.S. dollars/barrel)& Daily  & FRBSL\tnote{*} \\ 
Gold & \makecell[tl]{Gold fixing price 10:30 A.M. in London Bullion \\ market (U.S. dollars)}  & Daily     & FRBSL  \\ 
TSX  & S\&P/TSX composite index  & Daily  & Yahoo Finance  \\
S\&P 500 & Standard \& Poor's index & Daily     & Yahoo Finance\\
ED  & 3-Month London interbank offered rate (LIBOR)& Daily  & FRBSL  \\ 
IR & Short-term interest rate parity  & Monthly & OECD\tnote{**} \\ 
PPI & Producer price indices parity& Monthly      & OECD \\ 
M1& Money stock & Monthly  & OECD \\ 
UnEmp & Unemployment rate & Monthly      & FRBSL  \\ 
IndProd & Industrial production index & Monthly  & OECD    \\ 
\bottomrule
\end{tabular}
\begin{tablenotes}
        \small
        \item[*] Federal Reserve Bank of St. Louis
        \item[**] Organization for Economic Co-operation and Development
    \end{tablenotes}
\end{threeparttable}
}
\end{table}

In Table~\ref{tab:data}, ``Oil'' is the daily price for West Texas Intermediate as a benchmark price used by crude oil markets. 
``Gold'' is the daily fixing price in London Bullion Market, and ``TSX'' and ``S\&P 500'' are daily stock market indexes in Canada and the United States, respectively. 
``ED'' is a daily three-month Eurodollars contract as represented by the London Interbank Offered Rate (LIBOR). 
We collect monthly data for other variables, including ``IR'', which is the short-term interest rate, ``PPI'', which measures the average change over time in the selling prices received by domestic producers for their output. 
``M1'' is the money supply that includes currency, demand deposits, and other liquid deposits, including savings deposits. 
``UnEmp'' is the number of unemployed people as a seasonally adjusted percentage of the labour force, and finally, ``IndProd'' represents the real production output of manufacturing, mining, and utilities. 

Table~\ref{tab:sumstat} shows the descriptive statistics of the exchange rate and macroeconomic variables over the considered time period, 2009-2021. 
Among the variables, TSX has the largest standard deviation, followed by S\&P 500. 
Note that the negative minimum crude oil price corresponds to April 20, 2020, when the U.S. crude finished at -\$36.98 a barrel. 
The skewness values are almost close to zero, indicating that the distributions of the variables are symmetric except for UnEmp with the largest positive value of 2, and the only negative value for IndProd that represents a left-skewed distribution with more values concentrated on the right side (tail) of the distribution.

\begin{table}[!ht]
\centering
\caption{Summary statistics on the CAD/USD rate and the macroeconomic variables from 2009 to 2021.}
\label{tab:sumstat}
\resizebox{.88\textwidth}{!}{
\begin{tabular}{L{3cm}R{2.2cm}R{2.2cm}R{2.2cm}R{2.2cm}R{2.2cm}R{2.2cm}}
\toprule
\textbf{Data} & \textbf{Mean} & \textbf{SD} & \textbf{Min} & \textbf{Median} & \textbf{Max} & \textbf{Skew}\\
\midrule
CAD/USD  & 0.86	&0.10	&0.69&	0.80&	1.06&0.35\\ 
Oil& 68.77&	21.98&	-36.98&	66.50&	113.39 & 0.09\\ 
Gold & 1385.07&	256.29&	813.00&	1311.07&	2061.50 &0.40\\ 
TSX  & 14336.88&	2483.96&	7566.90&	14374.35&	21768.50  &0.41\\
S\&P 500 & 2154.34	&921.88&	676.53&	2049.15	&4793.06&0.77\\
ED  & 0.77&	0.75&	0.11&	0.36	&2.82 &1.34\\ 
IR & 1.03&	0.50	&0.18&	1.16	&2.19 & 0.14\\ 
PPI & 101.33	&7.89&	88.45&	100.08	&126.01 &0.94\\ 
M1& 107.59	&37.44	&57.68&	99.24&	204.29&0.92\\ 
UnEmp & 7.28&	1.23&	5.40	&7.10&	13.40 &2.00\\ 
IndProd & 98.10	&5.26	&74.64	&99.05	&107.11 & -1.02  \\ 
\bottomrule
\end{tabular}

}
\end{table}

%%%%%%%%%%%%%%%%%%%%%%%%%%%%%%%%%%%%%%%%%%%%%%%%%%%%%%%%%%%%%
% \subsection{Prediction Process}
\subsection{Experimental Setup}
%%%%%%%%%%%%%%%%%%%%%%%%%%%%%%%%%%%%%%%%%%%%%%%%%%%%%%%%%%%%%
% We consider an identical procedure for prediction and interpretability analysis for all models in different economic states over time to have consistent and unbiased results. 
In our analysis, we consider some specific subperiods marked by different economic cycles in addition to the entire period of study over the years 2009-2021. 
In particular, we divide the sample period into Economic Expansion (2009-2011), Economic Stagnation (2014-2016), and Covid (2019-2021). 
These subperiods are identified by OECD-based Indicators for Canada obtained from the Federal Reserve Bank of St. Louis\footnote{\href{https://fred.stlouisfed.org/series/CANRECDM}{https://fred.stlouisfed.org/series/CANRECDM}}. 
Note that this indicator can be considered to be a time series of dummy variables that shows the expansionary and stagnant periods, where stagnation is indicated by a value of 1, while an expansion is represented by a value of 0. The stagnation is assumed to start on the 15th day of the peak month and ends on the 15th day of the trough month.
The Economic Expansion (2009-2011) and Economic Stagnation (2014-2016) specifically reflect the business cycles and turning points that are observed and identified in the deviation-from-trend series of the GDP in Canada and are consequently bound up with the currency performance. 

To evaluate the methods using the entire sample period, not just the ending portion of a time series, we consider a time window of 2.5 years (30 months) and divide the data into train-test with a ratio of 80:20 in this window. 
Then we roll the window forward by 6 months each time until it covers the entire data (2009-2021). 
In this way, we end up with 22 consecutive test periods with overall prediction performance being reported by taking the average of the results over the individual test windows. 
Regarding the subperiods mentioned earlier, we divide them into a single train and test with the same ratio of 80:20.

\begin{figure}[!ht]
    \centering
    \subfloat{\includegraphics[width=.8\textwidth]{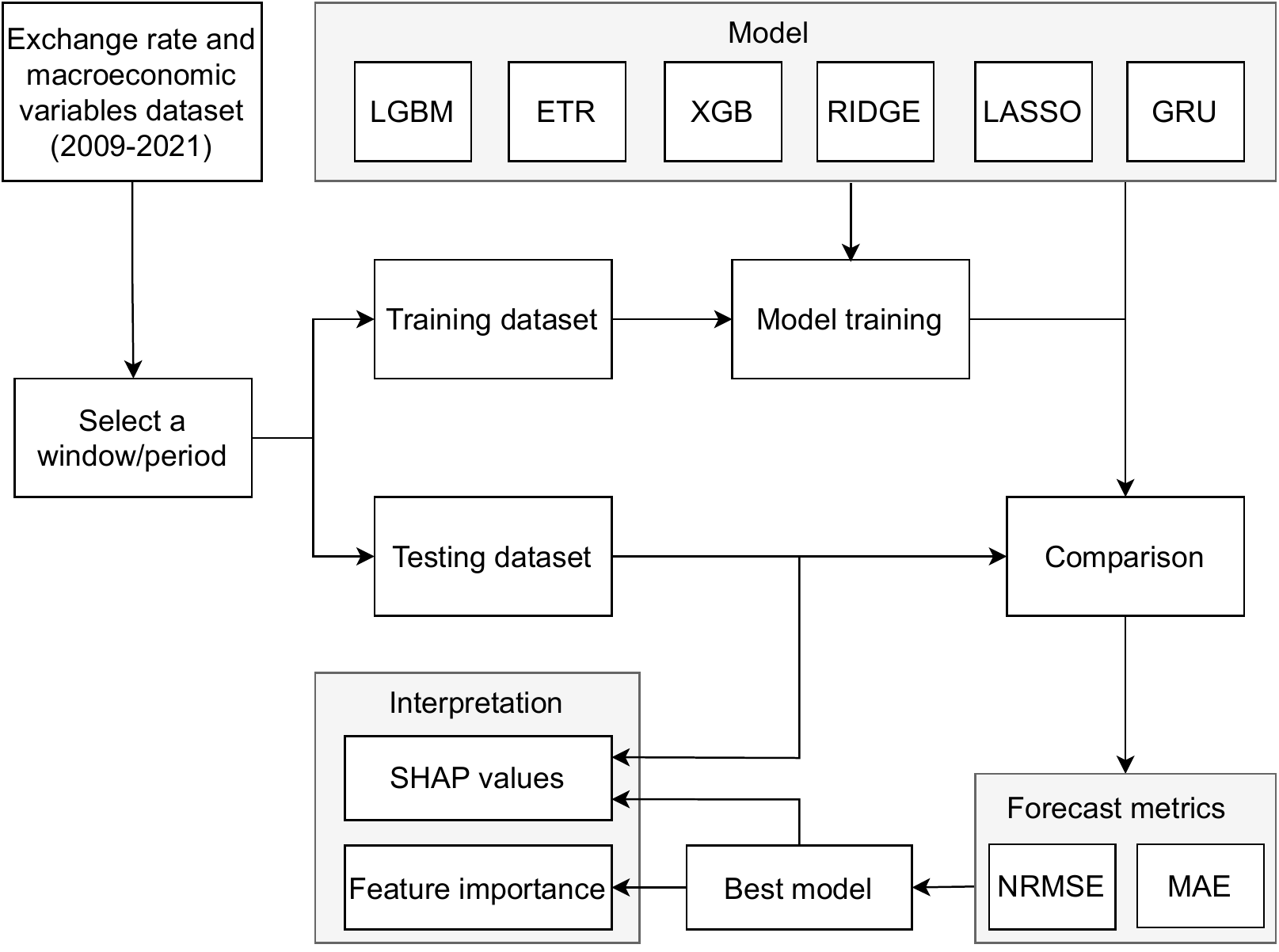}}
    \caption{Flow diagram of prediction process in models.}
    \label{fig:predictionProcess}
\end{figure}

Figure~\ref{fig:predictionProcess} shows the flowchart for the experimental setup, illustrating how the models are trained and evaluated. %the diagram of the prediction process. 
After selecting the period of interest (either a fixed window from the sample period or a specific subperiod), we divide the time interval into the train and test sets. 
The data in the train set is used to train the models. 

% Once the models are trained, they are used to forecast the exchange rate in the test set.
% We then compare the models based on the performance metrics using the predicted  and the real exchange values. 
% It is logical to 
We generate the interpretations using the best-performing models. Thus, before starting the interpretation step, we identify the best model that outperforms the others in terms of average forecasting performance values. 
To interpret the models and specify the contribution of the individual variables to the predictions, we consider feature importance and SHAP methods.
% two approaches (1) feature importance that is based on the trained models and (2) SHAP values that are obtained by the output of the trained models for the test values. 
We also perform hyperparameter tuning for each considered model based on a grid search over the hyperparameter values listed in Table~\ref{tab:hyperparam}.
% prevent the models from underfitting/overfitting. 
% Accordingly, to find the best parameter set for each model, we use a grid search over the values for the hyperparameters as listed in Table~\ref{tab:hyperparam}.
% Figure~\ref{fig:evaltime} depicts how we divide a training set in a sample window into smaller training and test sets for cross-validation. 
% We then train the model on the smaller training set to find the parameter values with minimum prediction error in the smaller test set. 
% We then use the best parameters to train the model on the larger primary training set and evaluate the performance of the model on the final test set~\citep{neghab2022integrated}. 

\begin{table}[!ht]
\centering
\caption{Hyperparameters and search space for model tuning.}
\label{tab:hyperparam}
\resizebox{0.98\textwidth}{!}{
\begin{threeparttable}
\begin{tabular}{L{3.5cm}L{3.5cm}L{11cm}}
\toprule
\textbf{Model} & \textbf{Hyperparameters} &  \textbf{Search space}\\
\midrule
LGBM, ETR, XGB&n\_estimators&[100, 300, 500, 700, 1000, 2000]\\
&max\_depth&  [3, 4, 5, 6, 7, 8, 9, 10]\\
&max\_features&[sqrt, log2]\\ [.2cm]
RIDGE, LASSO
&alpha&[0.001, 0.01, 0.1, 1, 10, 100, 1000, 2000, 3000, 4000, 5000, 10000]\\[.2cm]
GRU\tnote{*}
& \# of layers&[1, 2]\\
& \# of hidden units&[2, 4, 8, 16, 32, 64]\\
& activation function& [sigmoid, relu]\\
\bottomrule
\end{tabular}
\begin{tablenotes}
        \small 
        \item[*] The optimizer is chosen as adam and the number of epochs is set to 200.
        
\end{tablenotes}
\end{threeparttable}
}
\end{table}

% \begin{figure}[!ht]
%     \centering
%  \subfloat{\includegraphics[width=.9\textwidth]{Figure/evaltime.pdf}}
%     \caption{Data split for cross-validation and final training and evaluation. Source: \citet{neghab2022integrated}.} 
%     \label{fig:evaltime}
% \end{figure}

%%%%%%%%%%%%%%%%%%%%%%%%%%%%%%%%%%%%%%%%%%%%%%%%%%%%%%%%%%%%%
% \subsection{Performance Metrics}
%%%%%%%%%%%%%%%%%%%%%%%%%%%%%%%%%%%%%%%%%%%%%%%%%%%%%%%%%%%%%
We consider two different metrics for measuring forecasting performance, namely, normalized root mean square error (NRMSE) and mean absolute deviation (MAE), which can be defined as follows:% ~\citep{ilic2021explainable, park2022interpretable}. 
% The first metric is the normalized root mean squared error (NRMSE) for point forecasting performance as follows:
\begin{equation*}
    \text{NRMSE}(y, \hat{y})=\frac{\sqrt{\frac{1}{T}\sum_{i=1}^T(\hat{y}_t-y_t)^2}}{\frac{1}{T}\sum_{t=1}^T|y_t|}, \qquad \text{MAE}(y, \hat{y})=\frac{1}{T}\sum_{t=1}^T|\hat{y}_t-y_t|.
\end{equation*}
where $y_t$ and $\hat{y}_t$ denote the actual and predicted values of the exchange rate at time $t$, respectively. 
% The second metric is mean absolute deviation (MAE), which is a better measure in case of non-Gaussian errors~\citep{chai2014root}:
% \begin{equation*}
%     \text{MAE}(y, \hat{y})=\frac{1}{T}\sum_{t=1}^T|\hat{y}_t-y_t|.
% \end{equation*}

%%%%%%%%%%%%%%%%%%%%%%%%%%%%%%%%%%%%%%%%%%%%%%%%%%%%%%%%%%%%%
\subsection{Exchange Rate Prediction Results}
%%%%%%%%%%%%%%%%%%%%%%%%%%%%%%%%%%%%%%%%%%%%%%%%%%%%%%%%%%%%%
% In this section, we present the results of exchange rate prediction for different methods and compare their performance based on the metrics. 
Tables~\ref{tab:dailyPrediction} and~\ref{tab:weeklyPrediction} show the results for six different forecasting models for daily and weekly data, respectively. 
The results are presented for four different data subsets considering the forecast horizons of 1, 5, and 10 steps. 
% The performance values are marked in bold for the best-fit model with the minimum error between the predicted value and the actual value.
Over the entire time period, which is specified by the dataset ``All (2009-2021)'', LASSO and LGBM are the best-performing models for 1-day and 1-week forecast horizons, respectively. 
For the 5-day forecast horizon, the performance of LGBM is better than other methods, while for the 5-week predictions, RIDGE outperforms the others. However, XGB is superior to other methods when the forecast horizon gets longer, e.g., as seen in the 10-period ahead forecasts. 
Note that these results are based on average performance across 22 rolling windows over the time period between 2009 and 2021.

For the ``Economic Expansion (2009-2011)'' and ``Stagnation (2014-2016)'' datasets, we observe that the tree-based methods, particularly the ETR model, are more successful in providing accurate predictions (in terms of average performance values) than the linear models and GRU as the deep learning model. 
This observation is more evident for the ``Economic Stagnation (2014-2016)'' dataset, where ETR outperforms others in all of the forecast horizons for both daily and weekly data. For ``Covid (2019-2021)'' dataset, the results show that the linear model, RIDGE, leads to a significant decrease in errors, especially for 1- and 5-step forecast horizons. 
For example, for a 5-day forecast horizon in Table~\ref{tab:dailyPrediction}, the NRMSE for the RIDGE model is 0.0099, which is much lower than the NRMSE of other models (e.g., LGBM: 0.0328, ETR: 0.0258, XGB: 0.0231, LASSO: 0.0207, and GRU: 0.0627). 
Similarly, for the 1-week forecast horizon in Table~\ref{tab:weeklyPrediction}, the NRMSE for the RIDGE model is 0.0095, much lower than the NRMSE of other models (e.g., LGBM: 0.0213, ETR: 0.0252, XGB: 0.0210, LASSO: 0.0129, and GRU: 0.0620).

\begin{table}[!ht]
\centering
\caption{Comparison of the performance of the exchange rate prediction models for three different forecast horizons based on the \underline{daily} data. The values in bold show the minimum error in each column among the six models (lower is better).}
\label{tab:dailyPrediction}
\resizebox{.98\textwidth}{!}{
\begin{tabular}{L{3.8cm}L{2cm}C{2.3cm}C{2.3cm}C{2.3cm}C{2.3cm}C{2.3cm}C{2.3cm}}
\toprule
&&\multicolumn{2}{c}{1-day forecast horizon}&\multicolumn{2}{c}{5-day forecast horizon}&\multicolumn{2}{c}{10-day forecast horizon}\\
\cmidrule(lr){3-4} \cmidrule(lr){5-6} \cmidrule(lr){7-8}
Period& Method& NRMSE& MAE&NRMSE& MAE&NRMSE& MAE\\
\midrule
All (2009-2021)
& LGBM	&	0.0285	&	0.0237& \textbf{0.0323}	&	\textbf{0.0262}&0.0392	&	0.0319\\ 
& ETR&	0.0292	&	0.0241 & 0.0332	&	0.0269& 0.0347	&	0.0280\\ 
& XGB&	0.0290	&	0.0241 & 0.0326	&	0.0264 & \textbf{0.0343}	&	\textbf{0.0276}\\ 
& RIDGE	& 0.0294	&	0.0242 & 0.0438	&	0.0353 & 0.0474	&0.0377\\
& LASSO	& \textbf{0.0283}	&	\textbf{0.0235} & 0.0400	&	0.0324 & 0.0412	&0.0331\\
& GRU& 0.0363	&	0.0301  & 0.0511	&	0.0418 & 0.0529	&0.0434\\ 
\midrule
Economic Expansion
& LGBM	& 0.0186	&	0.0190 &\textbf{0.0159}	&\textbf{0.0151}&  \textbf{0.0136}&	\textbf{0.0119}\\ 
(2009-2011) 
& ETR&\textbf{0.0147}	&	\textbf{0.0147}&	0.0168&	0.0159 & 0.0169&	0.0154\\
& XGB&	0.0148	&	0.0150&0.0166	&0.0156&  0.0179	&0.0160\\ 
& RIDGE	&0.0169	&	0.0172&0.0174&	0.0164 &0.0207	&0.0180\\
& LASSO&0.0161	&	0.0162	&0.0164&	0.0154 & 0.0201	&0.0176\\
& GRU& 0.0191	&0.0192&0.0375&	0.0377 & 0.0362&	0.0359\\ 
\midrule
Economic Stagnation 
& LGBM& 	0.0505& 	0.0389& 0.0509	&0.0383&0.0599	&0.0441\\
(2014-2016)&ETR	 & \textbf{0.0276} & 	\textbf{0.0213}& \textbf{0.0325}	&\textbf{0.0241}&\textbf{0.0313}&	\textbf{0.0225}\\
&XGB & 	0.0358 & 	0.0276& 0.0581	&0.0440&0.0412&	0.0301\\
&RIDGE & 	 0.0584 & 	0.0446& 0.0678	&0.0512&0.0773	&0.0582\\
&LASSO & 	0.0504	 & 0.0385 & 0.0594&	0.0447&0.0651&	0.0488\\
&GRU	 & 0.0299	 & 0.0229& 0.0552&	0.0403&0.0580	&0.0417\\
\midrule
Covid (2019-2021) 
& LGBM	&0.0189	&0.0154& 0.0328&	0.0261&0.0644	&0.0518\\
&ETR&	0.0254&	0.0206& 0.0258&	0.0206&\textbf{0.0272}	&\textbf{0.0216}\\
&XGB&	0.0213&	0.0173& 0.0231	&0.0184&0.0275	&0.0219\\
&RIDGE&	\textbf{0.0157}&	\textbf{0.0125}& \textbf{0.0099}	&\textbf{0.0072}&0.0346	&0.0267\\
&LASSO&	0.0171&	0.0137& 0.0207&	0.0159&0.0390	&0.0300\\
&GRU	&0.0617&	0.0497& 0.0627&	0.0503&0.0641&	0.0514\\
\bottomrule
\end{tabular}
}
\end{table}

\begin{table}[!ht]
\centering
\caption{Comparison of the performance of the exchange rate prediction models for three different forecast horizons based on the \underline{weekly} data. The values in bold show the minimum error in each column among the six models (lower is better).}
\label{tab:weeklyPrediction}
\resizebox{.98\textwidth}{!}{
\begin{tabular}{L{3.8cm}L{2cm}C{2.3cm}C{2.3cm}C{2.3cm}C{2.3cm}C{2.3cm}C{2.3cm}}
\toprule
&&\multicolumn{2}{c}{1-week forecast horizon}&\multicolumn{2}{c}{5-week forecast horizon}&\multicolumn{2}{c}{10-week forecast horizon}\\
\cmidrule(lr){3-4} \cmidrule(lr){5-6} \cmidrule(lr){7-8}
Period& Method& NRMSE& MAE&NRMSE& MAE&NRMSE& MAE\\
\midrule
All (2009-2021) 
&LGBM	&\textbf{0.0317}	&\textbf{0.0265}&0.0563&	0.0458&0.0571&	0.0460\\
&ETR	&0.0325	&0.0267&0.0381&	0.0303&0.0408&	0.0318\\
&XGB	&0.0327	&0.0269&0.0375&	0.0296&\textbf{0.0399}&	\textbf{0.0306}\\
&RIDGE	&0.0419	&0.0344&\textbf{0.0364}&	\textbf{0.0276}&0.0503&	0.0376\\
&LASSO	&0.0379	&0.0315&0.0370&	0.0281&0.0495&	0.0365\\
&GRU	&0.0487	&0.0403&0.0575&	0.0470&0.0576&	0.0463\\
\midrule
Economic Expansion & LGBM&0.0188&	0.0187&0.0412	&0.0393&0.0455	&0.0424\\ 
(2009-2011) & ETR&	\textbf{0.0170}	&\textbf{0.0169}&\textbf{0.0191}	&\textbf{0.0154}&0.0306&	0.0259\\ 
& XGB&0.0178	&0.0178&0.0201&	0.0157&	\textbf{0.0294}&	\textbf{0.0256}\\
& RIDGE	&0.0174&	0.0173&0.0206&	0.0183&0.1004&	0.0886\\
& LASSO	&0.0191	&0.0192&0.0204	&0.0165&0.0449&	0.0402\\
& GRU&0.0318&	0.0326&0.0395&	0.0375&0.0436&	0.0406\\ 
\midrule
Economic Stagnation  & LGBM&0.0379	&0.0290	&0.0744	&0.0523 & 0.0772&	0.0523\\ 
(2014-2016)& ETR&\textbf{0.0233}	&\textbf{0.0179}	&\textbf{0.0294}	&\textbf{0.0174} & \textbf{0.0454}	&\textbf{0.0302}\\ 
& XGB&	0.0353&	0.0271  &0.0302	&0.0186 & 0.0452&	0.0301\\ 
& RIDGE& 0.0546&	0.0416&	0.0375&	0.0249&  0.0702	&0.0476\\
& LASSO	&0.0441	&0.0336  &0.0492&	0.0330 & 0.0665	&0.0438\\
& GRU&0.0654	&0.0480  &0.0740&	0.0522 & 0.0716	&0.0472	\\ 
\midrule	
Covid (2019-2021) & LGBM&0.0213	&0.0174&	0.0637&	0.0511 & 0.0648&	0.0511\\ 
& ETR&0.0252&	0.0205	&0.0307&	0.0244 & 0.0524	&0.0405\\ 
& XGB&	0.0210	&0.0171 & 0.0283&	0.0224&  0.0561&	0.0411\\ 
& RIDGE	&  \textbf{0.0095}&	\textbf{0.0076}&  \textbf{0.0157}&	\textbf{0.0108} & 0.1257	&0.0736\\
& LASSO	&  0.0129&	0.0102 & 0.0275	&0.0202 & \textbf{0.0378}&	\textbf{0.0251}\\
& GRU&  0.0620&	0.0500&  0.0656&	0.0526 & 0.0657	&0.0517\\ 
\bottomrule
\end{tabular}
}
\end{table}

Overall, tree-based models perform considerably better than other models on average, followed by linear models. 
This result supports the findings in \citep{su2022exchange}, which shows that tree-based models outperform deep learning models in other exchange rate prediction tasks. 
It is worth considering the efficiency of methods such as RIDGE and LASSO compared to other complex models, such as GRU, in exchange rate forecasting. 
Our finding of the competitive performance of the new linear models for the ``Covid (2019-2021)'' dataset, is consistent with the evidence reported in~\citep{abedin2021deep}, which shows that the performance of different models depends on the period of study.
%shown to achieve lower prediction errors compared to other methods such as tree-based and deep methods for the forecasting tasks associated with several exchange rates.
% They use different models including linear, tree-based, and deep methods to forecast several exchange rates and obtain lower predictive errors for the RIDGE and LASSO models.

We find that the performance of all models deteriorates along with (1)~increasing the forecast horizon from 1 to 5 and 10 steps and also with (2)~modelling long-term dependencies in weekly data rather than short-term daily ones. 
The former result might be caused by reducing the number of observations for longer forecast horizons. 
In addition, it is typically the case in forecasting tasks that the time steps which are far away from the input/training data are the most challenging ones to predict (e.g., for a 10-step forecast, the 10th step is more difficult to predict than the 1st step). 
The latter, i.e., better prediction performance in daily data, confirms the existence of a short-term relationship among the exchange rate and the macroeconomic variables. 
This result corroborates the evidence reported in~\citet{ferraro2015can}; that is, they observe a robust predictive power in currency exchange rate predictions using commodity prices at a daily frequency and find that such predictive ability is not evident at monthly and quarterly frequencies. 
Similarly,~\citet{zhang2016exchange} investigate the relationship among commodity prices and exchange rates and observe that causality is more pronounced at shorter horizons. 

These results also indicate that GRU has a weak performance in every period and all forecasting horizons. 
This result is in line with the findings of~\citet{abedin2021deep}. 
Their results show that the performance of deep learning models in exchange rate prediction during Covid-19 is not robust across different currencies compared to other methods, such as RF, regression tree, and linear models. 
In addition, \citet{hernandez2021using} empirically demonstrate that the performance of deep learning models in exchange rate forecasting is not superior to that of tree-based models. 
Some other studies provide evidence of the stronger performance of tree-based methods compared to deep learning models in time series forecasting in other application domains~\citep{elsayed2021we, januschowski2022forecasting, sprangers2022parameter}.

In terms of average training times, as expected, linear RIDGE and LASSO models are the fastest (e.g., taking less than one second to train).
The tree-based models take between 2.1 and 24.0 seconds to train, on average, with XGB being the slowest among these.
On the other hand, the deep learning model, GRU, is the most computationally expensive model, with average training times ranging from 384.1 to 1050.2 seconds, depending on the dataset and forecasting horizon. 

\subsection{Backward and Forward-looking Economic Interpretation of the Forecasting Models}\label{sec:interpretation}
%%%%%%%%%%%%%%%%%%%%%%%%%%%%%%%%%%%%%%%%%%%%%%%%%%%%%%%%%%%%%%%%%%%%%%%%%%%%%%%%%%%%%%%%%%%%%%%%%%%%%%%%%%%%%%%%%%%%%%%%%%
% It is important to interpret the models in addition to making a statistical assessment for prediction. 
% The reliability of prediction methods, especially machine learning models, heavily depends on the transparency of the information delivered along with the predictions. 
% With the complexity and uncertainty of the systems in financial and economic applications, this issue is of great importance, where the model's output is the departure point in the prospect of successful decision-making. 
% Therefore, in this section, 
We next focus on interpreting the well-performing prediction models to shed light on several analytical aspects of the exchange rate forecasting problem. 
First, we identify the variables that have strong forecasting ability by using the feature importance method and SHAP. 
Second, we assess the time-varying relationship between the crude oil price, as an important factor, and the CAD/USD exchange rate at some specific dates, and also over a specific time interval.

\begin{figure}[!ht]
    \centering
    \subfloat[\tiny FI: Daily, LASSO \label{fig:lassoFI}]{\includegraphics[width=.24\textwidth]{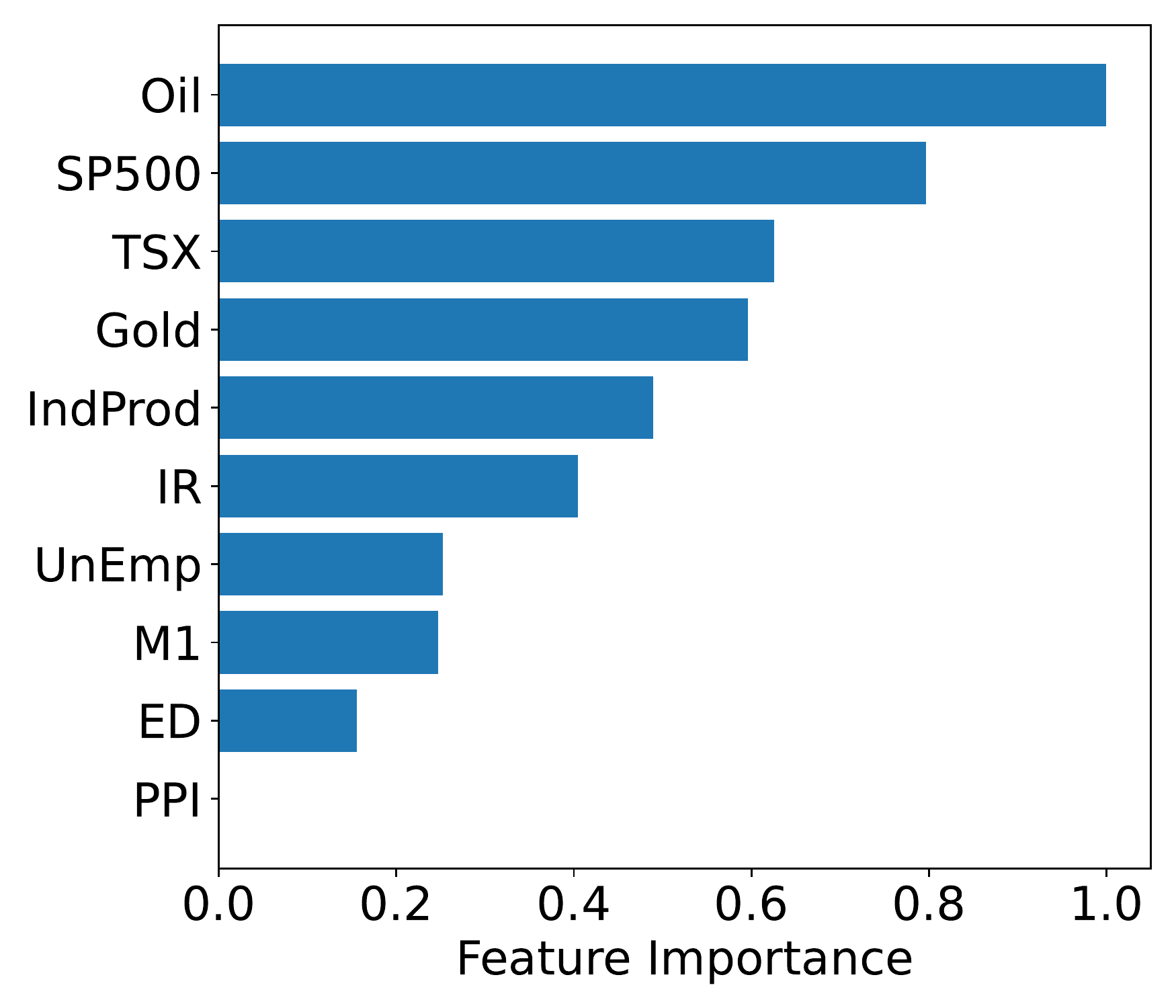}}
    \subfloat[\tiny SHAP: Daily, LASSO \label{fig:lassoSHAP}]{\includegraphics[width=.24\textwidth]{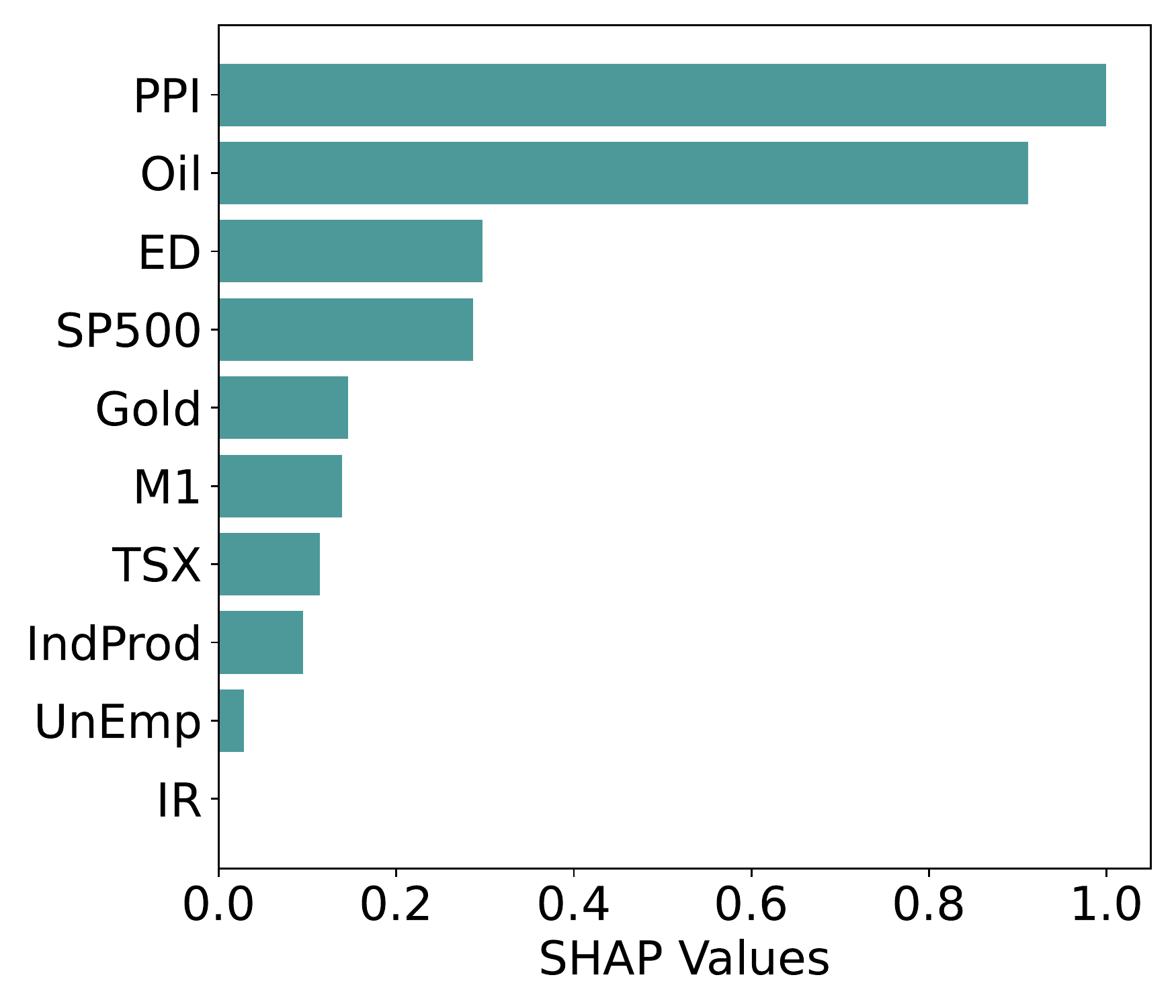}}
    \subfloat[\tiny FI: Weekly, LGBM \label{fig:lgbmFI}]{\includegraphics[width=.24\textwidth]{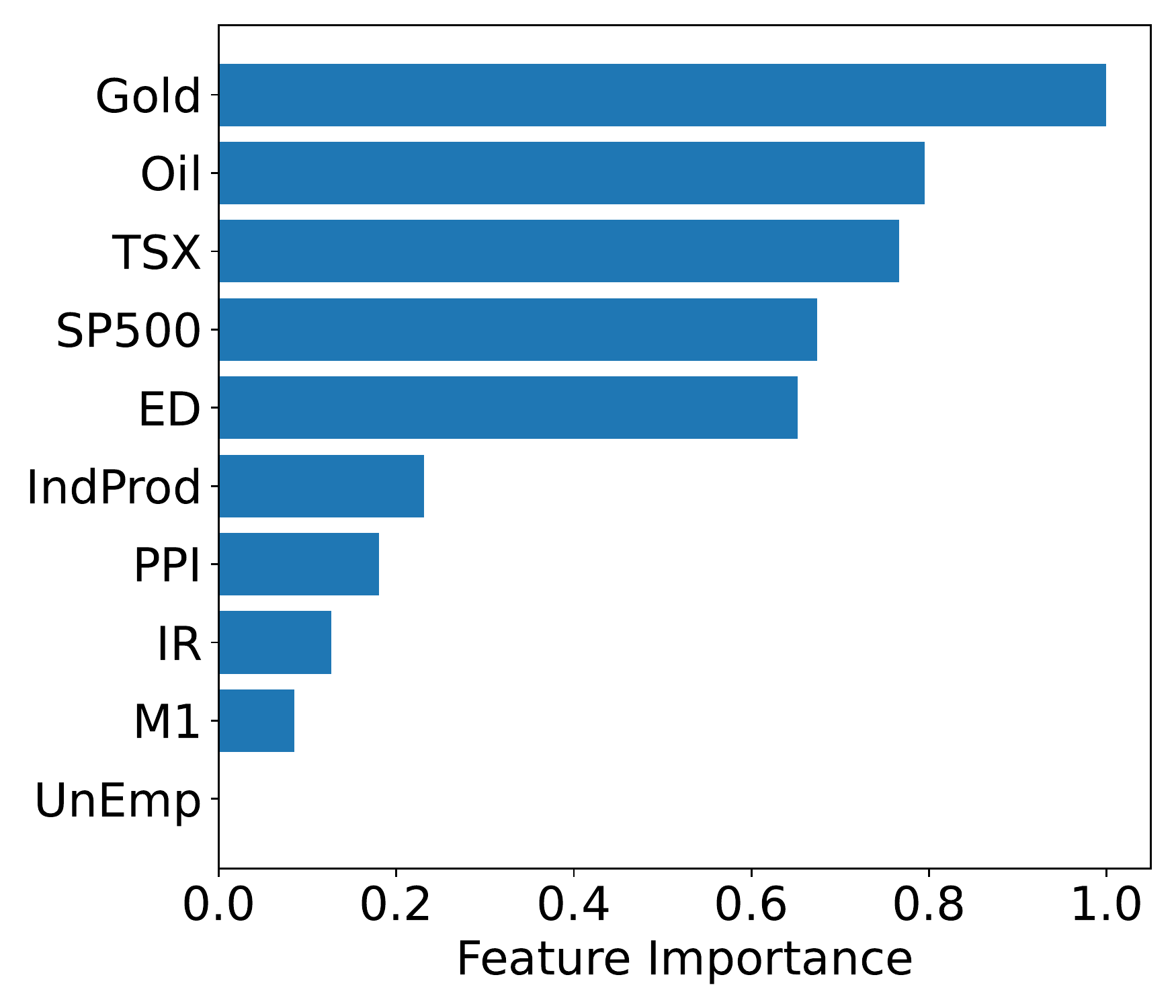}}
    \subfloat[\tiny SHAP: Weekly, LGBM \label{fig:lgbmSHAP}]{\includegraphics[width=.24\textwidth]{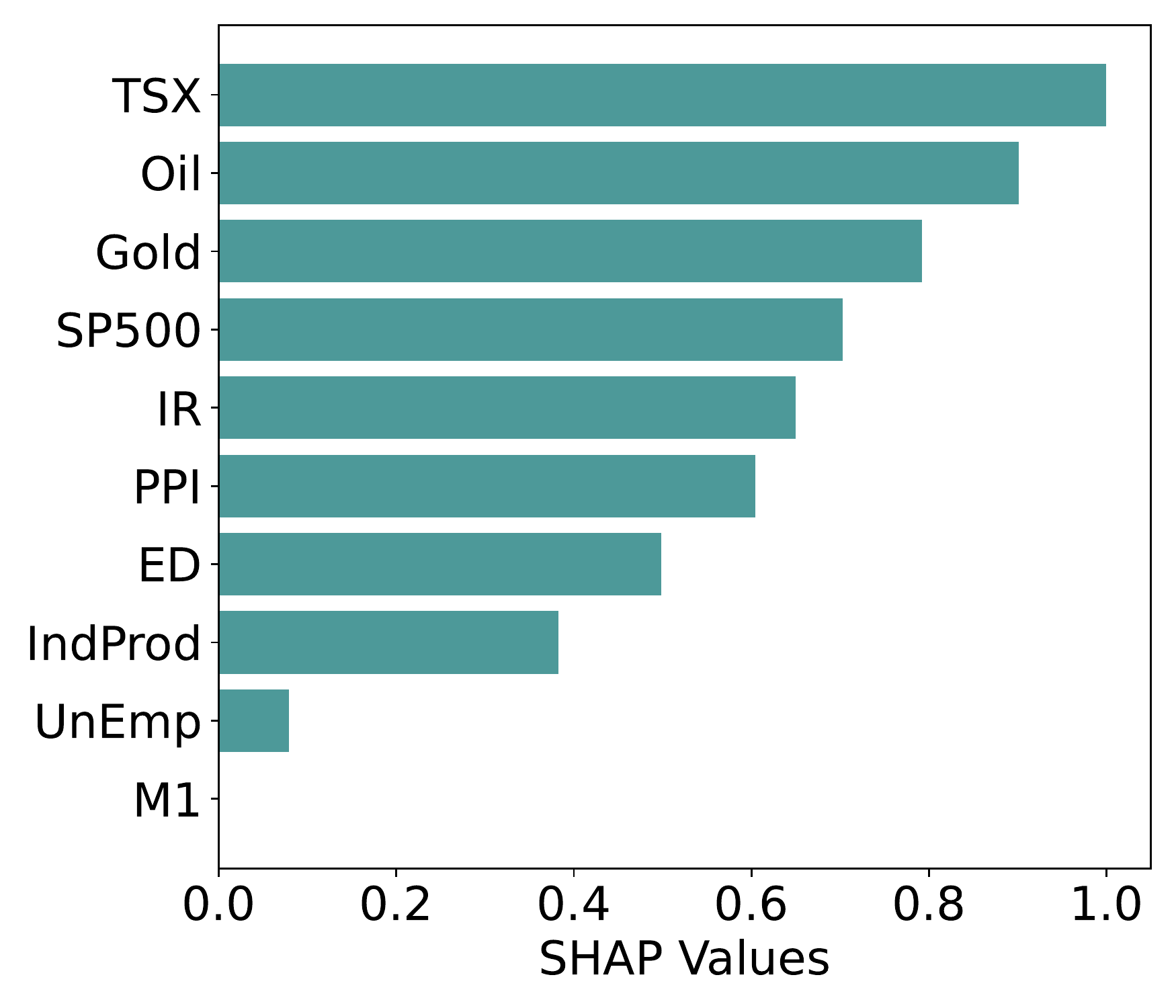}}\\\vspace{.3cm}
    \caption{Scores for feature importance (FI; blue bar plots) and SHAP (green bar plots) for 1-day and 1-week forecast horizons.}
    \label{fig:FI}
\end{figure}

We identify the best prediction models for the 1-day and 1-week ahead forecasts as LASSO and LGBM, respectively (see Tables \ref{tab:dailyPrediction} and \ref{tab:weeklyPrediction}). 
We compute feature importance values representing the impact of each variable on the exchange rate for each model. 
% These scores are obtained by taking the average of all values from the models over the training windows. 
We normalize the score values to the range of $[0,1]$ for the sake of easier comparison. 
Figure~\ref{fig:FI} shows the scores for daily and weekly modelling using the LASSO and LGBM models. 
We observe that crude oil, S\&P500, TSX, and gold are important features for the daily exchange rates. 
The importance of crude oil indicates that the CAD/USD exchange rate heavily depends on crude oil as Canada's dominant exporting commodity~\citep{zhang2016exchange}.

For the weekly LGBM forecasting model in Figure~\ref{fig:lgbmFI}, we find the same four features (i.e., crude oil, S\&P500, TSX, and gold) to be important. 
However, the rankings in these models are different. 
Crude oil, in particular, comes in second, confirming previous findings that the relationship between crude oil and the exchange rate is weaker at a longer horizon due to very short-term crude oil price shocks. 
Our results demonstrate that recently increased gold exports from Canada have made it a useful variable in exchange rate prediction. 
This observation corroborates the findings by~\citet{ferraro2015can} that show the predictive ability of gold prices using an out-of-sample fit measure in a simple linear model. 
Moreover, the importance of both U.S. and Canadian market indices in the exchange rate prediction is consistent with studies that employ statistical tests to investigate interdependencies among these economic indicators~\citep{kanas2000volatility, michelis2010dependence}. 
We see that the score for S\&P 500 is followed by the TSX in the daily data, while it is the fourth after the TSX in the weekly version. 
This implies the short-run effect of the U.S. market against the long-run effect of the Canadian stock market on the exchange rate. 
However, the comparable importance levels for both stock market indices indicate that capital markets are becoming increasingly integrated, leading to similar influences on exchange rates. 

It is essential to disentangle the in-sample and out-of-sample analyses and find a potentially strong link between past knowledge about relationships and adequate interpretations for the future. 
Since the feature importance method provides knowledge about historical relationships --the importance scores are obtained from trained models using the training data--, we use SHAP values to investigate whether the impact of different variables holds for out-of-sample analysis. 
Accordingly, we first compute the contribution of each variable to the predictions, then take the average of the absolute SHAP values over the test samples, and finally rescale computed values to the range of $[0,1]$. 
Figure~\ref{fig:lassoSHAP} indicates the values for daily forecasts. 
We observe that PPI, crude oil, and ED are found to be the top three variables. 
In particular, the prominent contribution of PPI to exchange rate prediction supports the fact that the relative purchasing power of two countries should be reflected in the exchange rate between them. 
Hence, if there is a difference in PPI between countries, the opportunity for arbitrage causes excess demand and, consequently, a change in foreign currency valuation, leading to a price balance. 
However, our finding of the short-term relationship contrasts with previous studies; it confirms that the two markets are integrated, and the countries foster price convergence.  

In weekly predictions, Figure~\ref{fig:lgbmSHAP} shows that the ranking of the variables based on SHAP values is more consistent with the in-sample feature importance scores in Figure~\ref{fig:lgbmFI}, and TSX, crude oil, and gold contribute to predictions more than other variables. 
Overall, from the results in Figures~\ref{fig:lassoFI}-\ref{fig:lgbmSHAP}, we find that crude oil, gold, and TSX have the highest impact based on the two interpretability methods. 
Moreover, using either the feature importance scores or the SHAP values, we find that crude oil is consistently important among these variables in both daily and weekly frequencies and makes a strong link between in-sample and out-of-sample analysis. 
Gold is the second commodity with a considerable effect on the exchange rate as a result of having a notable proportion of exported products from Canada (see Figure~\ref{fig:oil2009-2019}, which illustrates gold among the top Canadian exports in 2019).

%%%%%%%%%%%%%%%%%%%%%%%%%%%%%%%%%%%%%%%%%%%%%%%%%%%%%%%%%%%%%%%%%%%%%%%%%%%%%%%%%%%%%%%%%%%%%%%%%%%
\subsection{Local Economic Interpretation for Future Predictions}
%%%%%%%%%%%%%%%%%%%%%%%%%%%%%%%%%%%%%%%%%%%%%%%%%%%%%%%%%%%%%%%%%%%%%%%%%%%%%%%%%%%%%%%%%%%%%%%%%%%

\begin{figure}[!ht]
    \centering
    \subfloat[\tiny Crude oil price \label{fig:oilprice}]{\includegraphics[width=.45\textwidth]{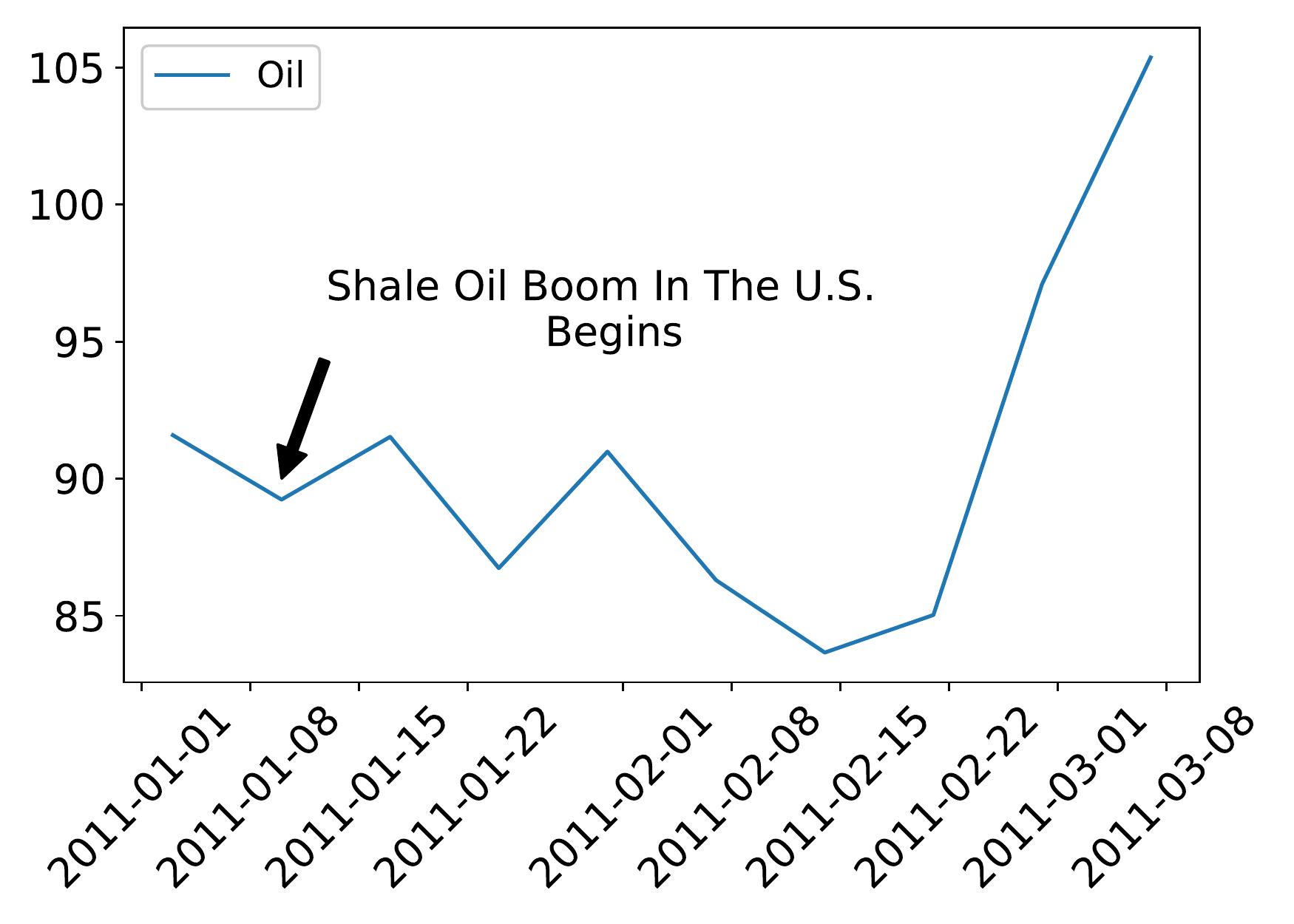}}
    \subfloat[\tiny Exchange rate \label{fig:exchangerate}]{\includegraphics[width=.45\textwidth]{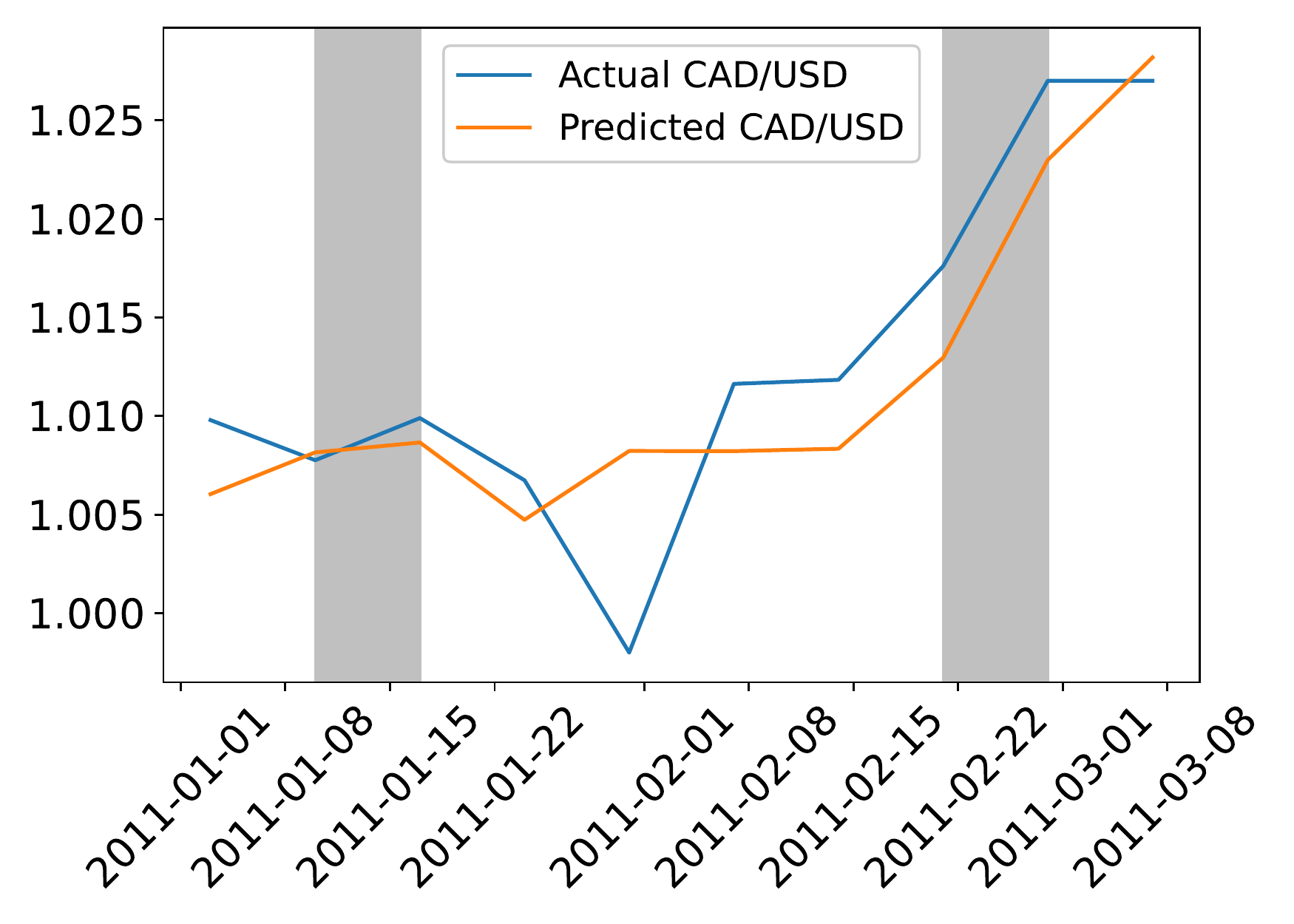}}\\ \subfloat[\tiny SHAP values for the predicted exchange rate in January, 2011  \label{fig:shapjan}]{\includegraphics[width=.9\textwidth]{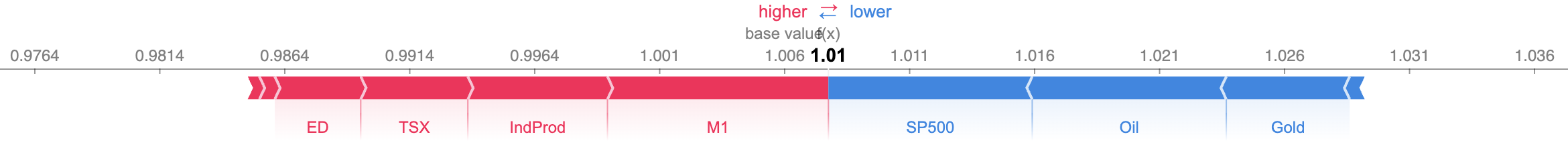}}\\
    \subfloat[\tiny SHAP values  for the predicted exchange rate in late February, 2011   \label{fig:shapfeb}]{\includegraphics[width=.9\textwidth]{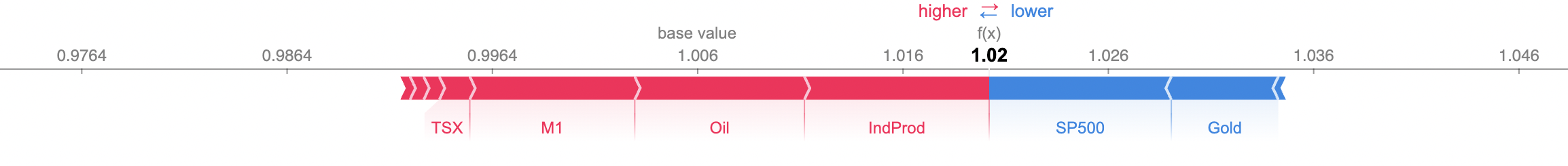}}\\\vspace{.3cm}
    \caption{The contribution of the macroeconomic variables to exchange rate prediction by LGBM around the time of the shale oil boom in the U.S. in January 2011.}
    \label{fig:BoomOil}
\end{figure}

% With time-varying relationships between macroeconomic variables, it is crucial to assess the impact of the variables in future exchange rate predictions. 
The SHAP values can be used to interpret the individual predictions by determining the negative or positive contribution of each variable to each forecast. 
% Retrieving from our previous results, in this section, we aim to analyze the contribution of oil and other variables to single instances of predicted exchange rates at specific dates and on time series forecasts over an important economic cycle.
Figure~\ref{fig:BoomOil} shows the weekly crude oil price time series and the actual and predicted exchange rate values from January to March 2011, along with SHAP plots for two specific dates.
With technological innovation and developments in the U.S. oil industry in January 2011, crude oil production surged in the U.S., contributing to the recovery of its economy. 
The SHAP values for the prediction of the exchange rate in January are shown in Figure~\ref{fig:shapjan}. 
We observe that crude oil is the second negative contributor to the exchange rate value predicted for the time, immediately after the boom. 
This negative relationship is attributable to the fact that the shale oil boom threatened Canada's crude oil exports to the south as the U.S. became less dependent on Canadian oil. 
% Canada must export oil to other countries, such as China, to alleviate the negative impact of the plummeting U.S. oil demand. 
% While the new untraditional oil supply by the U.S. has significant consequences on the global oil market, Canadian investors were advised to monitor other oil suppliers and their decisions, which may alter this artificially elevated oil price and adjust it to their long-term strategic interests. 

Figure~\ref{fig:shapfeb} depicts the impact of the crude oil price on exchange rates in February following this energy revolution. 
This figure shows that due to the rise in crude oil prices and the consequent benefits for Canada, crude oil positively contributes to the increase in the exchange rate. 
This outcome is in line with the claims made in support of the theoretical models about the relationship between the price of crude oil and exchange rate dynamics. 
That is, the increase in crude oil prices inflates the prices of tradable goods relative to nontradables in both U.S. and Canada. 
Since the U.S. is more reliant on crude oil imports than Canada, the increase in the price of tradables relative to nontradables is greater in the U.S. than in Canada. 
This leads to the depreciation of the U.S. dollar. Consequently, the CAD/USD exchange rate increases. 

\begin{figure}[!ht]
    \centering
    \subfloat[\tiny Crude oil price \label{fig:oilprice2}]{\includegraphics[width=.45\textwidth]{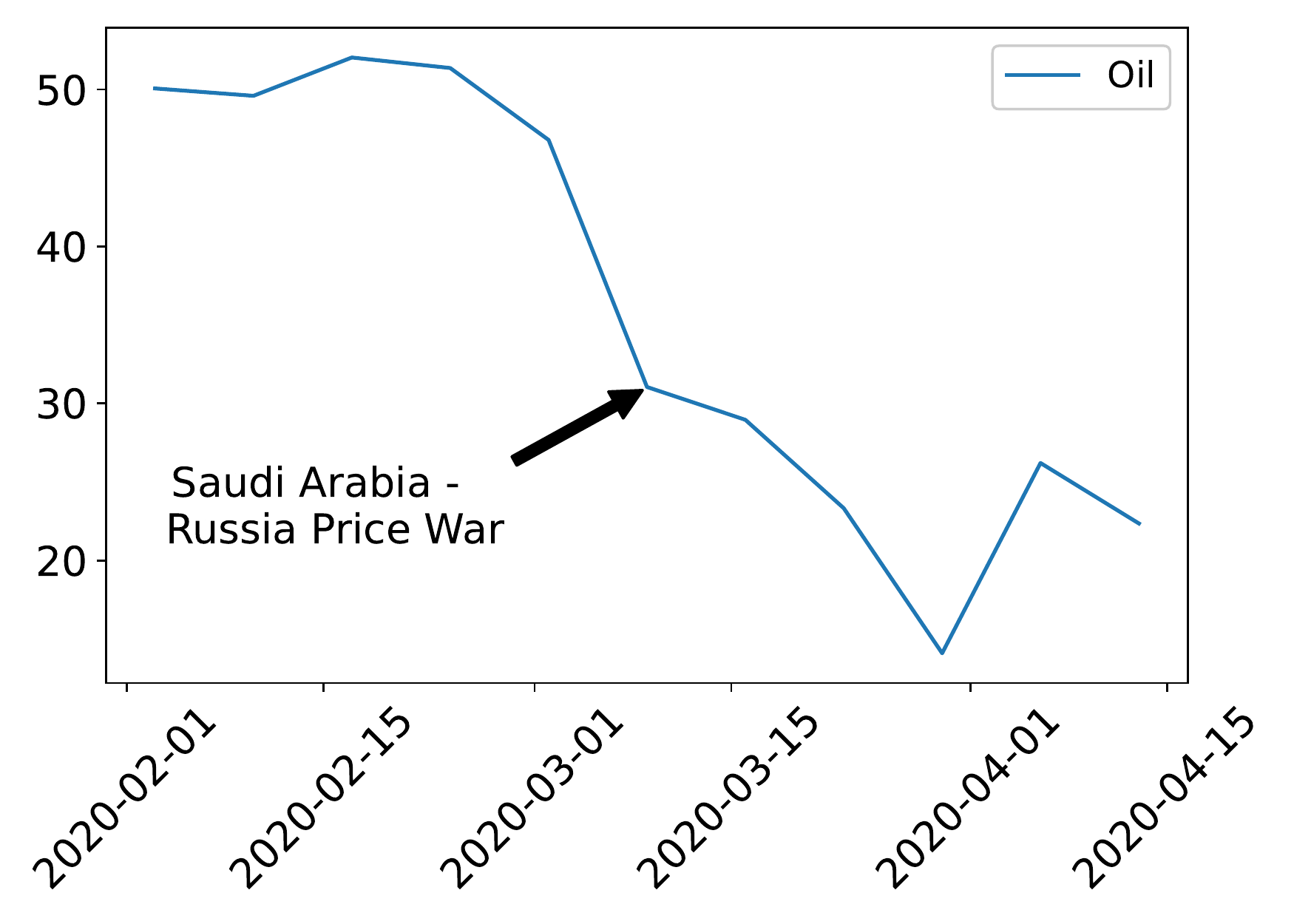}}
    \subfloat[\tiny Exchange rate \label{fig:exchangerate2}]{\includegraphics[width=.45\textwidth]{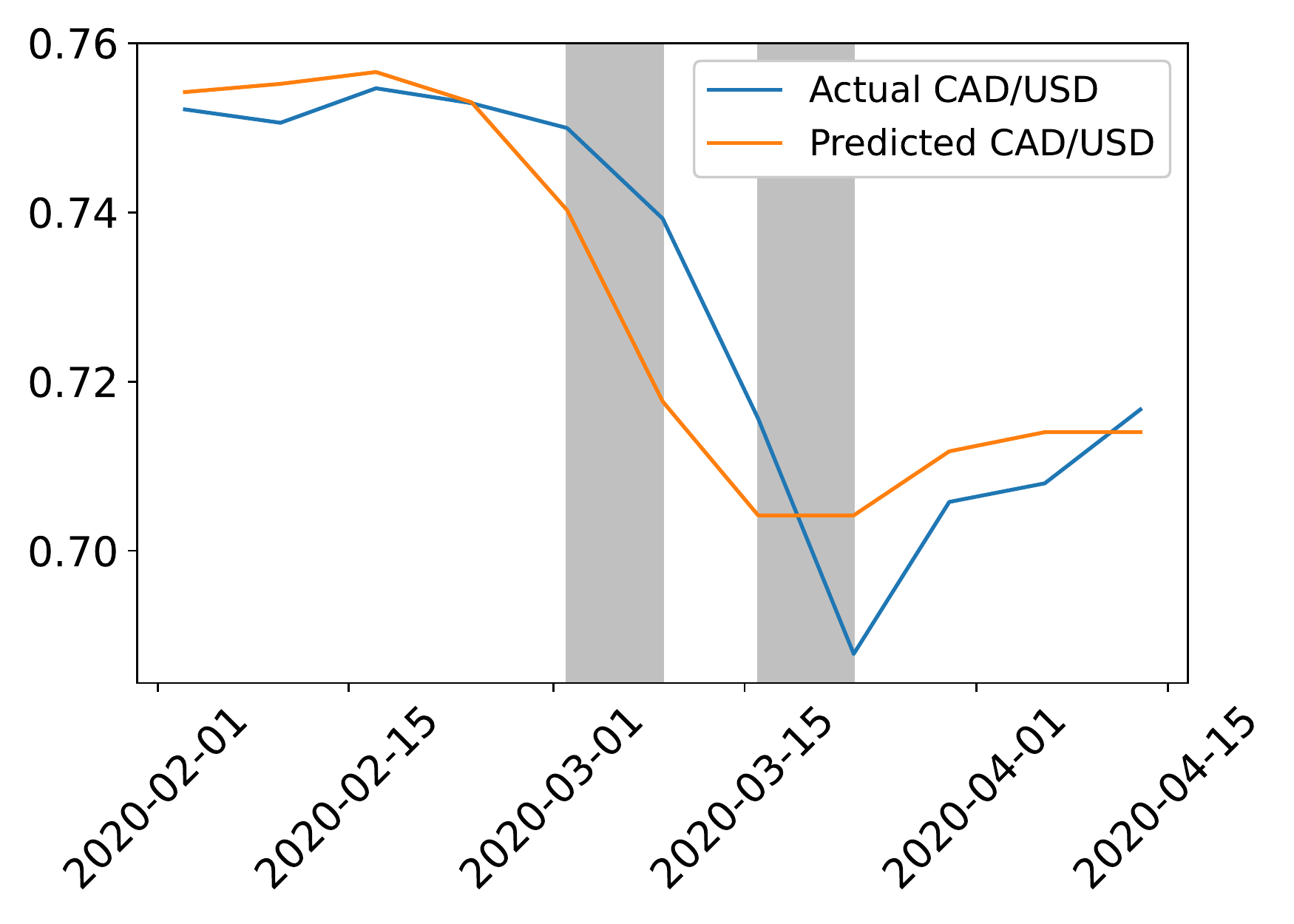}}\\ 
    \subfloat[\tiny SHAP values for  the predicted exchange rate in early March, 2020   \label{fig:shapmarch1}]{\includegraphics[width=.9\textwidth]{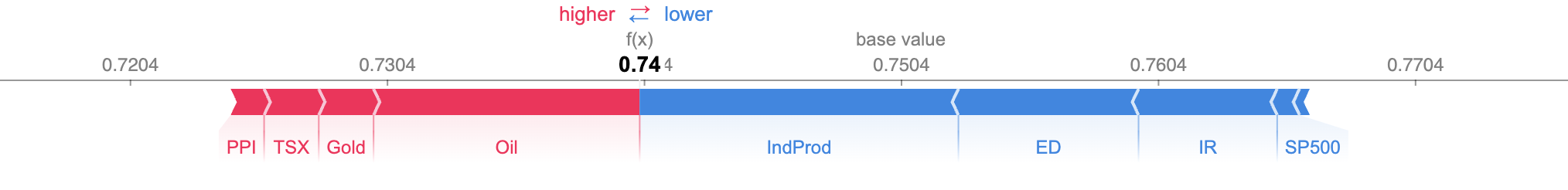}}\\
    \subfloat[\tiny SHAP values for  the predicted exchange rate in mid March, 2020   \label{fig:shapmarch2}]{\includegraphics[width=.9\textwidth]{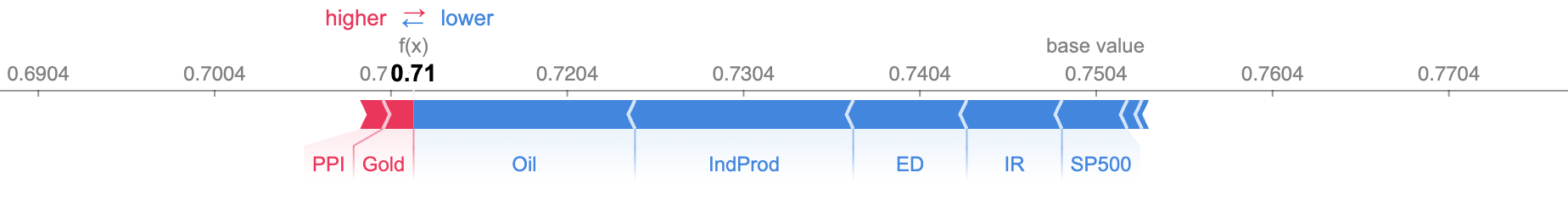}}\\
    \vspace{.3cm}
    \caption{The contribution of the macroeconomic variables to exchange rate prediction by LGBM for the week following the Saudi Arabia and Russia oil price war, which occurred on March 8, 2020.}
    \label{fig:PriceWar}
\end{figure}

We next look into the year 2020, which involves another economic shock due to Covid-19, and examine how the oil market impacts the exchange rate. 
Figure~\ref{fig:oilprice2} shows a sharp decline in the weekly crude oil price by 70\% from February to April 2020. 
This reduction stemmed from several reasons, including the price war between Saudi Arabia and Russia and the lockdown during the Covid-19 pandemic, which in turn led to a drop in crude oil demand. 
Figure~\ref{fig:shapmarch1} shows the SHAP values before the crude oil price shock in March 2020. 
Crude oil is found to positively contribute to the prediction despite a declining trend in the exchange rate.

\begin{figure}[!ht]
    \centering 
    \subfloat{\includegraphics[width=.75\textwidth]{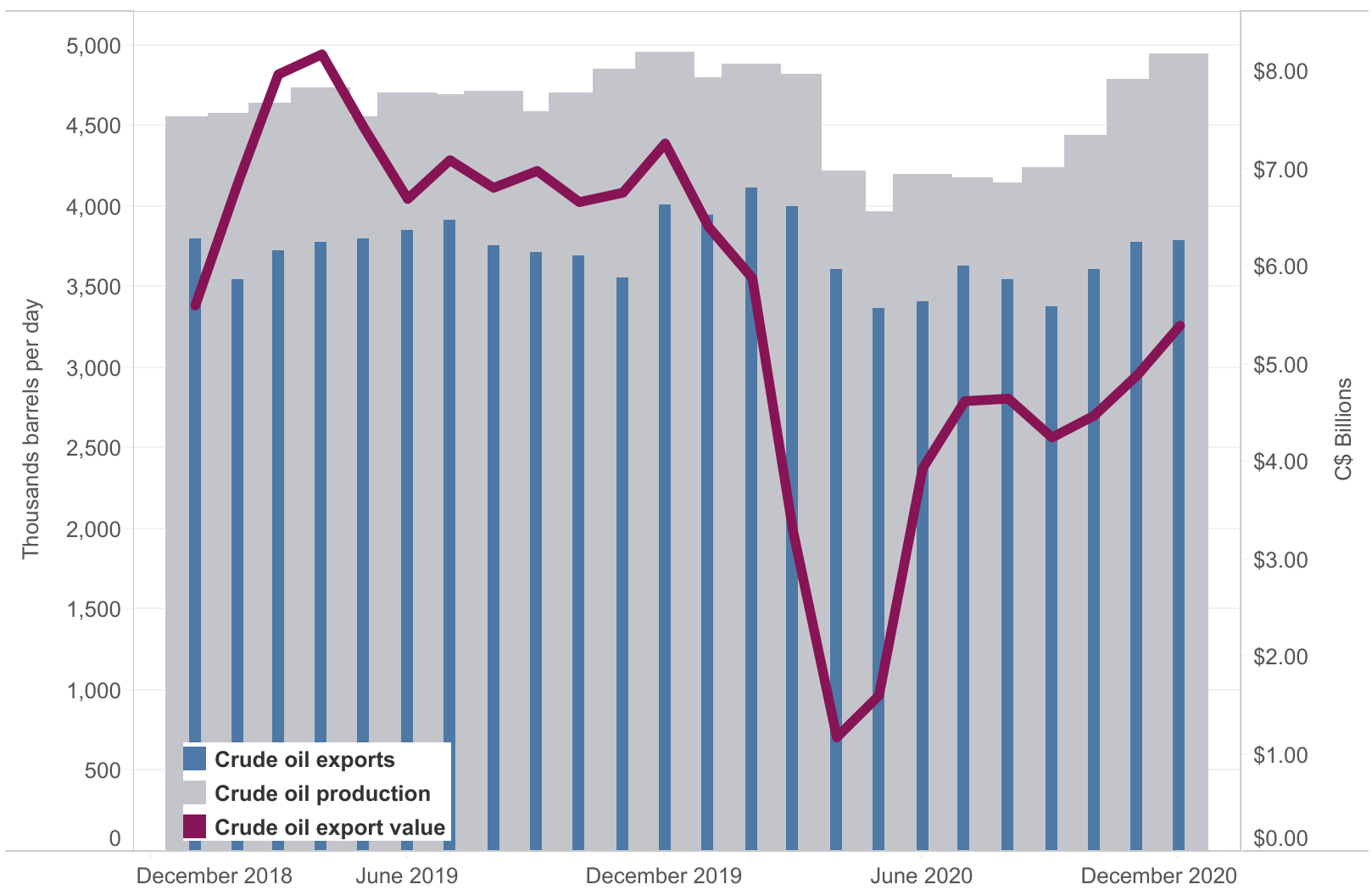}}\\
    \caption{Canadian crude oil production, exports and export value in 2019-2020  (Source:  \url{www.cer-rec.gc.ca}).} \label{fig:export2020}
\end{figure}

After the price war, crude oil exports have been greatly affected by declining prices and low demand. 
Figure~\ref{fig:export2020} shows how crude oil production cuts, due to dropping demand, decreased the crude oil export value sharply in March 2020. 
As a result, the Canadian economy was impacted significantly as it relies heavily on crude oil exports. 
The SHAP values in Figure~\ref{fig:shapmarch2} confirm this argument and indicate that crude oil is the first negative contributor to the exchange rate prediction at that time. 

\begin{figure}[!ht]
    \centering \subfloat{\includegraphics[width=.95\textwidth]{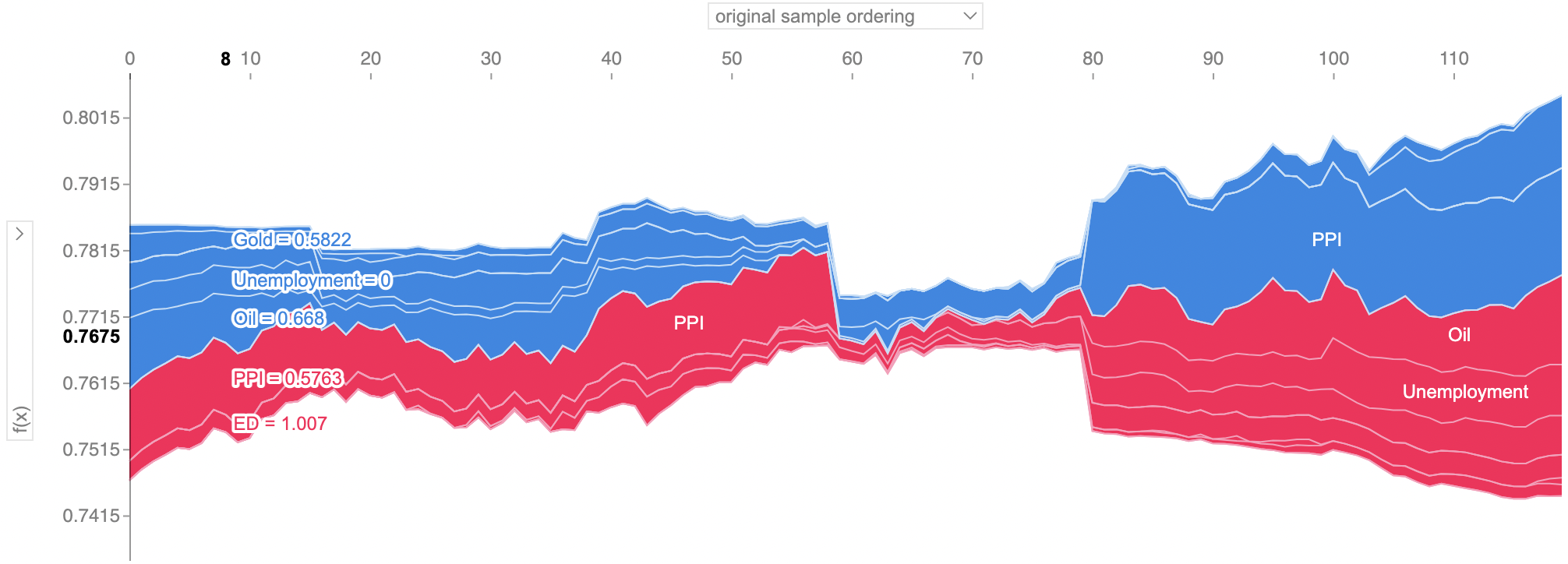}}\\
    \caption{The contribution of the macroeconomic variables to 120 daily exchange rate forecasts by LASSO for the period from July 10 to December 29, 2017.}
    \label{fig:SHAPtimeseries}
\end{figure}

\begin{figure}[!ht]
    \centering 
    \subfloat{\includegraphics[width=.75\textwidth]{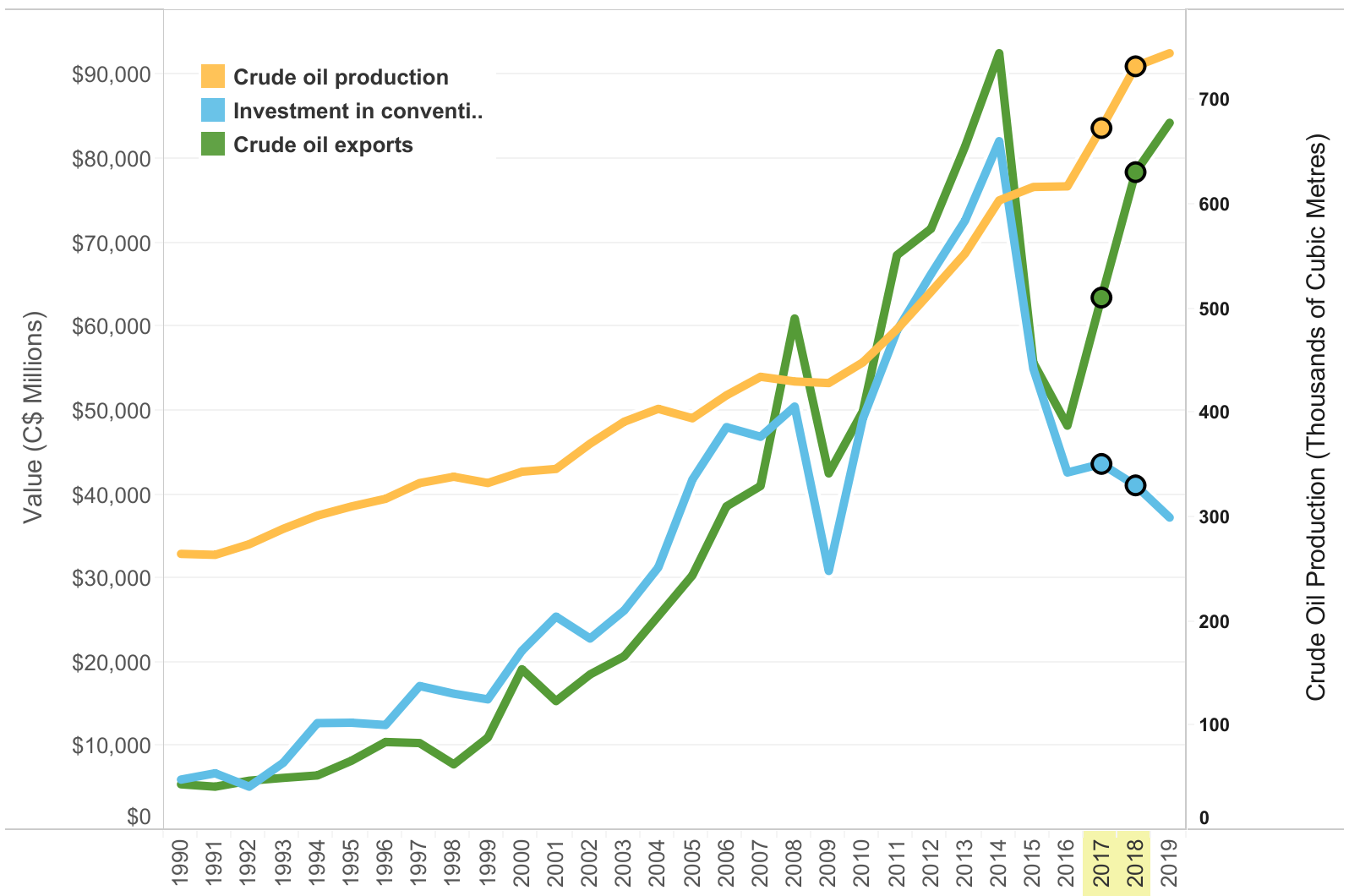}}\\
    \caption{Crude oil export value, oil and gas investment, and crude oil production quantities in Canada (Source: \url{www.cer-rec.gc.ca}).} \label{fig:export2017}
\end{figure}

We also depict the contribution level of the features/variables as a time series during a specific period to better illustrate their contributions for a range of time steps. 
In this respect, we focus on the second half of the year 2017 as it is marked by significant growth in total revenue of the Canadian oil and gas extraction industry. 
As such, it can be of great interest to see how the SHAP values reflect the time-varying contribution of macroeconomic variables to the corresponding exchange rate predictions. 
Figure~\ref{fig:SHAPtimeseries} illustrates the daily time series of the SHAP values in the predictions, consisting of 120 observations from July to December 2107. 
Based on the SHAP values in this figure, starting on July 10, we observe that the negative impact of crude oil steadily diminishes. 
By mid-September (around the $70^{th}$ observation), crude oil has its smallest negative impact, and then it shifts to a positive contributor. 
Eventually, crude oil is converted to the major determinant, among others, which positively affects exchange rate predictions by December 29, 2017, due to the improving situation in Canada's oil market. % that is described in the following.
This observation can be explained by Figure~\ref{fig:export2017}, which shows considerable declines in both investment and export values in 2015 and 2016, mostly due to a sharp decline in the price of crude oil globally. 
After these years, export values started to recover in 2017, while investment levels remained unchanged. 
Although stagnant investment hurt the Canadian economy at the beginning of 2017, earlier investments and growing efficiency have kept production quantities and export values stable. 
Finally, elevated export values have overcome the adverse effect of low investment and turned crude oil into a positive factor in strengthening the economy and exchange rate. 
Our results corroborate the findings in \citep{issa2008turning}, which is about the timing of changes in the sign of this relationship through the structural break tests, and are consistent with major changes in Canada's energy policies and energy-related cross-border trade and investment. 

%The rise in income in 2017 was also a result of higher pricing. Compared to 2016, when the average yearly price of crude oil decreased by 8.0\%, 2017 saw a 19.4\% surge.

%%%%%%%%%%%%%%%%%%%%%%%%%%%%%%%%%%%%%%%%%%%%%%%%%%%%%%%%%%%%%%%%%%%%%%%%%%%%%%%%%%%%%%%%%%%%%%%%%%%%%%%%%%%%%%%%%%%%%%%%%%
\subsection{Ablation Study}
%%%%%%%%%%%%%%%%%%%%%%%%%%%%%%%%%%%%%%%%%%%%%%%%%%%%%%%%%%%%%%%%%%%%%%%%%%%%%%%%%%%%%%%%%%%%%%%%%%%%%%%%%%%%%%%%%%%%%%%%%%
Lastly, we estimate the effectiveness of the forecasting models using different input parameter combinations to understand whether they can deliver more accurate forecasts and/or achieve competitive performance using fewer input variables that lead to less complex models. 
In this regard, we assess strategies for time-lagged exchange rate inclusion and incremental analysis with the most important features. % to reach these goals, respectively.

%%%%%%%%%%%%%%%%%%%%%%%%%%%%%%%%%%%%%%%%%%%%%%%%%%%%%%%%%%%%%%%%%%%%%%%%%%%%%%%%%%%%%%%%%%%%%%%%%%%%%%%%%%%%%%%%%%%%%%%%%%
\subsubsection{Time-lagged Input Inclusion}%\hfill\\
%%%%%%%%%%%%%%%%%%%%%%%%%%%%%%%%%%%%%%%%%%%%%%%%%%%%%%%%%%%%%%%%%%%%%%%%%%%%%%%%%%%%%%%%%%%%%%%%%%%%%%%%%%%%%%%%%%%%%%%%%%
The existing theories and empirical studies suggest that previous observations contain useful information with predictive power for future values in time series forecasting~\citep{ince2006hybrid, su2022exchange}.
Accordingly, we incorporate lagged exchange rates into the input set for each model.
% Accordingly, we follow the intuition of similar studies that incorporate lagged exchange rates into the input set of the model to improve forecasting~\citep{ince2006hybrid, su2022exchange}. 
% To implement this experiment, we update the input data by adding the lagged exchange rates to the  macroeconomic variables. 
Table~\ref{tab:lag_effect} shows the result for the experiments with the time-lagged inputs. 
Specifically, we report the percentage reduction in NRMSE gained by models with additional inputs of time-lagged exchange rate values compared to the corresponding performance of models without the time-lagged information from Tables~\ref{tab:dailyPrediction} and~\ref{tab:weeklyPrediction}. 
We choose the number of lagged exchange rate values as five by following the earlier studies that determine the best value based on autocorrelation analysis and different criteria such as the Akaike information criterion. 
% In this experiment, we incorporate 5 lag values of exchange rate that comprise entire calendar features for the preceding period (i.e., cover one week for daily data and one month for weekly data). 
We find most values in this table to be positive, indicating that the performance of models improves if additional inputs of lagged data are given to the models. 
The average NRMSE reduction across all models is only negative for 5-week and 10-week forecast horizons. 
For the dataset ``All (2009-2021)'', we observe that the RIDGE and LASSO models benefit from the lagged inputs more than any other model. 
For instance, in the 1-day forecast horizon, LASSO performs about 84.90\% better than the same model without lagged data, while the performance of GRU increases only by 19.10\%. 

\begin{table}[!ht]
\centering
\caption{Result of prediction models with different inputs. The values are the percentage reduction in NRMSE for the new input set containing lagged target variable compared to the corresponding results without lagged data. A higher positive value shows that incorporating lags to the inputs improves the forecast more.}
\label{tab:lag_effect}
\resizebox{.98\textwidth}{!}{
\begin{tabular}
{L{3.8cm}L{2cm}R{2.3cm}R{2.3cm}R{2.3cm}R{2.3cm}R{2.3cm}R{2.3cm}}
\toprule
&&\multicolumn{3}{c}{Daily}&\multicolumn{3}{c}{Weekly}\\
\cmidrule(lr){3-5} \cmidrule(lr){6-8}
Period& Method& 1-day& 5-day&10-day& 1-week&5-week& 10-week\\
\midrule 
All (2009-2021)
&LGBM & 45.24 &  22.71 &  10.92 &  27.26 &   1.91 &  -0.83 \\
&ETR &33.08 &  28.50 &  25.52 &  23.04 &  14.05 &   8.03 \\
&XGB&  44.50 &  34.41 &  28.32 &  29.82 &  10.01 &   3.01 \\
&RIDGE &  79.84 &  59.64 &  55.27 &  69.27 &  15.87 & -51.13 \\
&LASSO &  84.90 &  75.97 &  59.01 &  88.38 &  25.60 & -15.71 \\
&GRU & 19.10 &   8.61 & 3.50 &   9.02 &   6.25 &   1.15 \\
\rowcolor{Gainsboro!60}  &  Average & 51.11 & 38.31&  30.42&  41.137 & 12.287&  -9.25\\
\midrule
Economic Expansion &  LGBM  &  55.63 &   9.61 & -17.96 &  35.78 &   3.86 &   7.07 \\
(2009-2011)&ETR &29.59 &  25.88 &  14.51 &  24.17 &   1.27 &   2.66 \\
&XGB&38.10 &  27.45 &  20.95 &  34.31 & -15.65 &   5.13 \\
&RIDGE &56.59 &  20.89 &  23.86 &  30.50 & -15.92 &  16.41 \\
&LASSO &61.52 &  22.76 &  28.29 &  60.92 &  -4.22 & -52.76 \\
&GRU &-16.64 &   2.50 &   0.99 &   1.37 &   6.52 &   7.88 \\ 
\rowcolor{Gainsboro!60} & Average&  37.47 &  18.18 &  11.77 &  31.18 &  -4.02 &  -2.27 \\
\midrule
Economic Stagnation&
LGBM & 35.65 &  17.96 &   5.11 &  11.25 &   7.45 &  -2.33 \\
(2014-2016) &ETR & 27.89 &  13.37 &   4.11 &  -1.95 &  -8.84 &  12.01 \\
&XGB&15.83 &  25.67 &  -7.52 &  19.27 &   0.29 &  14.54 \\
&RIDGE &81.24 &  42.76 &  34.43 &  82.64 &  10.61 &   1.10 \\
&LASSO &85.80 &  63.80 &  24.92 &  86.13 & -48.69 &  25.19 \\
&GRU &13.21 &  17.51 &   9.59 &  12.46 &  12.44 &  -4.95 \\
 \rowcolor{Gainsboro!60}  &Average &43.27 &  30.18 &  11.77 &  34.97 &  -4.46 &   7.59 \\
\midrule
Covid (2019-2021)
&LGBM & -7.89 &  20.76 &   0.71 &  -8.25 &   -2.06 &  -6.80 \\
&ETR &15.03 &  14.92 &  14.50 &   7.59 &  -24.73 & -30.18 \\
&XGB&-3.67 &  -4.82 &  10.69 &   1.28 & -116.62 & -50.10 \\
&RIDGE &74.76 &   1.47 &  59.15 &  18.28 &  -13.43 &  61.05 \\
&LASSO & 82.98 &   6.06 &  47.78 &  79.65 &   -2.81 & -21.57 \\
&GRU &40.89 &   2.78 &   0.85 &   3.87 &   -9.29 &  -9.27 \\
\rowcolor{Gainsboro!60}  & Average&33.68 &   6.86 &  22.28 &  17.07 &  -28.16 &  -9.48 \\
\bottomrule
\end{tabular}
}
\end{table}

%Figure~\ref{fig:transformation} denotes an example for time instance $t$ and how previous $\ell$ lagged target variables $(y_{t-\ell}, y_{t-\ell+1}, \cdots, y_{t})$ and macroeconomic feature vector $(x_{1,t}, x_{2,t}, \cdots, x_{N,t})$ are concatenated to use as a single vector of inputs to forecast exchange rates for the  next $h$ periods, i.e., $(y_{t+1}, y_{t+2}, \cdots,y_{t+h})$.   

% \begin{figure}[!ht]
%     \centering
%     {\includegraphics[width=1\textwidth]{Figure/transformation.pdf}}\\\vspace{.3cm}
%     \caption{Data transformation for an example input. Preceding $\ell$ time-lagged target variables are concatenated with other inputs at time $t$ to predict the following $h$ periods ahead.}
%     \label{fig:transformation}
% \end{figure}

The positive effect of incorporating lagged exchange rate on the prediction performance decreases in two directions: (1)~increasing the forecast horizon, and (2)~decreasing the frequency of data from daily to weekly. 
The first evidence supports our previous finding and confirms that it is intuitive that the explanatory power between consecutive values of exchange rate decreases in the long term. 
While the second piece of evidence is related to overfitting. 
In our dataset, as we resample data weekly, the number of observations is divided by 5, which means that the number of training data decreases by 80\% in each train window compared to daily observations. 
On the other hand, when we incorporate more inputs, the number of parameters increases in the models. 
Therefore, the ratio of samples to inputs gets lower for these new models compared to corresponding previous models with no lag, leading to no improvement. 
The overfitting issue is not observed for daily data, as the number of observations is sufficient even after making the models more complex by feeding lagged input to the models.

In accordance with different economic cycles, our results demonstrate that it is worth analyzing and considering lagged input for the ``Economic Stagnation (2014-2016)'' dataset as the exchange rate series possesses greater sequential impact than in the other subperiods. 
For example, the average of NRMSE reductions according to 1-day and 1-week forecast horizons for ``Economic Stagnation (2014-2016)'' are 43.27\% and 34.97\%, respectively, which is much higher than in other datasets/subperiods (e.g., ``Economic Expansion (2009-2011)'' has NRMSE reductions of 37,47\% and 31.18\%, and ``Covid (2019-2021)'' has NRMSE reductions of  33.68\% and 17.07\%).

%%%%%%%%%%%%%%%%%%%%%%%%%%%%%%%%%%%%%%%%%%%%%%%%%%%%%%%%%%%%%%%%%%%%%%%%%%%%%%%%%%%%%%%%%%%%%%%%%%%%%%%%%%%%%%%%%%%%%%%%%%
% \subsubsection{Incremental Effect of Top Variables}%\hfil\\
\subsubsection{Incremental Analysis with External Covariates}
\begin{figure}[!ht]
    \centering
    \vspace{-3.5cm} {\includegraphics[width=.70\textwidth]{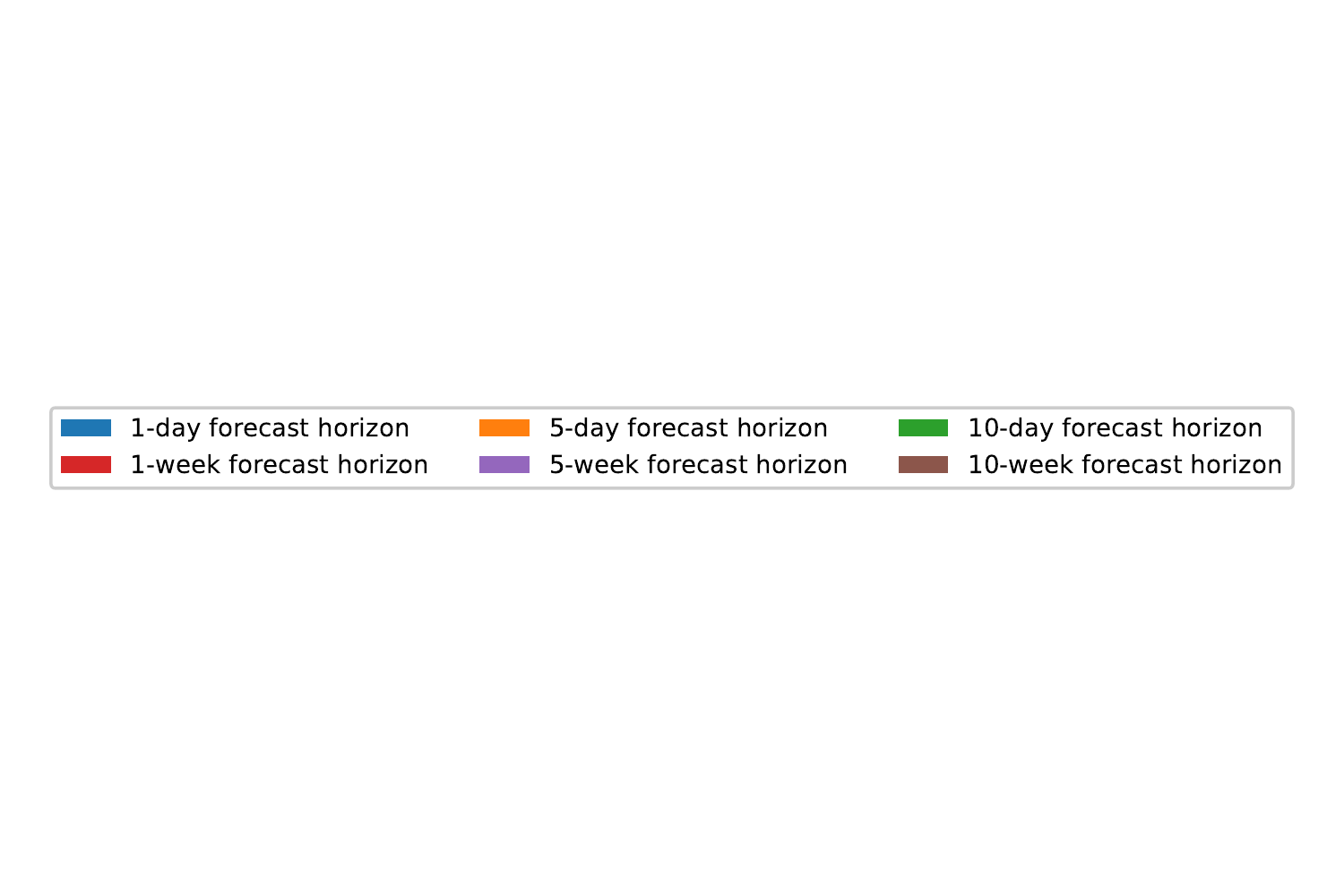}} \\\vspace{-3.5cm}
    \subfloat[\tiny All-CalFt \label{fig:allcalft}]{\includegraphics[width=.24\textwidth]{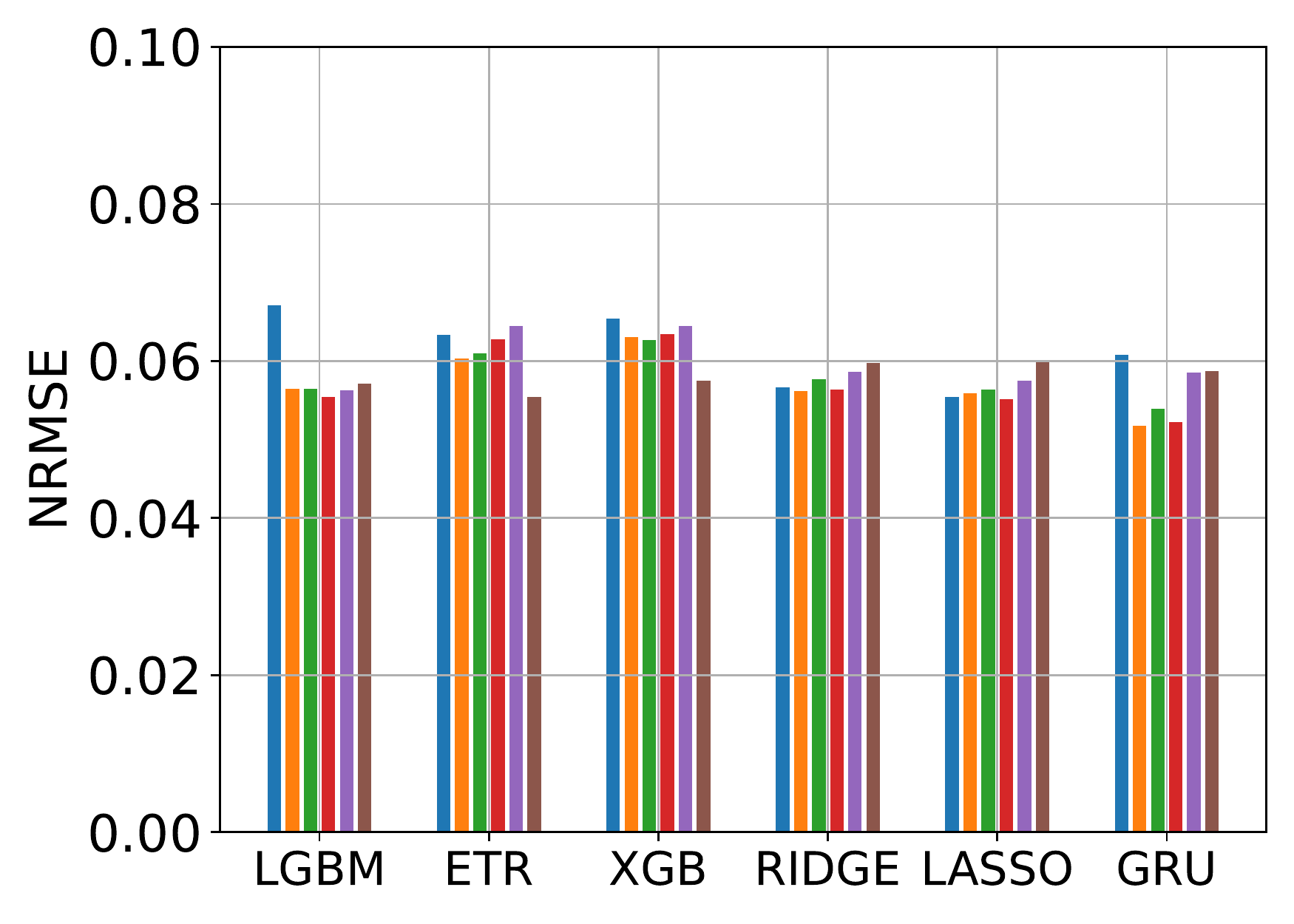}}
    \subfloat[\tiny All-CalFt,Oil \label{fig:allcalftoil}]{\includegraphics[width=.24\textwidth]{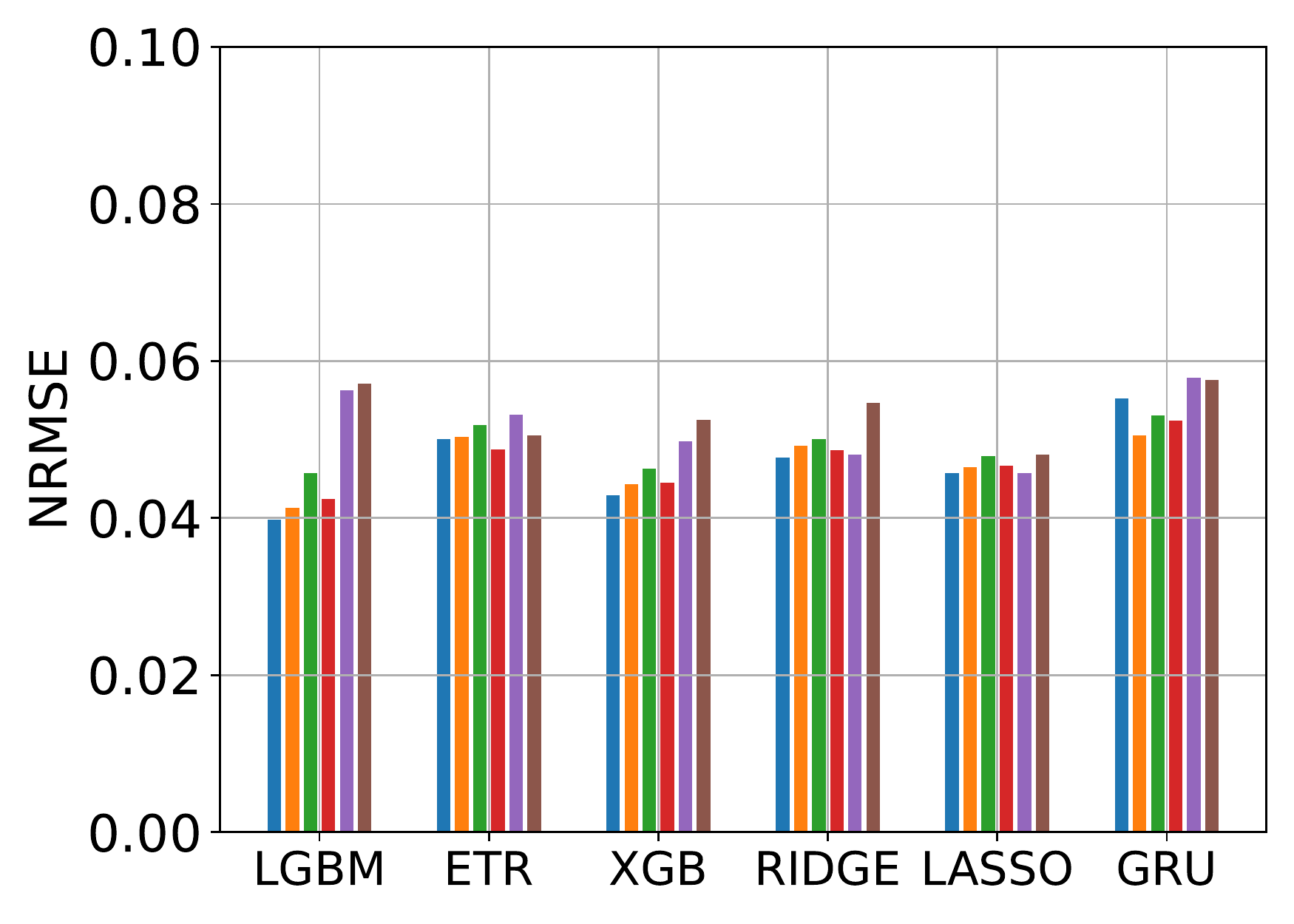}}
    \subfloat[\tiny All-CalFt,Oil,Gold \label{fig:allcalftoilgold}]{\includegraphics[width=.24\textwidth]{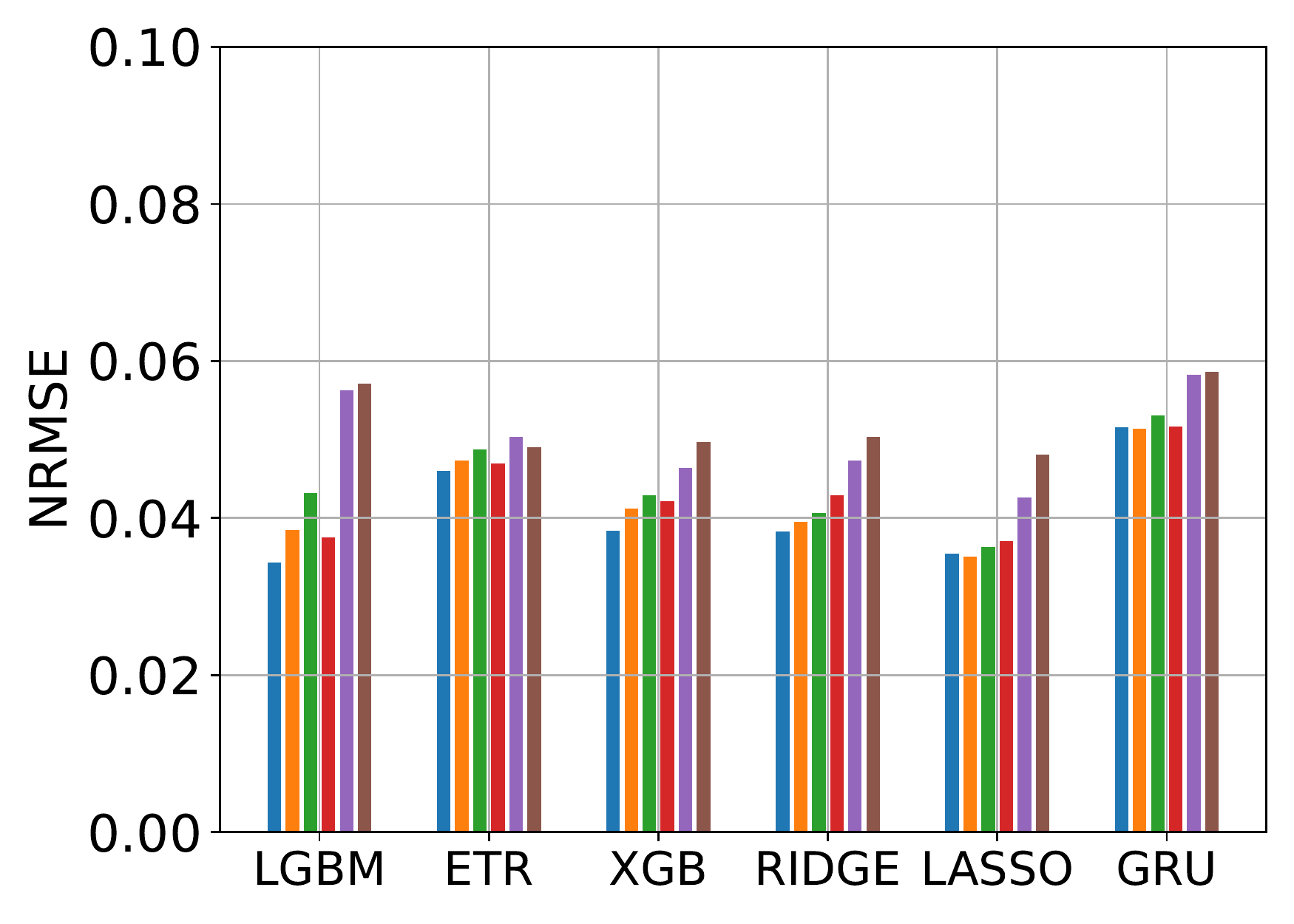}}
    \subfloat[\tiny All-CalFt,Oil,Gold,TSX\label{fig:allcalftoilgoldtsx}]{\includegraphics[width=.24\textwidth]{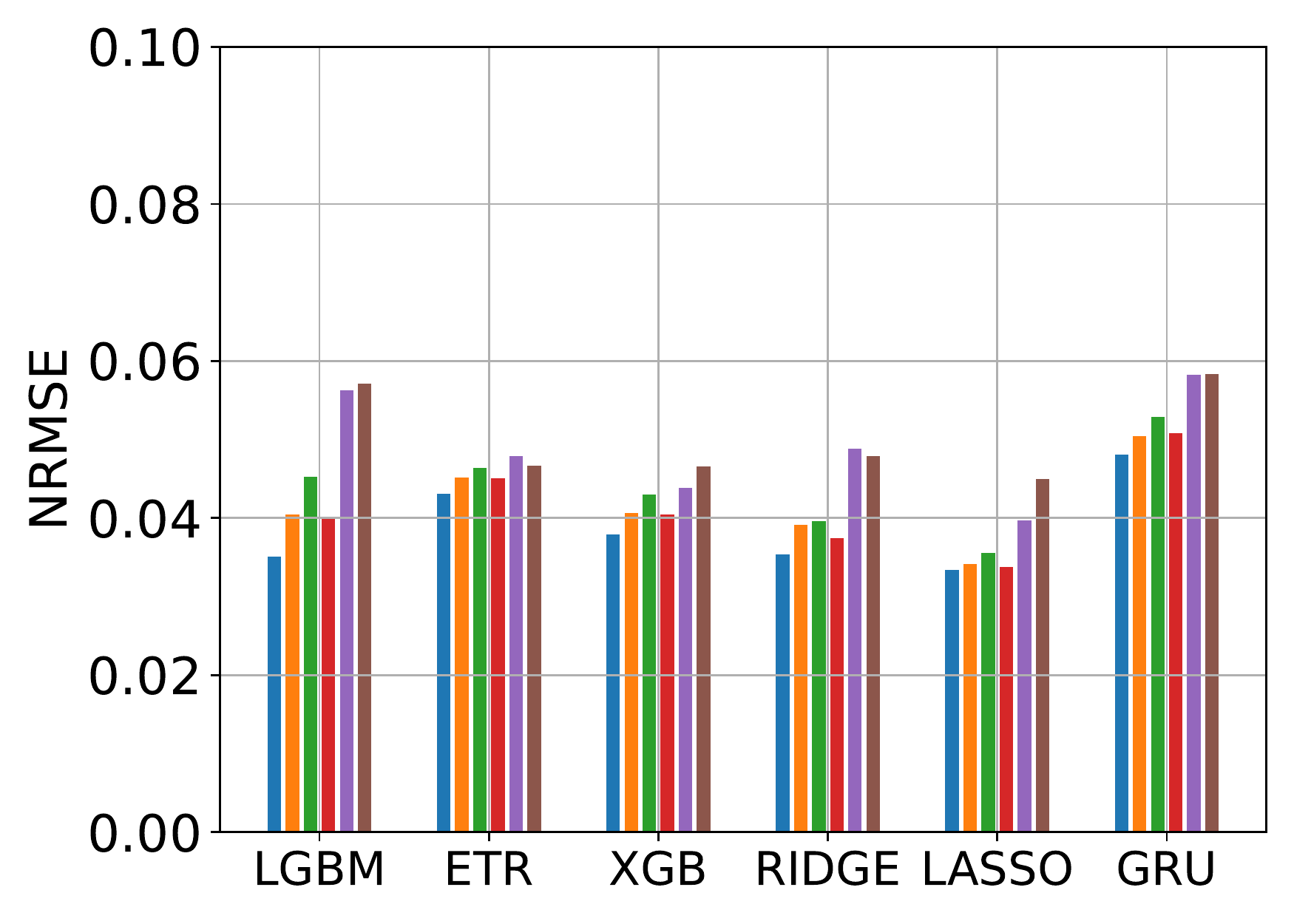}}\\ \vspace{-.2cm}
    \subfloat[\tiny EcoE-CalFt \label{fig:ecoecalft}]{\includegraphics[width=.24\textwidth]{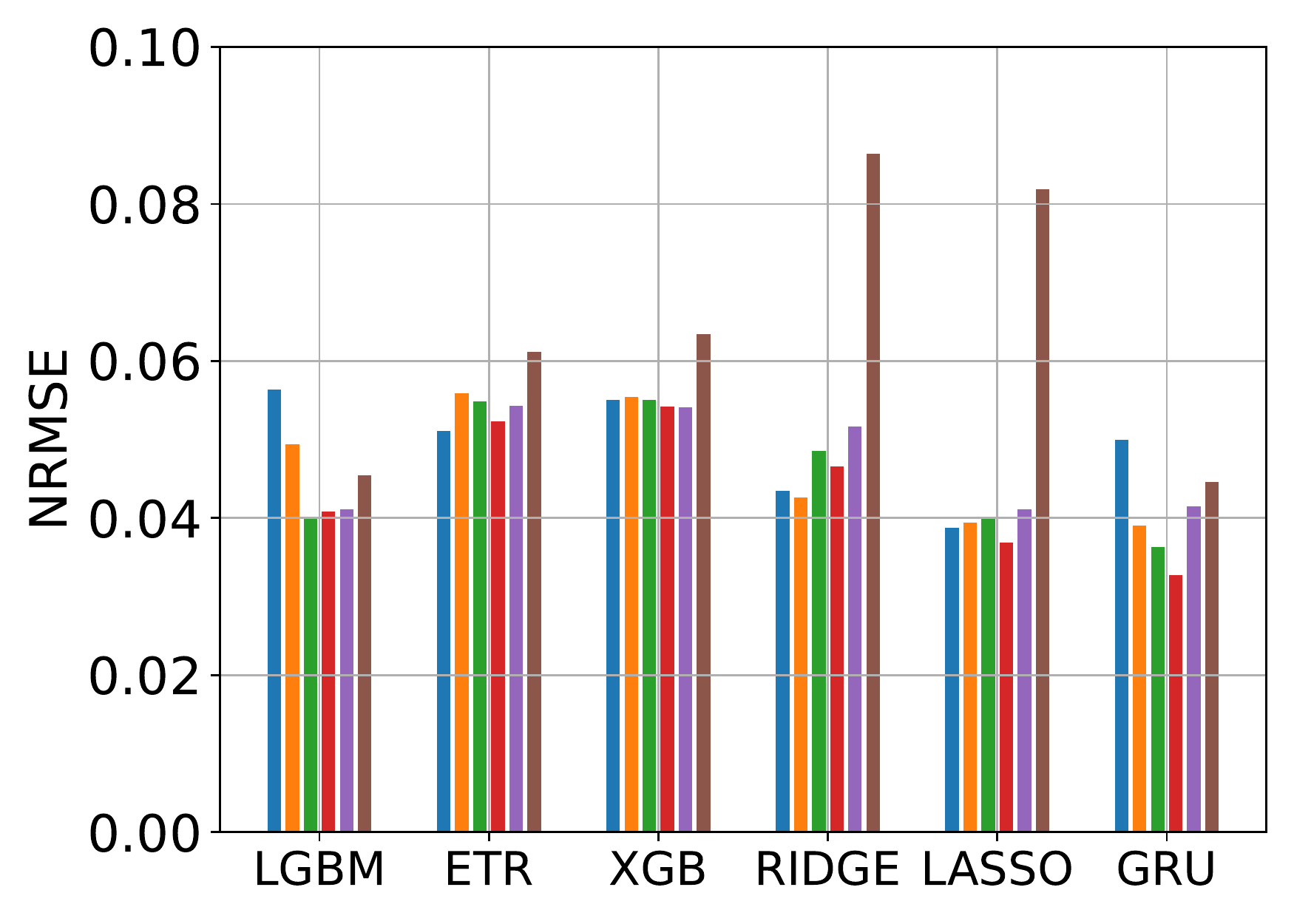}}
    \subfloat[\tiny EcoE-CalFt,Oil \label{fig:ecoecalftoil}]{\includegraphics[width=.24\textwidth]{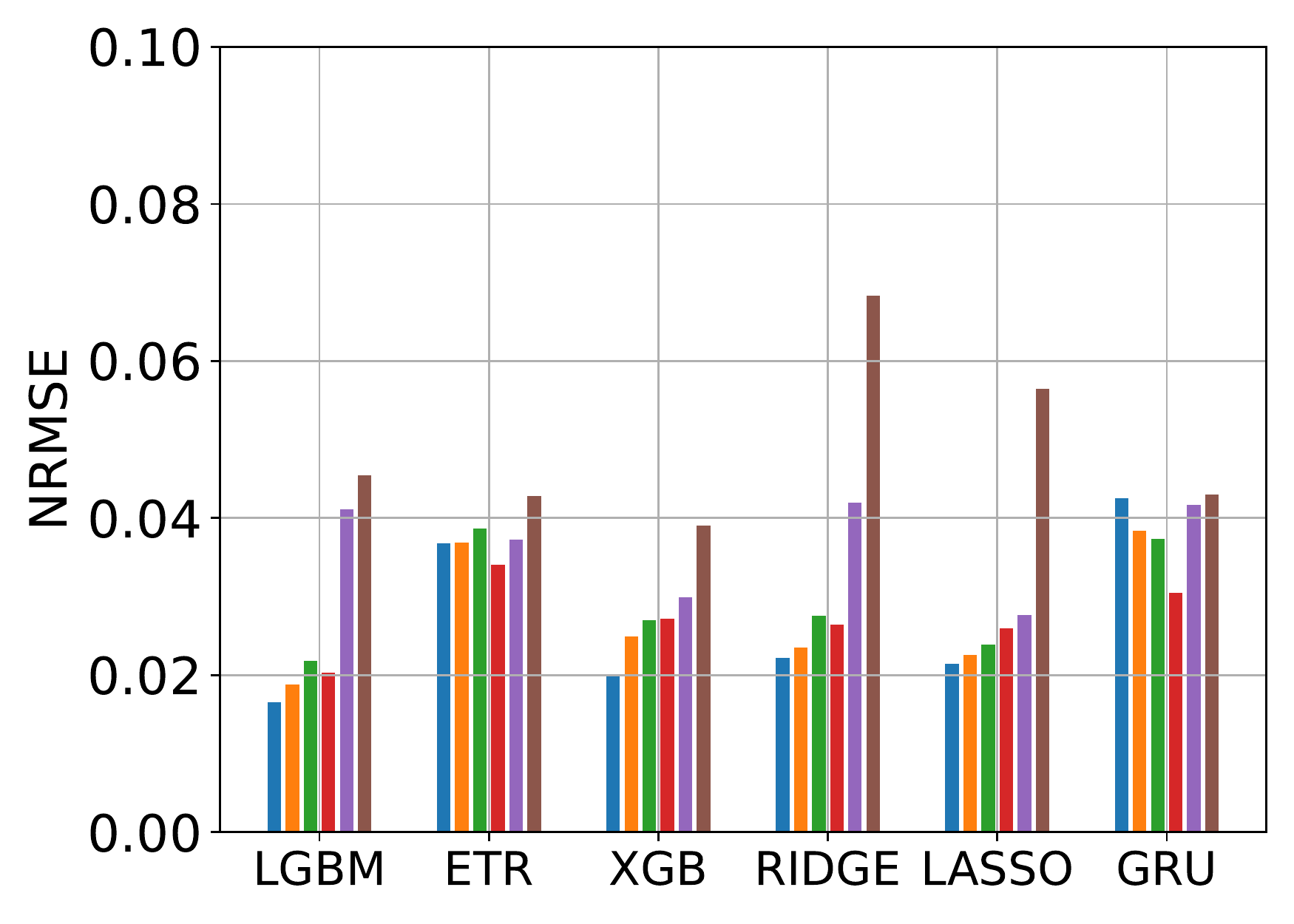}}
    \subfloat[\tiny EcoE-CalFt,Oil,Gold \label{fig:ecoecalftoilgold}]{\includegraphics[width=.24\textwidth]{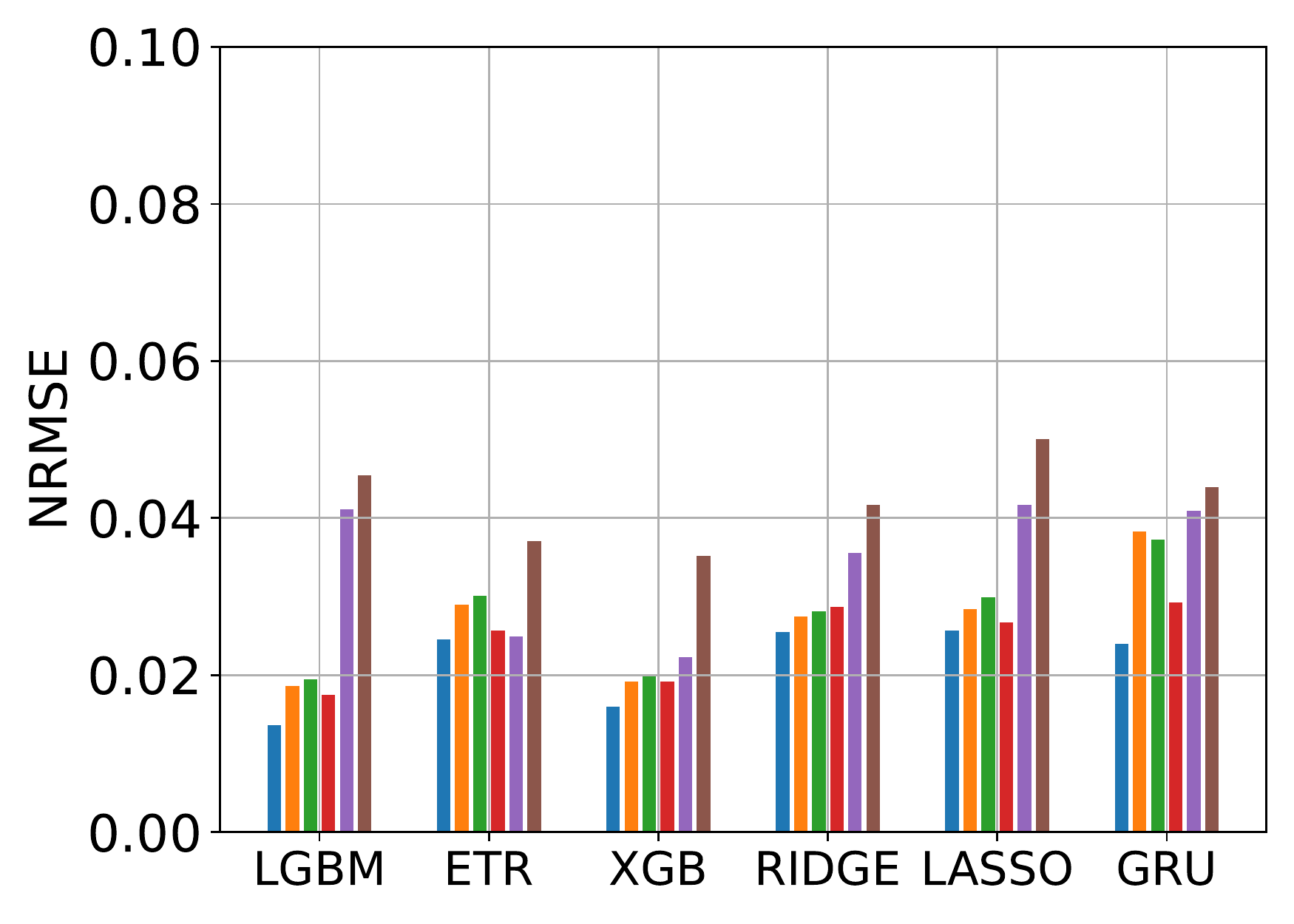}}
    \subfloat[\tiny EcoE-CalFt,Oil,Gold,TSX \label{fig:ecoecalftoilgoldtsx}]{\includegraphics[width=.24\textwidth]{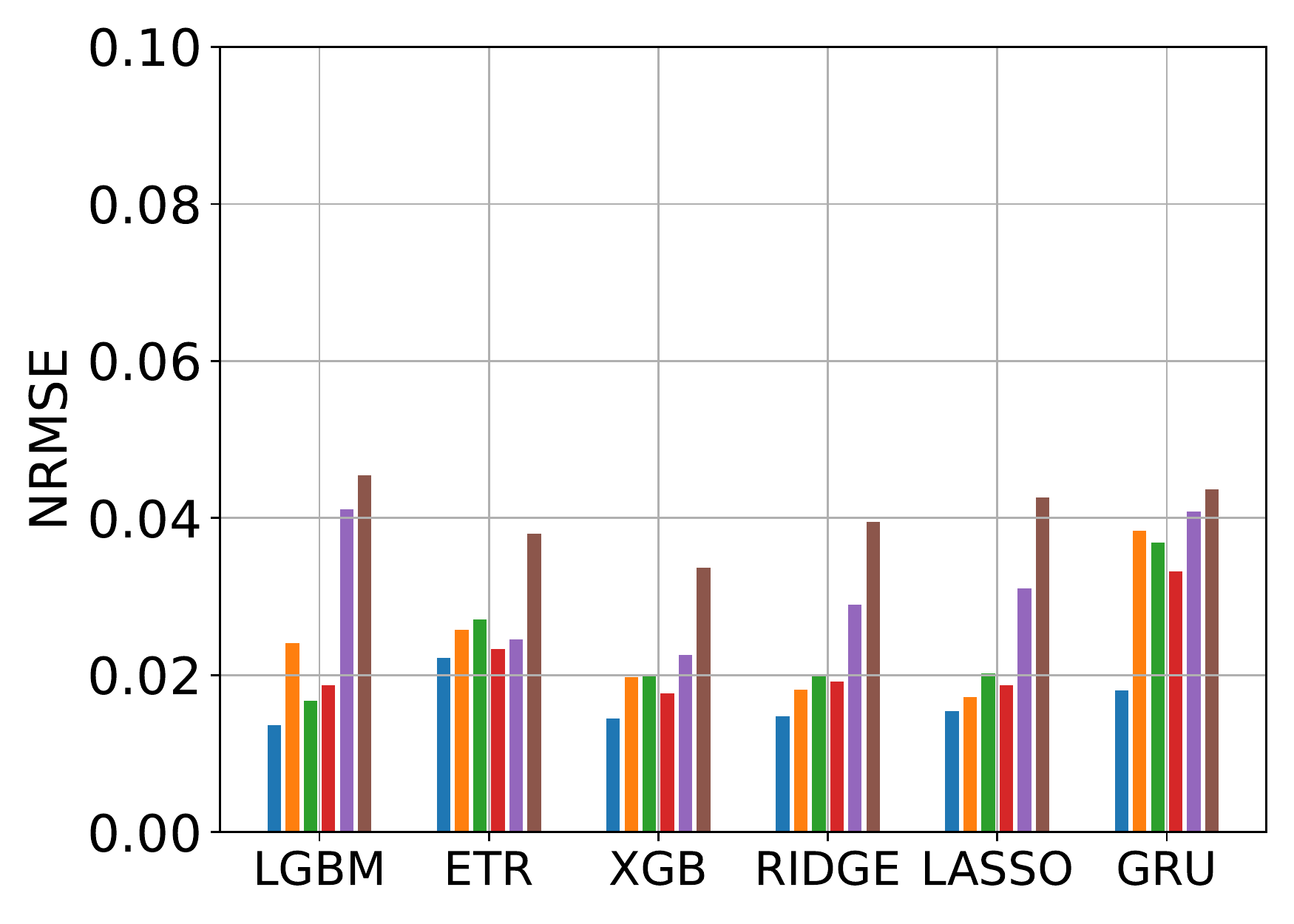}}\\ \vspace{-.2cm}
    \subfloat[\tiny EcoS-CalFt \label{fig:ecorcalft}]{\includegraphics[width=.24\textwidth]{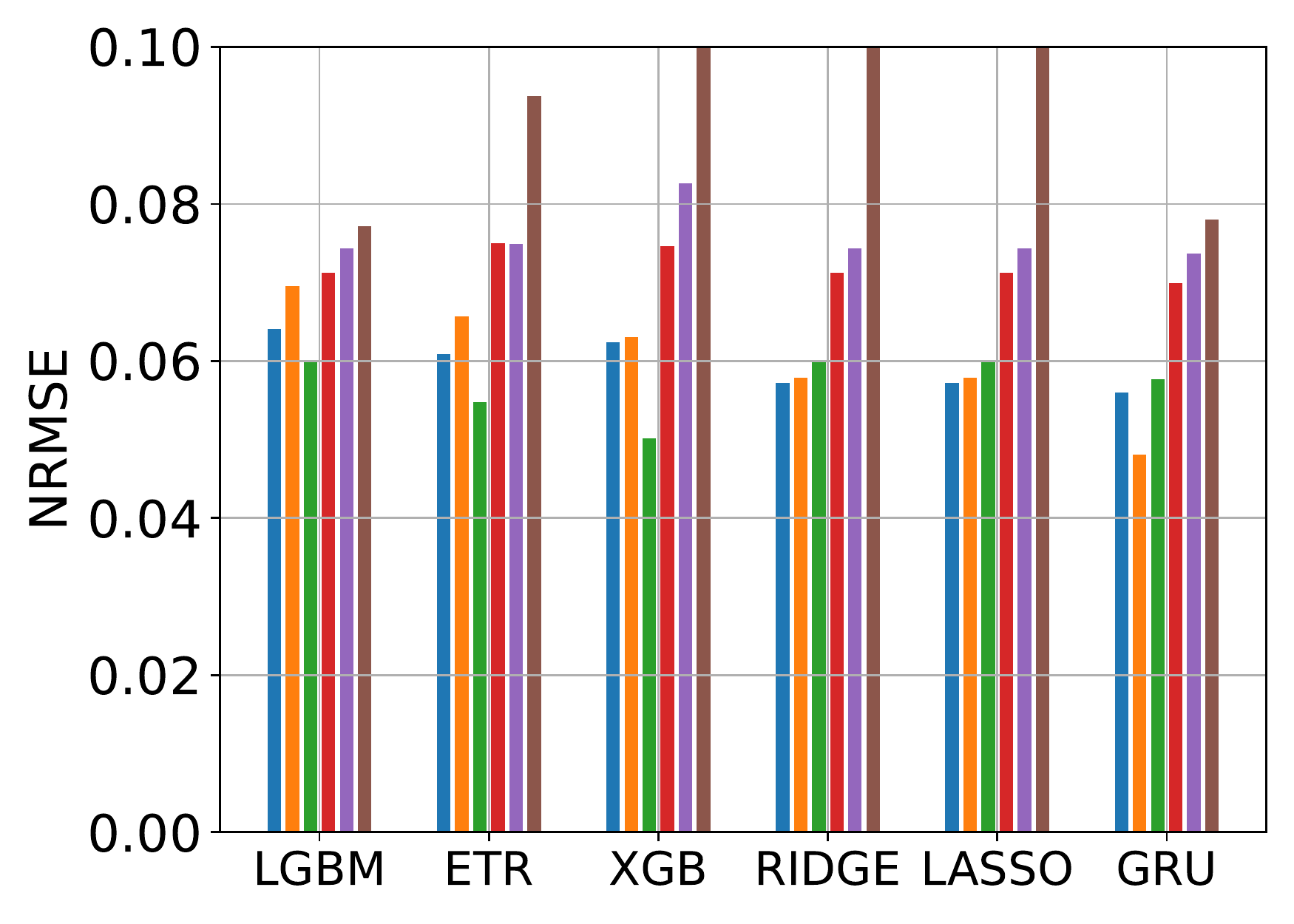}}
    \subfloat[\tiny EcoS-CalFt,Oil \label{fig:ecorcalftoil}]{\includegraphics[width=.24\textwidth]{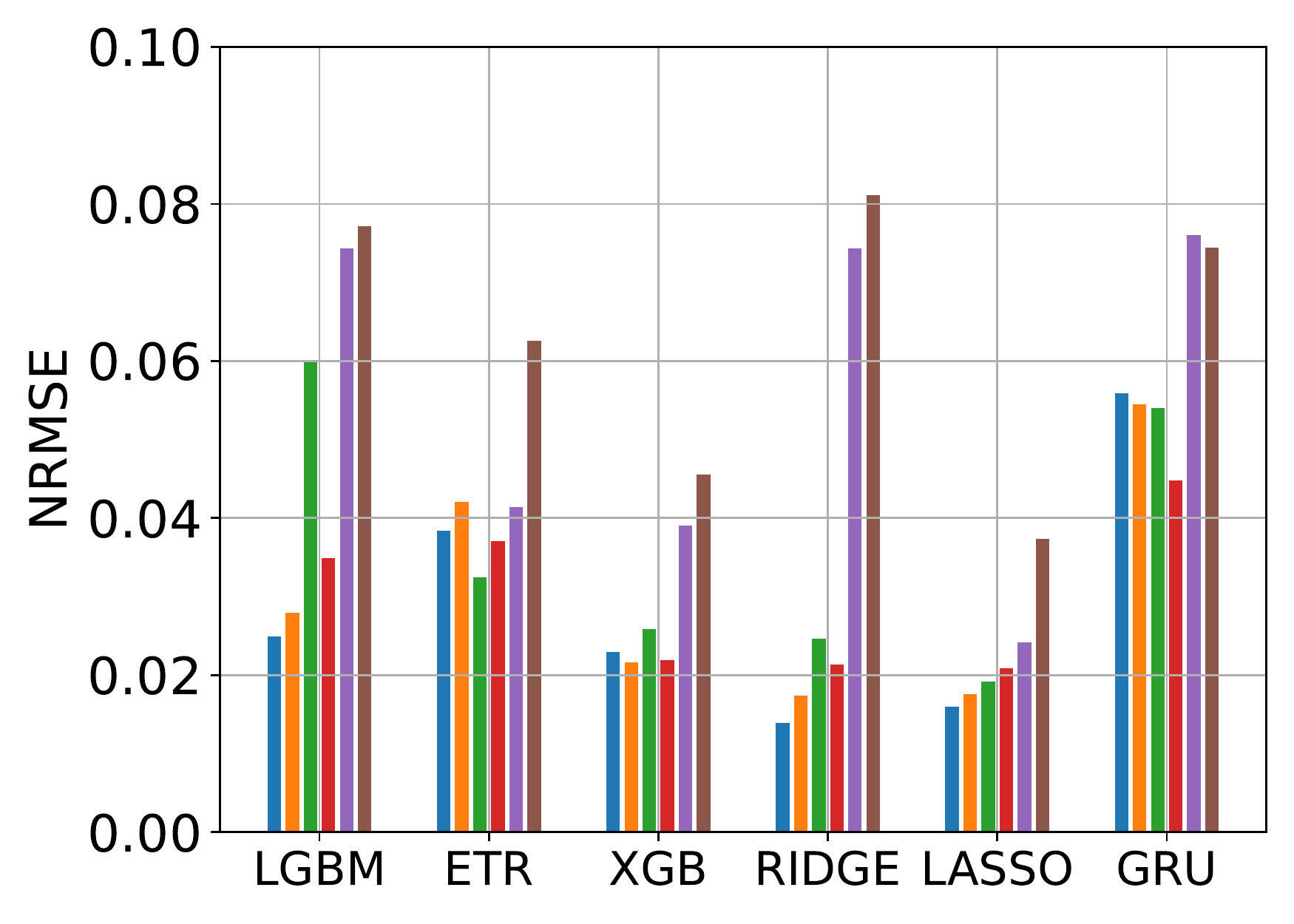}}
    \subfloat[\tiny EcoS-CalFt,Oil,Gold \label{fig:ecorcalftoilgold}]{\includegraphics[width=.24\textwidth]{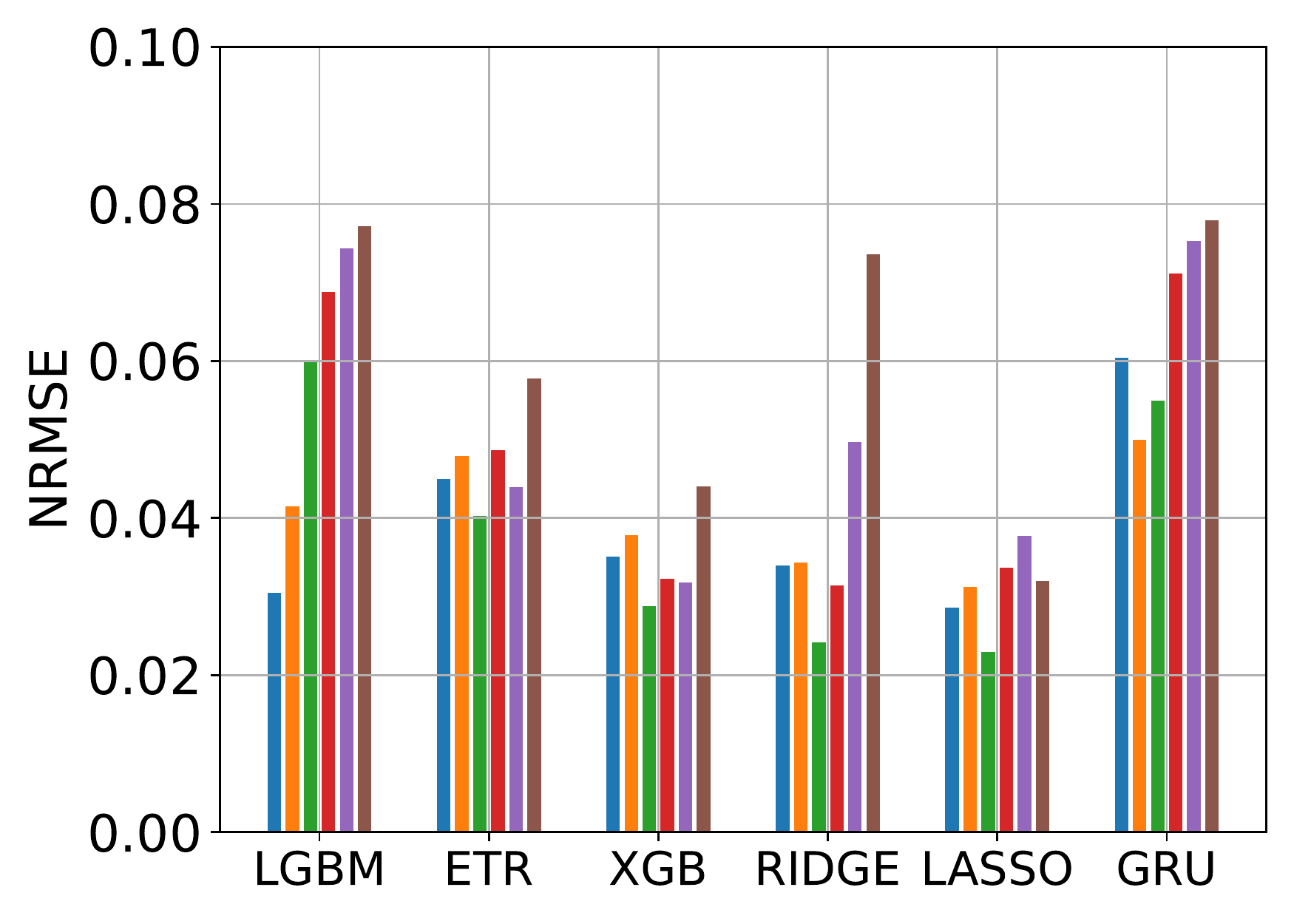}}
    \subfloat[\tiny EcoS-CalFt,Oil,Gold,TSX \label{fig:ecorcalftoilgoldtsx}] {\includegraphics[width=.24\textwidth]{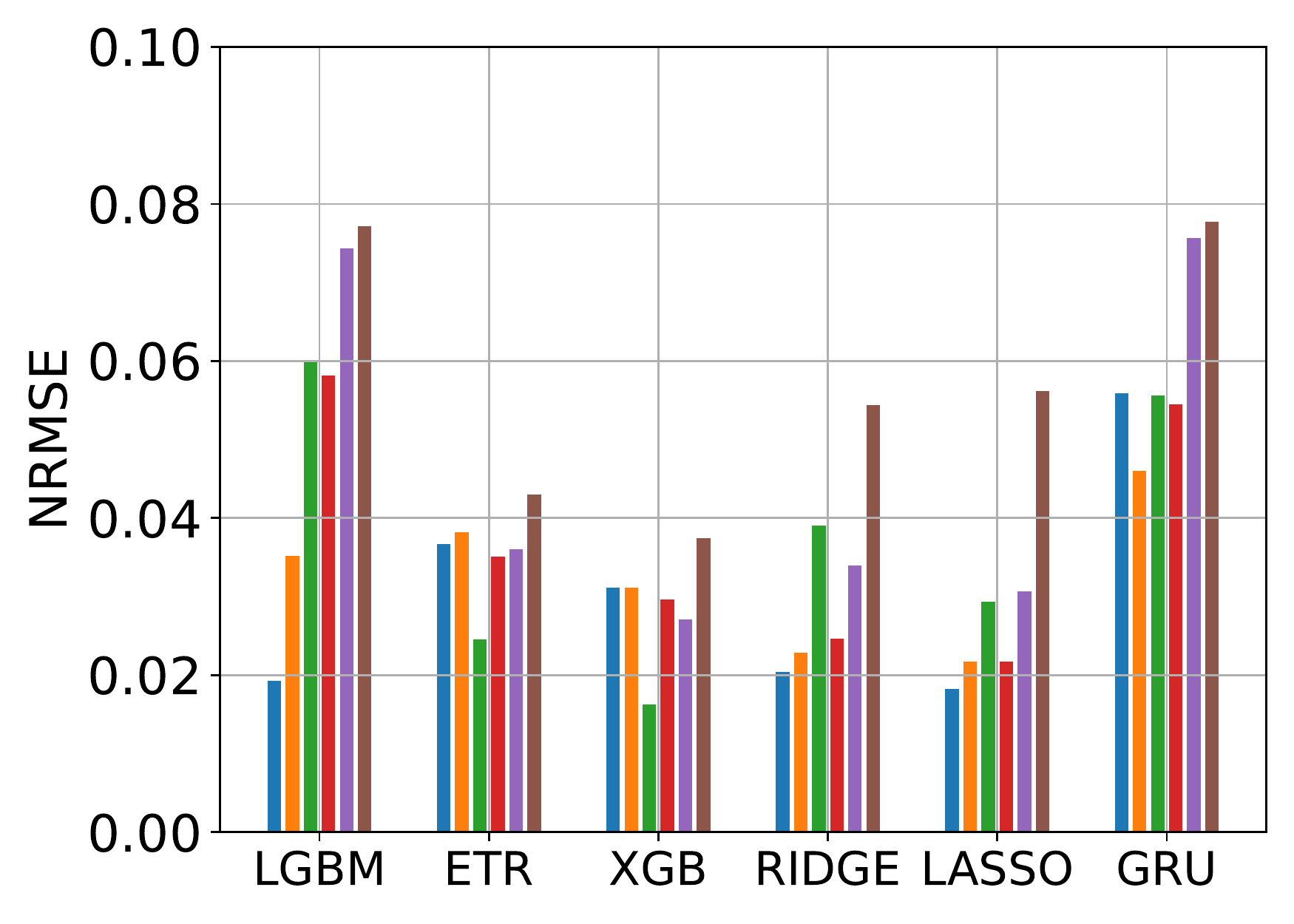}}\\ \vspace{-.2cm}
    \subfloat[\tiny Covid-CalFt \label{fig:covidcalft}]{\includegraphics[width=.24\textwidth]{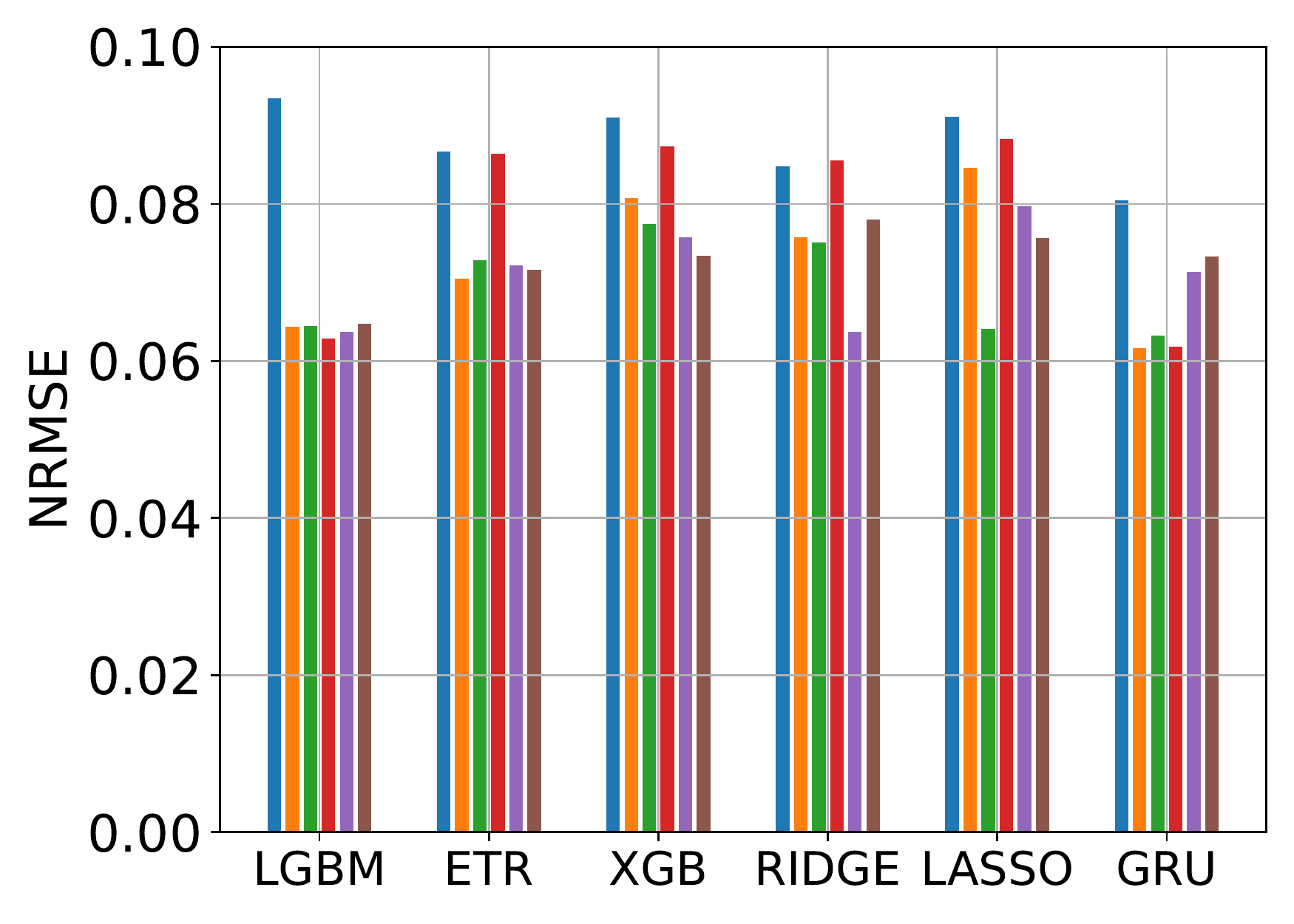}}
    \subfloat[\tiny Covid-CalFt,Oil \label{fig:covidcalftoil}]{\includegraphics[width=.24\textwidth]{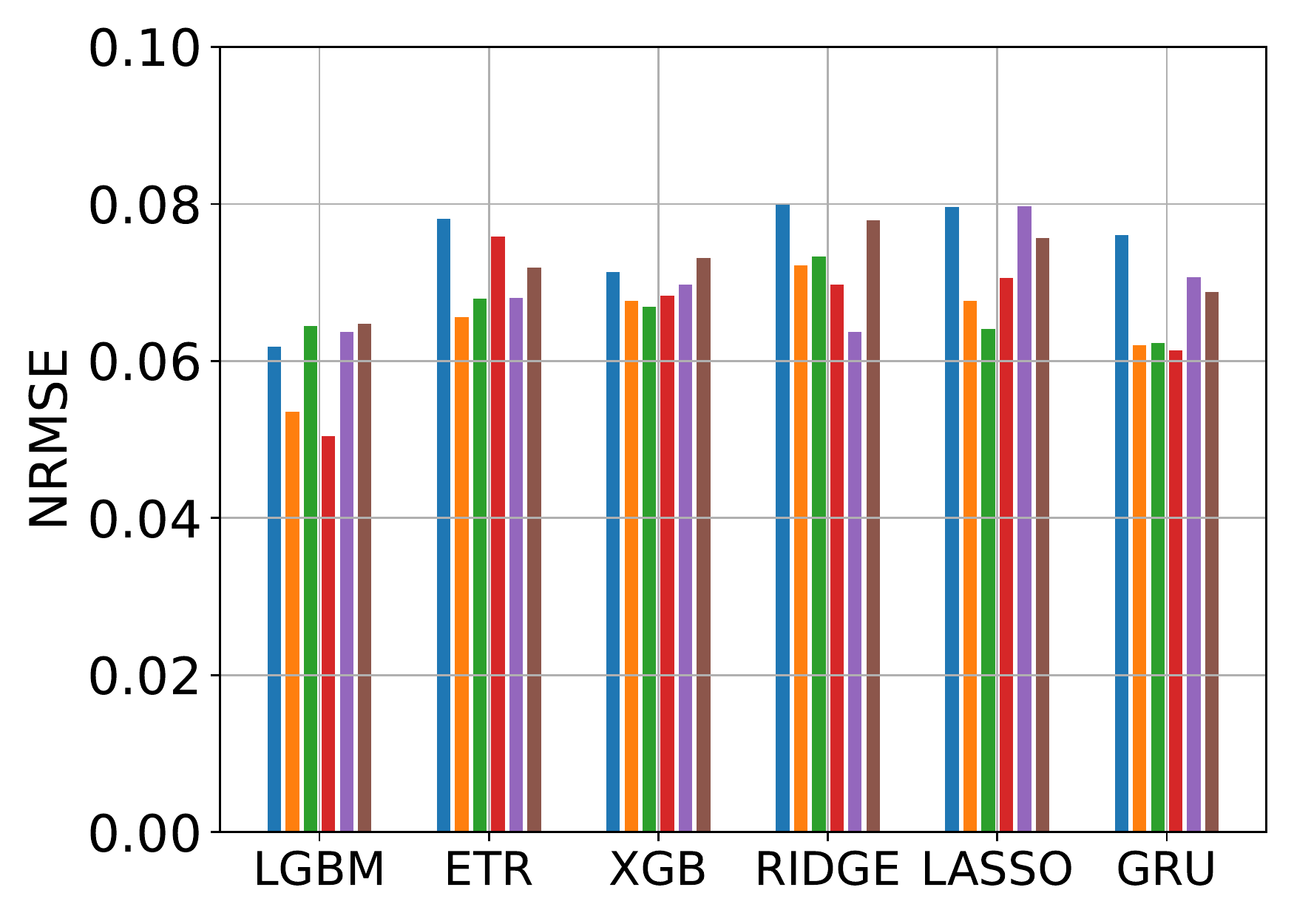}}
    \subfloat[\tiny Covid-CalFt,Oil,Gold \label{fig:covidcalftoilgold}]{\includegraphics[width=.24\textwidth]{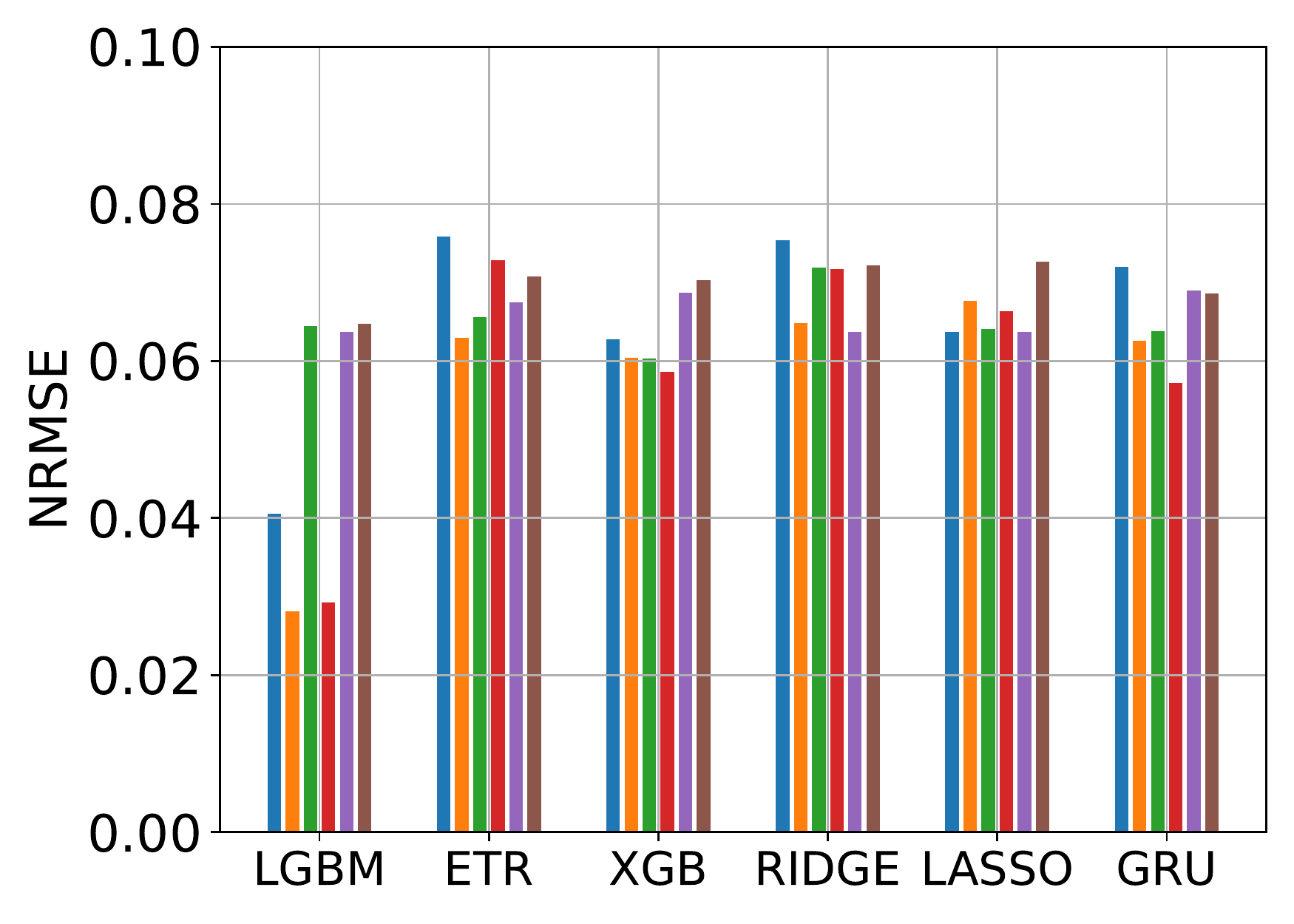}}
    \subfloat[\tiny Covid-CalFt,Oil,Gold,TSX \label{fig:covidcalftoilgoldtsx}]{\includegraphics[width=.24\textwidth]{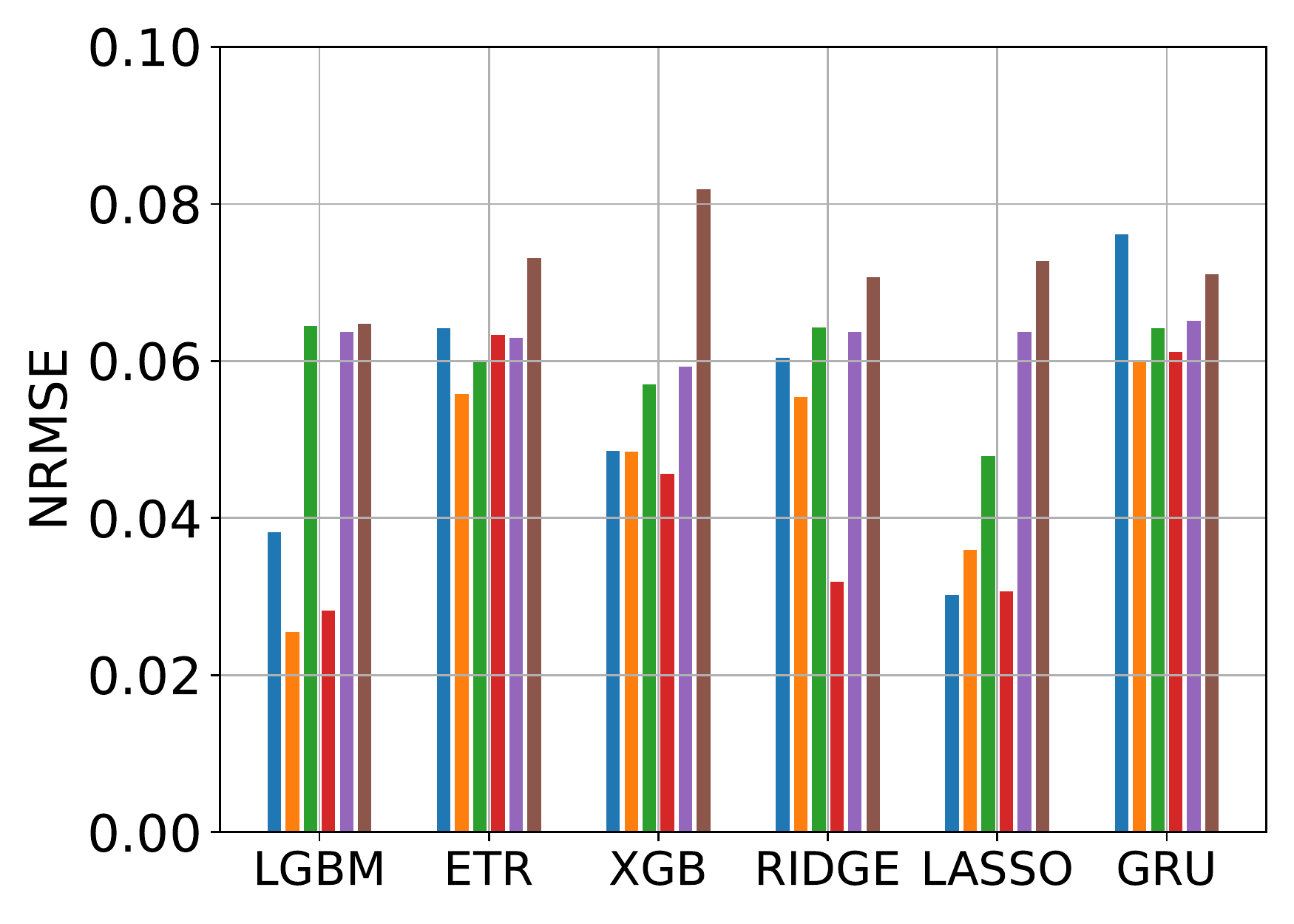}} \\ \vspace{.3cm}
    \caption{Result of ablation study. These figures show the incremental effect of incorporating important variables into a simple model that includes only time-specific data as dummy inputs. (CalFt: Calendar Feature, EcoE: Economic Expansion, EcoS: Economic Stagnation). Note that in Subplot~\ref{fig:ecorcalft}, the values for XGB, RIDGE, and LASSO exceed the y-axis upper value of 0.10.}
    \label{fig:factors}
\end{figure}

%%%%%%%%%%%%%%%%%%%%%%%%%%%%%%%%%%%%%%%%%%%%%%%%%%%%%%%%%%%%%%%%%%%%%%%%%%%%%%%%%%%%%%%%%%%%%%%%%%%%%%%%%%%%%%%%%%%%%%%%%%
We next investigate the isolated effect of the most important variables -- as indicated by feature importance and SHAP values -- and their incremental contribution to the prediction performance. 
Before incorporating the important variables into the input set, we first consider a baseline model. 
This model provides a simple benchmark to which we compare the performance of the models that leverage the information from other variables. 
The baseline model consists of calendar features as baseline variables. 
Hence, for daily frequency, the input of the baseline model includes day-of-week, week-of-month, and month-of-year, and for weekly frequency, these are week-of-month and month-of-year. 
Afterward, we add three variables (crude oil, gold, and TSX) to the input set to understand their effects on prediction performance.
Figure~\ref{fig:factors} shows the results from this analysis. 
The subplots in rows and columns of this figure correspond to different periods/datasets and input sets, respectively. 
For example, the first row contains the subplots for the dataset ``All (2009-2021)'' and the first column includes the subplots for the baseline model with dummy calendar features (CalFt). 
In other columns, we add crude oil, gold, and TSX cumulatively to the baseline model (e.g., ``CalFt, Oil'' in the second column). In each subplot, the NRMSE values are separately presented for the prediction models, and the grouped bars on each model name correspond to daily and weekly data with different forecast horizons.

In Figure~\ref{fig:allcalft}, which shows the results for the baseline model for ``All (2009-2021)'' dataset, we observe that the performance of all prediction models is similar to an NRMSE value of around 0.06 on average. 
Dummy variables lead to better performance in the ``Economic Expansion (2009-2011)'' dataset compared to other subperiods (i.e., the NRMSE values in Figure~\ref{fig:ecoecalft} are lower compared to Figures~\ref{fig:ecorcalft} and~\ref{fig:covidcalft}).
The second column of the subplots indicates the incremental effect of incorporating crude oil into the input set. 
Figure~\ref{fig:allcalftoil} shows that crude oil information improves the prediction power of the models for ``All (2009-2021)'' dataset and reduces the error to the low value of 0.04 in some cases, such as the 1-day forecast horizon by LGBM. 
However, considering crude oil in the input decreases the forecasting error in the ``Economic Expansion (2009-2011)'' and ``Economic Stagnation (2014-2016)'' periods compared to the corresponding baseline model, its effect is negligible during the ``Covid (2019-2021)'' period. 
This implies that, during the last subperiod, the prediction performance is impacted to a higher degree by other macroeconomic variables.

In the third column of the subplots, we consider additional information about gold in the input set. 
Our findings indicate that the difference in performance for ``All (2009-2021)'' is not significant compared to the model that only includes crude oil. 
However, gold prices deteriorate the forecasts during the ``Economic Stagnation (2014-2016)'' period, and it improves the results only slightly during ``Economic Expansion (2009-2011)'' and ``Covid (2019-2021)'' periods, especially when using LGBM (e.g., compared to LGBM in Figures~\ref{fig:covidcalftoilgold} and~\ref{fig:covidcalftoil}).
In the fourth column of the subplots, we add the information about TSX to the input set. 
We observe that the errors do not decrease in the ``All (2009-2021)'' and ``Economic Expansion (2009-2011)'' datasets. 
On the other hand, TSX has explanatory power during the ``Economic Stagnation (2014-2016)'' period, especially when using ETR, XGB, RIDGE, and LASSO models (i.e., comparing Figures~\ref{fig:ecorcalftoilgoldtsx} and~\ref{fig:ecorcalftoilgold}). Similarly, in the ``Covid (2019-2021)'' period, the RIDGE and LASSO models benefit from TSX information more than other models in 1-day and 1-week forecast horizons.

Overall, these results demonstrate that the isolated effect of important macroeconomic variables in prediction accuracy mainly depends on the period of study. 
In particular, in the expansionary period, taking only these three top features into account barely leads to information loss as the performance is close to those obtained in Tables~\ref{tab:dailyPrediction} and~\ref{tab:weeklyPrediction}. However, for the ``Covid (2019-2021)'' dataset, this effect is not consistent and the information loss depends on the model and forecast horizon.

%%%%%%%%%%%%%%%%%%%%%%%%%%%%%%%%%%%%%%%%%%%%%%%%%%%%%%%%%%%%%%%%%%%%%%%%%%%%%%%%%%%%%%%%%%%%%%%%%%%%%%%%%%%%%%%%%%%%%%%%%%
%%%%%%%%%%%%%%%%%%%%%%%%%%%%%%%%%%%%%%%%%%%%%%%%%%%%%%%%%%%%%%%%%%%%%%%%%%%%%%%%%%%%%%%%%%%%%%%%%%%%%%%%%%%%%%%%%%%%%%%%%%
\section{Conclusion}\label{sec:conclusion}
%%%%%%%%%%%%%%%%%%%%%%%%%%%%%%%%%%%%%%%%%%%%%%%%%%%%%%%%%%%%%%%%%%%%%%%%%%%%%%%%%%%%%%%%%%%%%%%%%%%%%%%%%%%%%%%%%%%%%%%%%%
%%%%%%%%%%%%%%%%%%%%%%%%%%%%%%%%%%%%%%%%%%%%%%%%%%%%%%%%%%%%%%%%%%%%%%%%%%%%%%%%%%%%%%%%%%%%%%%%%%%%%%%%%%%%%%%%%%%%%%%%%%

In an uncertain economic environment, policymakers and economists regularly face ambiguity in their expectations of exchange rate movements. 
Neglecting the underpinnings and knowledge of the models' outputs might lead to false decisions, especially with respect to the dynamics of the relationships.
In this study, we consider an interpretative machine learning approach to extend the traditional exchange rate modelling and employ a wide variety of methods from recent machine learning literature. 
We assess the performance of machine learning techniques subject to theory consistency with statistical analysis and economic interpretation.  

We specifically focus on modelling the Canadian exchange rate dynamics by capturing the impact of crude oil prices along with other macroeconomic indicators such as PPI and money supply.
The results from our detailed empirical analysis show that tree-based models outperform other methods on average, and new linear models excel in explaining the relationships during the recent pandemic. 
We efficiently address contemporaneous issues with the relationships among the input features and the target variable and find a strong link between in-sample and out-of-sample analyses. 
The contributions of variables to the exchange rate are also found to be time-varying and consistent with significant events in the related markets and economic trends.  
%The exchange rate forecasts are strongly influenced by oil prices, with time-varying effects that are consistent with major events in the oil market and the main evolution of the oil trend in Canada. Gold and the stock market index, along with oil, are the most important variables and make a strong link between in-sample and out-of-sample analysis. 
To this end, we show how the theories and empirical evidence verify the accuracy of the interpretability methodologies, and also how the interpretations improve the statistical prediction performance of the models. 
Overall, accurate interpretations with respect to both historical observations and future predictions enhance the practicality and reliability of machine learning in financial and economic modelling. 
% However, its credibility relies on being cautiously justified by theoretical considerations. 

This work can be extended in different directions.
First, the methodology considered in our analysis can be applied to other exchange rate forecasting tasks, especially those of developing countries that tend to be visibly impacted by economic shocks and major global affairs.
Interpretations provided by the explainability methods can help validate the machine learning models in those settings as well.
Secondly, time series-modeling specific interpretability methods can be employed to achieve a higher level of interpretability and connection among input features and the outputs.
Lastly, by considering a larger collection of exchange rate datasets from a variety of countries, global forecasting methods can be employed to achieve interpretability on a macro level.

\bibliographystyle{chicago} % outcomment this and next line in Case 1
\bibliography{ref} % if more than one, comma separated

\end{document}